\documentclass[12pt]{article}
\usepackage{report}
\title{\textbf{Topological Vibration Analysis of Elastic Lattices via Bloch Sphere Mapping}}

\author{Kazi T. Mahmood}
\author{M. Arif Hasan\thanks{Corresponding author: \texttt{\href{mailto:hasan.arif@wayne.edu}{hasan.arif@wayne.edu}}}}
\affil{Department of Mechanical Engineering, Wayne State University, Detroit, MI 48202}
\affil{\texttt{\href{mailto:kazi.mahmood@wayne.edu}{kazi.mahmood@wayne.edu}, \href{mailto:hasan.arif@wayne.edu}{hasan.arif@wayne.edu}}}

\date{} 

\begin{document}
\maketitle
\begin{abstract}
    Mechanical lattices support topological wave phenomena governed by geometric phases. We develop a compact Hilbert space description for one-dimensional elastic chains, expressing intra-cell motion as a normalized superposition of orthogonal eigenstates and tracking complex amplitudes as trajectories on a Bloch sphere. For diatomic lattices, this construction makes inversion symmetry protection explicit: the relative phase between in-phase and out-of-phase modes is piecewise locked, and the Zak phase is quantized with band-dependent jumps at symmetry points. Extending the framework to triatomic lattices shows that restoring inversion retains quantization, whereas breaking it dequantizes the geometric phase while leaving the spectrum origin invariant. Viewing norm-preserving transformations of the modal coefficient pair as Bloch-sphere rotations, we demonstrate classical analogues of single-qubit logic gates: a $\pi$-phase rotation about a transverse axis swaps the modal poles, and a longitudinal-axis phase flip maps balanced superpositions to their conjugates. These gate-like operations are realized by controlled evolution across wavenumber space and can be driven or reprogrammed through spatiotemporal stiffness modulation. Introducing space-time modulation hybridizes carrier and sideband harmonics, producing continuous phase winding and open-path geometric phases accumulated along the Floquet trajectory. Across static and modulated regimes, the framework unifies algebraic and geometric viewpoints, is robust to gauge and basis choices, and operates directly on amplitude–phase data. Results clarify how symmetry, modulation, and topology jointly govern dispersion, modal mixing, and phase accumulation, providing tools to analyze and design vibration and acoustic functionalities in engineered structures.
\end{abstract}

\section{Introduction}

Topological mechanics has revealed that mechanical lattices—assemblies of masses and springs obeying Newton’s laws—can emulate the robust wave phenomena long associated with electronic, photonic, and atomic systems \cite{deymier2017sound,huber2016}. A central organizing idea is the geometric phase accumulated by eigenstates as a control parameter (e.g., the Bloch wavenumber) is varied. In crystalline solids, the one-dimensional (1D) Berry—or Zak—phase is quantized (to $0$ or $\pi$) by inversion symmetry and encodes bulk information with interfacial consequences, establishing an early bridge between band topology and material response \cite{PhysRevLett.62.2747,PhysRevLett.89.077002}. Within classical and quantum wave systems alike, the geometric phase extends beyond the original adiabatic, cyclic setting: Samuel and Bhandari showed that neither cyclic evolution nor unitarity is required, enabling the definition of open-path phases determined by the Pancharatnam connection between nonorthogonal states \cite{PhysRevLett.60.2339}. These ideas have catalyzed a broader program in topological phononics and mechanics, where symmetry-protected transport, boundary localization, and defect-bound modes have been designed and observed in architected structures \cite{kane2014,mousavi2015}.

In this context, elastic mass-spring chains serve a transparent platform to investigate how symmetry, dispersion, and modulation collaborate to create nontrivial topology. For 1D bands, inversion symmetry constrains the Zak phase to 0 or $\pi$, making it a sensitive diagnostic of how a unit cell is arranged and where its origin is chosen \cite{PhysRevLett.62.2747,PhysRevB.108.035403}. Recent work has also clarified that time-dependent, space–time modulated lattices break reciprocity and hybridize carrier harmonics with Floquet sidebands, enriching dispersion and enabling direction-dependent gaps \cite{doi:10.1098/rspa.2017.0188}. Together, these developments lead to a description that treats state geometry—amplitudes and phases—as primary entities, together with the eigenfrequencies themselves.

Our earlier studies introduced the Spectral Analysis of Amplitudes and Phases (SAAP) method, which uses molecular-dynamics (MD) simulations under tailored initial conditions to extract complex modal amplitudes, phases, and associated Berry (Zak) phases in periodic elastic superlattices \cite{10.1121/1.5114911,HASAN2019114843}. SAAP revealed how inversion-symmetric unit cells enforce Berry-phase quantization, and how varying spring arrangements or changing the unit-cell origin toggles between 0 and $\pi$ without affecting the spectrum. Operationally, SAAP first determines the band structure $\omega_j(k)$ and then, in a second MD pass with traveling-wave initial conditions, projects the displacement field onto plane waves to obtain band-indexed complex amplitudes $a_n^j(k)$; a normalized amplitude vector per band is then used to compute a discrete Berry connection and the Zak phase. Consequently, SAAP yields geometric phases and amplitude profiles indexed by band over the Brillouin zone but does not provide compact, gauge-robust state-coefficient trajectories that expose how a single lattice state navigates an internal basis across $k$ or time; it also presumes periodicity and closed paths in wavenumber $k$, making open-path phases and space–time modulation cumbersome to treat. The present manuscript addresses these limitations directly. We recast intra-cell motion as a normalized superposition of orthogonal eigenstates and use the resulting complex coefficients as intrinsic coordinates for the state on a Bloch sphere. In this representation, phases and amplitude hierarchies become geometry; Berry/Zak phases appear as windings of state trajectories; and the effects of symmetry and modulation are read off from paths on $S^2$. Thus, our approach differs from common eigenvector-inner-product routes. Rather than inferring Berry/Zak phases indirectly, we construct the state coefficients directly and map their evolution to the Bloch sphere. This “algebraic (coefficients)–geometric (sphere trajectories) unification” makes symmetry-enforced plateaus, jumps, and windings in phases geometrically transparent and provides a gauge-invariant way to treat static and time-modulated lattices within a single language. By moving beyond band-level phase extraction to explicit, gauge-robust state-coefficient trajectories on the Bloch sphere, we provide a compact, symmetry-aware, and experimentally tractable framework that links intra-cell state geometry to Berry/Zak and open-path phases in both static and space–time-modulated elastic chains.

The remainder of the paper is organized as follows. Section \ref{elastic_lattice} defines the mass–spring lattice, boundary conditions, and unit-cell choices underpinning the analysis. Section \ref{lattice_with_time_Indep_Stiff} develops the Bloch-sphere formulation for time-independent lattices, highlighting inversion-symmetry-protected quantization in diatomic cells and its controlled breakdown in triatomic cells. Section \ref{Time Dependent Stiffness} extends the framework to space–time modulated lattices, demonstrating open-path geometric phases and modulation-induced windings consistent with Floquet hybridization. Section \ref{Conclusion} concludes by situating the contributions relative to the broader literature on space–time modulation and topological mechanics.

\section{Elastic Lattice Under Consideration}
\label{elastic_lattice}
The objective of this study is to represent the eigenstates that carry Berry phases as classical superpositions of normal modes in a one-dimensional elastic lattice. The lattice comprises identical point masses connected by linear (harmonic) springs in a periodic chain. Each lattice contains $N_{c}$ unit cells; each unit cell contains $N_{m}$ masses. We impose Born-von Kármán boundary conditions so that $e^{i k L N_{c}}=1$, where $L$ is the unit-cell length and $k$ is the Bloch wave number. Consequently, the first Brillouin zone is discretized as $k \in[-\pi, \pi)$ with spacing $2 \pi / N_{c}$. The Brillouin zone is the reciprocal-space domain of unique wave vectors that encode the periodic wave dynamics.

Two prototypical cells are analyzed in detail: Diatomic cell $\left(N_{m}=2\right)$---the minimal configuration that admits acoustic and optical bands. Triatomic cell $\left(N_{m}=3\right)$---the simplest system in which Berry-phase quantization can be lifted when inversion symmetry is broken. Throughout, we treat time-independent and, later in the manuscript, time-dependent stiffness modulations; the former builds the foundation for the superposition formalism and its geometric interpretation on the Bloch sphere, while the latter extends the approach beyond strict periodicity. The latter case extends the accessible Hilbert-space trajectories from a great-circle arc to the full Bloch sphere, thereby demonstrating a classical analogue of qubit control.

\section{Lattice with Time-Independent Stiffness}
\label{lattice_with_time_Indep_Stiff}
\subsection{Diatomic Unit Cell}
\label{diatomic unit}
We consider a diatomic unit cell composed of identical masses coupled by nearest-neighbor springs of stiffness $\psi_{1}$ and $\psi_{2}$, arranged periodically and subject to Born-von Kármán boundary conditions. Our aim is twofold: (i) to obtain closed-form acoustic/optical traveling-wave solutions and their complex amplitudes, and (ii) to recast the lattice motion as a normalized superposition of two orthogonal eigenstates, thereby exposing how inversion symmetry fixes the relative phase (0 or $\pi$) across the Brillouin zone---the 1D Zak-phase quantization that we later visualize on a Bloch sphere. The quantization of the Zak phase in inversion-symmetric 1D lattices and its origin dependence are well established and provide the topological backdrop for our superposition analysis \cite{PhysRevLett.62.2747}.

\subsubsection{Traveling-Wave Ansatz and Dispersion}

Let $u_{1}(t)$ and $u_{2}(t)$ denote the displacements of the two masses within a given unit cell $N_{i}$. The linear equations of motion are
\begin{equation}
\begin{aligned}
m \ddot{u}_{1, N_{i}}(t) &= \psi_{2}\left[u_{2, N_{i-1}}(t)-u_{1, N_{i}}(t)\right] - \psi_{1}\left[u_{1, N_{i}}(t)-u_{2, N_{i}}(t)\right], \\
m \ddot{u}_{2, N_{i}}(t) &= \psi_{1}\left[u_{1, N_{i}}(t)-u_{2, N_{i}}(t)\right] - \psi_{2}\left[u_{2, N_{i}}(t)-u_{1, N_{i+1}}(t)\right]
\end{aligned}
\label{eq1}
\end{equation}
We seek Bloch traveling-wave solutions with the compact ansatz (periodic in reciprocal space).
\begin{equation}
u_{1, N_{i}}(t) = A_{1} e^{\mathrm{i}\left(k N_{i} L + \omega t\right)}, \quad u_{2, N_{i}}(t) = A_{2} e^{\mathrm{i}\left(k N_{i} L + \omega t\right)}.
\label{eq2}
\end{equation}

Substitution of (\ref{eq2}) into (\ref{eq1}) yields the dispersion relation and the band-dependent complex amplitude ratio. In particular, the amplitudes satisfy,
\begin{equation}
\frac{A_{1}}{A_{2}} = \pm \frac{\sqrt{\tau}}{\sqrt{\tau^{*}}}, \quad \tau = \psi_{1} + \psi_{2} e^{-\mathrm{i} k L}
\label{eq3}
\end{equation}
and the acoustic/optical branch selection can be expressed compactly as \cite{10.1121/1.5114911}:
\begin{equation*}
\begin{aligned}
& \text{Acoustic branch:} \quad \begin{pmatrix} A_{1} \\ A_{2} \end{pmatrix} = \begin{pmatrix} \sqrt{\tau} \\ \sqrt{\tau^{*}} \end{pmatrix}, \\
& \text{Optical branch:} \quad \begin{pmatrix} A_{1} \\ A_{2} \end{pmatrix} = \begin{pmatrix} \sqrt{e^{-\mathrm{i} \pi} \tau} \\ \sqrt{e^{+\mathrm{i} \pi} \tau^{*}} \end{pmatrix}.
\end{aligned}
\end{equation*}
Together with the familiar acoustic $(+)$ and optical $(-)$ branches dispersion relation:
\begin{equation*}
\omega_{\pm}^{2} = \frac{\psi_{1} + \psi_{2}}{m} \left[ 1 \pm \sqrt{1 - \frac{4 \psi_{1} \psi_{2}}{(\psi_{1} + \psi_{2})^{2}} \sin^{2} \frac{k L}{2}} \right].
\end{equation*}

These are the standard diatomic chain results cast in the compact ansatz; they are unitarily equivalent to those obtained with a general ansatz and therefore will yield identical Berry connections and Berry phases, consistent with prior analyses of ansatz invariance \cite{10.1121/1.5114911}.

\subsubsection{Superposition Basis and Bloch Sphere Representation}

Because the dynamics are linear, the two independent eigenstates within a unit cell may be chosen orthonormally as an in-phase mode 
$\ket{E_{1}} = \frac{1}{\sqrt{2}} \begin{pmatrix} 1 \\ 1 \end{pmatrix}$ and an out-of-phase mode $\ket{E_{2}} = \frac{1}{\sqrt{2}} \begin{pmatrix} 1 \\ -1 \end{pmatrix}$ (a convenient pseudospin basis). The cell-wise displacement vector can then be written as a normalized superposition,
\begin{equation}
\vec{U} \equiv \begin{bmatrix} u_{1} \\ u_{2} \end{bmatrix} = \begin{bmatrix} A_{1} \\ A_{2} \end{bmatrix} e^{\mathrm{i}\left(k N_{i} L + \omega t\right)} \equiv \left( \alpha \ket{E_{1}} + \beta \ket{E_{2}} \right) e^{\mathrm{i} \omega t}
\label{eq4}
\end{equation}
with complex coefficients $\alpha, \beta$. Using (\ref{eq3})–(\ref{eq4}) and the relation $A_{2} = A_{1}^{*}$, one convenient normalization leads to
\begin{equation}
\begin{aligned}
\hat{\alpha} &= \frac{1}{\sqrt{(|A_1|)^2+(|A_1^*|)^2}} \frac{1}{\sqrt{2}} \left(A_{1} + A_{1}^{*}\right) e^{\mathrm{i} k N_{i} L}, \\
\hat{\beta} &= \frac{1}{\sqrt{(|A_1|)^2+(|A_1^*|)^2}} \frac{1}{\sqrt{2}} \left(A_{1} - A_{1}^{*}\right) e^{\mathrm{i} k N_{i} L}.
\end{aligned}
\label{eq5}
\end{equation}
so that

\begin{equation*}
\arg(\hat\alpha) - \arg(\hat\beta) = \arg \left[ \frac{1}{\sqrt{2}} (A_{1} + A_{1}^{*}) e^{\mathrm{i} k N_{i} L} \right] - \arg \left[ \frac{1}{\sqrt{2}} (A_{1} - A_{1}^{*}) e^{\mathrm{i} k N_{i} L} \right].
\end{equation*}

Here the coefficients $(\hat{\alpha},\hat{\beta})$ are normalized using the conventional $L^2$ condition $|\hat{\alpha}|^2 + |\hat{\beta}|^2 = 1$, consistent with standard state‐vector representations in quantum mechanics.

Because $A_{1}+A_{1}^{*}$ is real,
$
\arg\!\Big[\tfrac{1}{\sqrt{2}}(A_{1}+A_{1}^{*})e^{\mathrm{i}kN_iL}\Big]
  = kN_iL + \arg\!\big(\mathbb{R}(A_1+A_1^*)\big),
$
whose argument is $0$ when $\mathbb{R}(A_1+A_1^*)>0$ and $\pi$ when
$\mathbb{R}(A_1+A_1^*)<0$. Likewise, because $A_{1}-A_{1}^{*}$ is purely
imaginary,
$
\arg\Big[\tfrac{1}{\sqrt{2}}(A_{1}-A_{1}^{*})e^{\mathrm{i}kN_iL}\Big]
  = -\tfrac{\pi}{2}\,\operatorname{sgn}\!\bigl(\mathrm{i}(A_{1}-A_{1}^{*})\bigr)
    + kN_iL.
$

Two consequences follow. First, inversion symmetry restricts the relative phase between $\hat\alpha$ and $\hat\beta$ to $0$ or $\pi$ across the Brillouin zone, directly reflecting the quantized Zak phase of inversion-symmetric 1D lattices. Second, the assignment of $\arg(\hat\alpha)-\arg(\hat\beta)\in\{0,\pi\}$ is branch dependent: for the acoustic branch the phase follows the $k$-parity dictated by~(\ref{eq4}), with $0$ for $k<0$ and $\pi$ for $k>0$, while the optical branch exhibits the complementary pattern. These selection rules rationalize the observed $0/\pi$ Berry phases for $\psi_{1}\gtrless\psi_{2}$ and are consistent with the invariance of Berry quantities under unitary changes of ansatz and origin\cite{10.1121/1.5114911}.

Interpreting $(\hat\alpha,\hat\beta)$ as a normalized two-component state allows the band-resolved evolution to be represented on a Bloch sphere, providing a compact geometric picture of the inversion-enforced phase locking and its corresponding $0/\pi$ Zak topology. This viewpoint connects directly to established two-mode wave frameworks \cite{PhysRevLett.62.2747,ALTEPETER2005105,PhysRevA.86.033830} and will serve as the basis for visualizing state trajectories and for extending the analysis to cases with broken periodicity or time-dependent stiffness.

These coefficients ($\alpha,\beta$) can be realized experimentally in classical elastic networks. Previous experimental studies have already demonstrated that the coefficients of the superposed modal states $(\alpha,\beta)$ can be directly measured in coupled masses \cite{Mahmood2022ExperimentalStates}. The responses of the individual masses were recorded in the time domain and processed using FFT to obtain the complex amplitudes and phases of the dominant spectral components. Projecting these experimentally measured components onto the orthonormal basis $\{\ket{E_{1}},\ket{E_{2}}\}$ yields the state coefficients in the same manner used in the present Bloch sphere analysis. The identical approach can be implemented in multi-cell systems using a scanning laser Doppler vibrometer to acquire non-contact measurements across the structure. Therefore, the Bloch sphere trajectories and the associated Hilbert space evolution described in this section are fully accessible through experimental observation.

\subsubsection{Acoustic and Optical Branches: Explicit Coefficient Forms}

The explicit expressions for the superposition coefficients $\alpha$ and $\beta$ on the acoustic and optical branches provide a real-space view of how inversion symmetry constrains the relative phase of the two basis states.

\subsubsection*{Acoustic Branch}

For the first unit cell, the normalized superposition coefficients $(\hat\alpha, \hat\beta)$ that weight the in-phase and out-of-phase basis states are written in terms of the complex scalar $\tau$ as

\begin{equation}
\begin{pmatrix}\hat{\alpha}\\ \hat{\beta}\end{pmatrix}
=
\frac{1}{\sqrt{2\bigl(|\tau|^{2}+|\tau^{*}|^{2}\bigr)}}
\,e^{\mathrm{i}kL}
\begin{pmatrix}
\tau+\tau^{*}\\[2pt]
\tau-\tau^{*}
\end{pmatrix}
=
\frac{1}{2|\tau|}\,e^{\mathrm{i}kL}
\begin{pmatrix}
\tau+\tau^{*}\\[2pt]
\tau-\tau^{*}
\end{pmatrix}.
\label{eq6}
\end{equation}
In the following discussion, we introduce the stiffness-ordering parameter.
\begin{equation*}
\Delta \equiv \psi_{1} - \psi_{2}
\end{equation*}
so that $\Delta > 0$ corresponds to $\psi_{1} > \psi_{2}$ and $\Delta < 0$ to $\psi_{1} < \psi_{2}$. Define also the small offset from the Brillouin-zone (BZ) edge $k L = \pm \pi$ by $k L = \mp \pi + \varepsilon$ with $\varepsilon \rightarrow 0^{\pm}$.

Using the explicit form of $\tau$ from the two-mass solution and expanding at the BZ edges, one finds the following limiting expressions. First, for
\begin{equation*}
k L = -\pi + \varepsilon, \quad \varepsilon \rightarrow 0,
\end{equation*}
the coefficient vector reduces to
\begin{equation*}
\begin{pmatrix}\hat\alpha\\ \hat\beta\end{pmatrix}
=
\frac{e^{ikL}}{2\sqrt{|\Delta|}}
\begin{pmatrix}
\sqrt{\Delta e^{-i\varepsilon}}+\sqrt{\Delta e^{+i\varepsilon}}\\[2pt]
\sqrt{\Delta e^{-i\varepsilon}}-\sqrt{\Delta e^{+i\varepsilon}}
\end{pmatrix},
\qquad (\Delta\neq 0)
\end{equation*}
since $\left(1 + e^{-\mathrm{i}(-\pi + \varepsilon)}\right) \approx 0$. Taking $\varepsilon \rightarrow 0$ gives two cases:

\noindent\textbf{Case I} $\left(\Delta > 0\right.$, i.e., $\left.\psi_{1} > \psi_{2}\right)$:
\begin{equation*}
\begin{pmatrix} \hat\alpha \\ \hat\beta \end{pmatrix} = \begin{pmatrix} -1 \\ 0 \end{pmatrix}.
\end{equation*}
\textbf{Case II} $\left(\Delta < 0\right.$, write $\Delta = -\delta$ with $\delta > 0$, i.e., $\left.\psi_{1} < \psi_{2}\right)$:
\begin{equation*}
\begin{pmatrix} \hat\alpha \\ \hat\beta \end{pmatrix} = \begin{pmatrix} 0 \\ -\mathrm{i} \end{pmatrix}.
\end{equation*}
Similarly, for
\begin{equation*}
k L = \pi - \varepsilon, \quad \varepsilon \rightarrow 0,
\end{equation*}
one obtains
\begin{equation*}
\begin{pmatrix}\hat\alpha\\ \hat\beta\end{pmatrix}
=
\frac{e^{ikL}}{2\sqrt{|\Delta|}}
\begin{pmatrix}
\sqrt{\Delta e^{+i\varepsilon}}+\sqrt{\Delta e^{-i\varepsilon}}\\[2pt]
\sqrt{\Delta e^{+i\varepsilon}}-\sqrt{\Delta e^{-i\varepsilon}}
\end{pmatrix},
\qquad (\Delta\neq 0)
\end{equation*}
again because $\left(1 + e^{-\mathrm{i}(-\pi + \varepsilon)}\right) \approx 0$. Taking the limit $\varepsilon \rightarrow 0$:

\noindent\textbf{Case I} $\left(\Delta > 0, \psi_{1} > \psi_{2}\right)$:
\begin{equation*}
\begin{pmatrix} \hat\alpha \\ \hat\beta \end{pmatrix} = \begin{pmatrix} -1 \\ 0 \end{pmatrix}.
\end{equation*}
\textbf{Case II} $(\Delta < 0, \Delta = -\delta < 0)$:
\begin{equation*}
\begin{pmatrix} \hat\alpha \\ \hat\beta \end{pmatrix} = \begin{pmatrix} 0 \\ \mathrm{i} \end{pmatrix},
\end{equation*}
which differs from $\begin{pmatrix} 0 \\ -\mathrm{i} \end{pmatrix}$ by a global $\pi$ phase, i.e., a physically irrelevant gauge choice for the spinor.

On the acoustic branch, the phase of the ratio $\hat\alpha / \hat\beta$ is pinned to $0$ or $\pi$ on each side of the BZ, and flips upon crossing $k=0$. Crucially, the sign of $k$ (not the ordering of $\psi_{1}, \psi_{2}$) selects which of the two real relative phases is realized---this is the real-space manifestation of the inversion-protected, quantized Zak phase in 1D. In other words, the acoustic-branch spinor toggles between two diametrically opposite longitudes on the Bloch sphere as $k$ traverses the zone. This is exactly the $0/\pi$ pattern expected for inversion-symmetric lattices (SSH-type physics) \cite{PhysRevB.108.035403,asboth2016short,doi:10.1098/rspa.1984.0023}.

\subsubsection*{Optical Branch}

An analogous calculation applies to the optical branch, for which the expression reads
\begin{equation}
\begin{pmatrix}\hat\alpha\\ \hat\beta\end{pmatrix}
=\frac{e^{ikL}}{2\sqrt{|\tau|}}
\begin{pmatrix}
\sqrt{e^{-i\pi}\tau}+\sqrt{e^{-i\pi}\tau^{*}}\\[2pt]
\sqrt{e^{-i\pi}\tau}-\sqrt{e^{-i\pi}\tau^{*}}
\end{pmatrix},
\qquad \tau=\psi_1+\psi_2 e^{-ikL}.
\label{eq7}
\end{equation}

Evaluating the same edge limits:

\noindent For $k L = -\pi + \varepsilon, \varepsilon \rightarrow 0$,
\begin{equation*}
\begin{pmatrix} \hat{\alpha} \\ \hat{\beta} \end{pmatrix}
= \frac{e^{ikL}}{2\sqrt{|\Delta|}}
\begin{pmatrix}
\sqrt{\Delta e^{-i\pi}e^{-i\varepsilon}} + \sqrt{\Delta e^{+i\pi}e^{+i\varepsilon}} \\
\sqrt{\Delta e^{-i\pi}e^{-i\varepsilon}} - \sqrt{\Delta e^{+i\pi}e^{+i\varepsilon}}
\end{pmatrix},
\qquad (kL=-\pi+\varepsilon).
\end{equation*}

\noindent\textbf{Case 1} $\left(\Delta > 0, \psi_{1} > \psi_{2}\right)$:
\begin{equation*}
\begin{pmatrix} \hat{\alpha} \\ \hat{\beta} \end{pmatrix} = \begin{pmatrix} -\mathrm{i} \\ 0 \end{pmatrix}.
\end{equation*}

\noindent\textbf{Case 2} $\left(\Delta < 0\right.$, put $\Delta = -\delta, \delta > 0$, i.e., $\left.\psi_{1} < \psi_{2}\right)$:
\begin{equation*}
\begin{pmatrix} \hat{\alpha} \\ \hat{\beta} \end{pmatrix} = \begin{pmatrix} -1 \\ 0 \end{pmatrix}.
\end{equation*}

\noindent For $k L = \pi - \varepsilon, \varepsilon \rightarrow 0$, the same algebra yields:

\noindent\textbf{Case 1} $(\Delta > 0)$:
\begin{equation*}
\begin{pmatrix} \hat{\alpha} \\ \hat{\beta} \end{pmatrix} = \begin{pmatrix} -\mathrm{i} \\ 0 \end{pmatrix}.
\end{equation*}

\noindent\textbf{Case 2} $(\Delta < 0)$:
\begin{equation*}
\begin{pmatrix} \hat{\alpha} \\ \hat{\beta} \end{pmatrix} = \begin{pmatrix} -1 \\ 0 \end{pmatrix}.
\end{equation*}

Collecting the four edge limits for the acoustic and optical branches shows that the relative phase between the two coefficients is piecewise constant on each side of the Brillouin zone and flips only at $kL = 0$. Thus the spinor $(\hat\alpha,\hat\beta)$ lies on a fixed great-circle segment of the Bloch sphere for $k<0$ or $k>0$, undergoing a discrete azimuthal jump at the zone center. This is the defining hallmark of inversion-protected $0/\pi$ Zak phases in one dimension and reflects the familiar gauge freedom that exchanges the two quantized values under a shift of the unit-cell origin \cite{asboth2016short,PhysRevLett.127.147401}.

The explicit coefficient forms make the classical two-state structure fully transparent: each band is represented by a normalized spinor $\hat\alpha\lvert E_1\rangle+\hat\beta\lvert E_2\rangle$, whose evolution on the Bloch sphere encodes the Berry/Zak phase as $k$ winds the Brillouin zone. This geometric representation, standard in qubit and two-level wave systems, provides a clean classical analogue and underpins the Bloch-sphere visualizations used throughout the paper\cite{PhysRevLett.62.2747,ALTEPETER2005105,PhysRevA.86.033830}.

In inversion-symmetric 1D lattices, the Zak phase of an isolated band is restricted to $\{0,\pi\}$, with the two possibilities interchanged by moving the unit-cell origin. The limiting spinors derived here reproduce exactly this behavior and connect the present elastic model to the SSH-family topological classification and its recent acoustic and photonic realizations \cite{PhysRevB.108.035403,asboth2016short,nano13243152}. The closed-form expressions for the zone-edge spinors therefore offer a compact algebraic route to the Zak class, complementing overlap-based Berry-phase methods, and supply the starting point for our subsequent geometric analysis of broken periodicity and time-dependent stiffness via non-cyclic phases \cite{doi:10.1098/rspa.1984.0023}.

\newcommand{\rowshiftleft}[2]{%
  \noindent\makebox[\linewidth][l]{\hspace*{#1}#2}%
}
\newlength{\panelH}
\setlength{\panelH}{0.2\linewidth}   
\newcommand{\panelbox}[1]{%
  \begin{minipage}[t][\panelH][t]{\linewidth}\centering #1\end{minipage}%
}

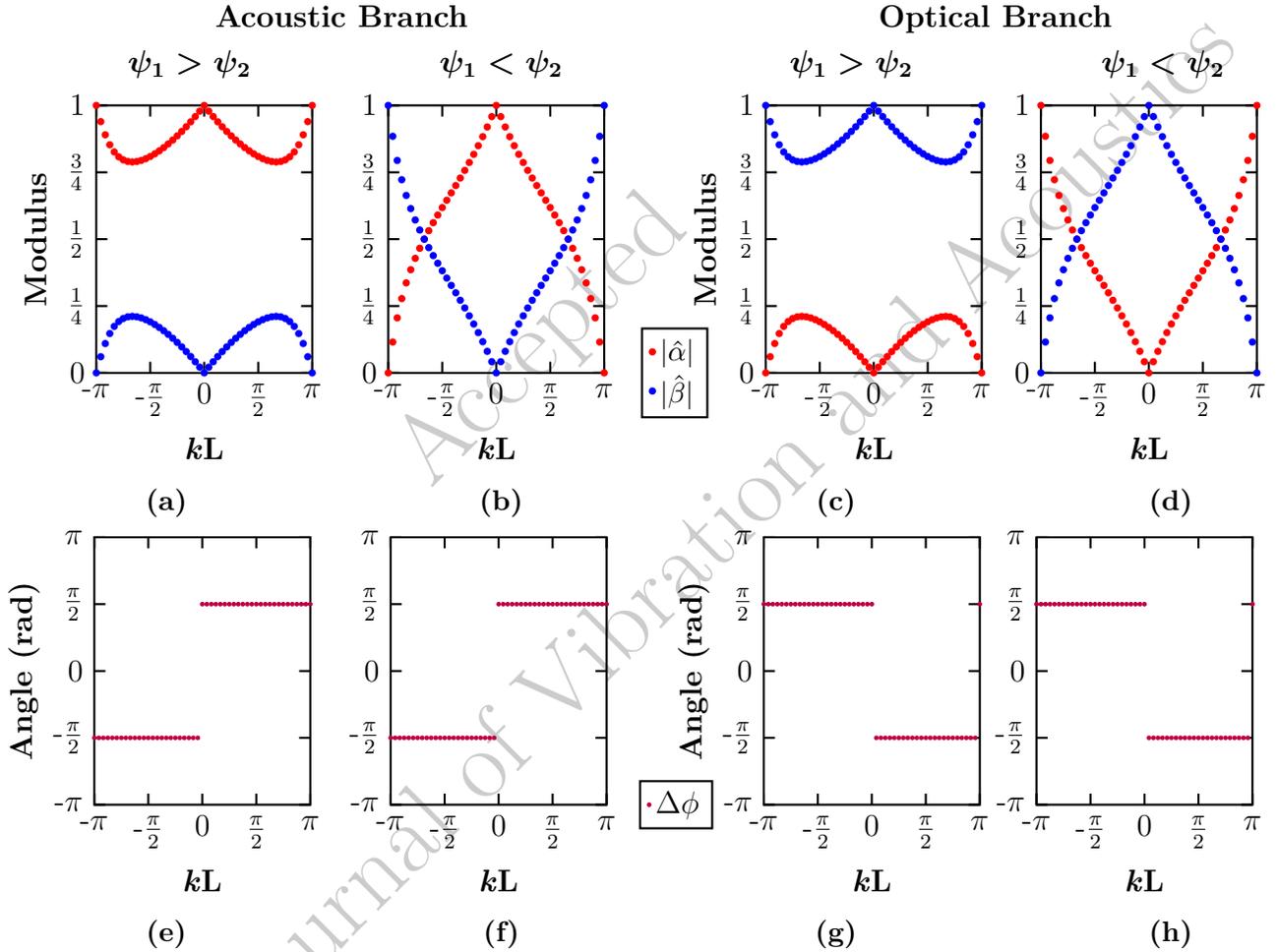
\begin{figure}[H]
\centering
\setlength{\tabcolsep}{6pt}
\renewcommand{\arraystretch}{1.2}

\begin{tabular}{@{}p{.24\textwidth} p{.24\textwidth} p{.24\textwidth} p{.24\textwidth}@{}}
  \multicolumn{2}{c}{\textbf{Acoustic Branch}} &
  \multicolumn{2}{c}{\textbf{Optical Branch}} \\
  \multicolumn{1}{c}{\boldmath$\quad\psi_1>\psi_2$} &
  \multicolumn{1}{c}{\boldmath$\psi_1<\psi_2$} &
  \multicolumn{1}{c}{\boldmath$\quad\psi_1>\psi_2$} &
  \multicolumn{1}{c}{\boldmath$\psi_1<\psi_2$} \\
  
  \begin{subfigure}[t]{\linewidth}
    \centering \panelbox{\begin{tikzpicture}
\begin{axis}[
    height=5.2cm,
width=4.5cm,
    thick,
    xlabel={\textbf{\textit{k}L}},
    ylabel={\textbf{Modulus}},
    xmin=-3.1416, xmax=3.1416,
    ymin=0, ymax=1,
    every x tick/.style={color=black, thick},
    xtick={-3.1416,-1.5708,0,1.5708,3.1416},
    xticklabels={-$\pi$,-$\frac{\pi}{2}$,0,$\frac{\pi}{2}$,$\pi$},
    every y tick/.style={color=black, thick},
    ytick={0,0.25,0.5,0.75,1},
    yticklabels={$0$,$\frac{1}{4}$,$\frac{1}{2}$,$\frac{3}{4}$,$1$},
]

\addplot[
    only marks,
    color=red,
    mark=*,
    mark size=1pt,
]
coordinates {
(-3.141592654,	1)
(-3.01069296,	0.939460433)
(-2.879793266,	0.890278163)
(-2.748893572,	0.852867432)
(-2.617993878,	0.82598335)
(-2.487094184,	0.807774976)
(-2.35619449,	0.796435532)
(-2.225294796,	0.790458496)
(-2.094395102,	0.788675135)
(-1.963495408,	0.790208217)
(-1.832595715,	0.794408113)
(-1.701696021,	0.800795582)
(-1.570796327,	0.809016994)
(-1.439896633,	0.818811158)
(-1.308996939,	0.829985283)
(-1.178097245,	0.842397667)
(-1.047197551,	0.855945181)
(-0.916297857,	0.870554138)
(-0.785398163,	0.886173513)
(-0.654498469,	0.90276981)
(-0.523598776,	0.920323021)
(-0.392699082,	0.938823305)
(-0.261799388,	0.958268065)
(-0.130899694,	0.978659157)
(0,	1)
(0.130899694,	0.978659157)
(0.261799388,	0.958268065)
(0.392699082,	0.938823305)
(0.523598776,	0.920323021)
(0.654498469,	0.90276981)
(0.785398163,	0.886173513)
(0.916297857,	0.870554138)
(1.047197551,	0.855945181)
(1.178097245,	0.842397667)
(1.308996939,	0.829985283)
(1.439896633,	0.818811158)
(1.570796327,	0.809016994)
(1.701696021,	0.800795582)
(1.832595715,	0.794408113)
(1.963495408,	0.790208217)
(2.094395102,	0.788675135)
(2.225294796,	0.790458496)
(2.35619449,	0.796435532)
(2.487094184,	0.807774976)
(2.617993878,	0.82598335)
(2.748893572,	0.852867432)
(2.879793266,	0.890278163)
(3.01069296,	0.939460433)
(3.141592654,	1)
};

\addplot[
    only marks,
    color=blue,
    mark=*,
    mark size=1pt,
]
coordinates {
(-3.141592654,	6.12323E-17)
(-3.01069296,	0.060539567)
(-2.879793266,	0.109721837)
(-2.748893572,	0.147132568)
(-2.617993878,	0.17401665)
(-2.487094184,	0.192225024)
(-2.35619449,	0.203564468)
(-2.225294796,	0.209541504)
(-2.094395102,	0.211324865)
(-1.963495408,	0.209791783)
(-1.832595715,	0.205591887)
(-1.701696021,	0.199204418)
(-1.570796327,	0.190983006)
(-1.439896633,	0.181188842)
(-1.308996939,	0.170014717)
(-1.178097245,	0.157602333)
(-1.047197551,	0.144054819)
(-0.916297857,	0.129445862)
(-0.785398163,	0.113826487)
(-0.654498469,	0.09723019)
(-0.523598776,	0.079676979)
(-0.392699082,	0.061176695)
(-0.261799388,	0.041731935)
(-0.130899694,	0.021340843)
(0,	0)
(0.130899694,	0.021340843)
(0.261799388,	0.041731935)
(0.392699082,	0.061176695)
(0.523598776,	0.079676979)
(0.654498469,	0.09723019)
(0.785398163,	0.113826487)
(0.916297857,	0.129445862)
(1.047197551,	0.144054819)
(1.178097245,	0.157602333)
(1.308996939,	0.170014717)
(1.439896633,	0.181188842)
(1.570796327,	0.190983006)
(1.701696021,	0.199204418)
(1.832595715,	0.205591887)
(1.963495408,	0.209791783)
(2.094395102,	0.211324865)
(2.225294796,	0.209541504)
(2.35619449,	0.203564468)
(2.487094184,	0.192225024)
(2.617993878,	0.17401665)
(2.748893572,	0.147132568)
(2.879793266,	0.109721837)
(3.01069296,	0.060539567)
(3.141592654,	6.12323E-17)
};
\end{axis}

\end{tikzpicture}}
    \caption{}\label{AC_M_1>2}
  \end{subfigure} &
  \begin{subfigure}[t]{\linewidth}
    \centering \panelbox{\begin{tikzpicture}
\begin{axis}[
    height=5.2cm,
width=4.5cm,
    thick,
    xlabel={\textbf{\textit{k}L}},
    xmin=-3.1416, xmax=3.1416,
    ymin=0, ymax=1,
    every x tick/.style={color=black, thick},
    xtick={-3.1416,-1.5708,0,1.5708,3.1416},
    xticklabels={-$\pi$,-$\frac{\pi}{2}$,0,$\frac{\pi}{2}$,$\pi$},
    every y tick/.style={color=black, thick},
    ytick={0,0.25,0.5,0.75,1},
    yticklabels={$0$,$\frac{1}{4}$,$\frac{1}{2}$,$\frac{3}{4}$,$1$},
    legend style={at={(1.17,0)}, anchor=west}, 
    legend cell align={left},
]

\addplot[
    only marks,
    color=red,
    mark=*,
    mark size=1pt,
]
coordinates {
(-3.141592654,	1.22465E-16)
(-3.01069296,	0.115463492)
(-2.879793266,	0.205782518)
(-2.748893572,	0.277783385)
(-2.617993878,	0.33654572)
(-2.487094184,	0.385811881)
(-2.35619449,	0.428283136)
(-2.225294796,	0.465886838)
(-2.094395102,	0.5)
(-1.963495408,	0.531618685)
(-1.832595715,	0.561478499)
(-1.701696021,	0.59013769)
(-1.570796327,	0.618033989)
(-1.439896633,	0.645523766)
(-1.308996939,	0.672909556)
(-1.178097245,	0.700460023)
(-1.047197551,	0.728425126)
(-0.916297857,	0.757048386)
(-0.785398163,	0.786577607)
(-0.654498469,	0.817275122)
(-0.523598776,	0.849428515)
(-0.392699082,	0.883362757)
(-0.261799388,	0.919454908)
(-0.130899694,	0.958152834)
(0,	1)
(0.130899694,	0.958152834)
(0.261799388,	0.919454908)
(0.392699082,	0.883362757)
(0.523598776,	0.849428515)
(0.654498469,	0.817275122)
(0.785398163,	0.786577607)
(0.916297857,	0.757048386)
(1.047197551,	0.728425126)
(1.178097245,	0.700460023)
(1.308996939,	0.672909556)
(1.439896633,	0.645523766)
(1.570796327,	0.618033989)
(1.701696021,	0.59013769)
(1.832595715,	0.561478499)
(1.963495408,	0.531618685)
(2.094395102,	0.5)
(2.225294796,	0.465886838)
(2.35619449,	0.428283136)
(2.487094184,	0.385811881)
(2.617993878,	0.33654572)
(2.748893572,	0.277783385)
(2.879793266,	0.205782518)
(3.01069296,	0.115463492)
(3.141592654,	1.22465E-16)
};
    \addlegendentry{\small $|\hat{\alpha}|$}

\addplot[
    only marks,
    color=blue,
    mark=*,
    mark size=1pt,
]
coordinates {
(-3.141592654,	1)
(-3.01069296,	0.884536508)
(-2.879793266,	0.794217482)
(-2.748893572,	0.722216615)
(-2.617993878,	0.66345428)
(-2.487094184,	0.614188119)
(-2.35619449,	0.571716864)
(-2.225294796,	0.534113162)
(-2.094395102,	0.5)
(-1.963495408,	0.468381315)
(-1.832595715,	0.438521501)
(-1.701696021,	0.40986231)
(-1.570796327,	0.381966011)
(-1.439896633,	0.354476234)
(-1.308996939,	0.327090444)
(-1.178097245,	0.299539977)
(-1.047197551,	0.271574874)
(-0.916297857,	0.242951614)
(-0.785398163,	0.213422393)
(-0.654498469,	0.182724878)
(-0.523598776,	0.150571485)
(-0.392699082,	0.116637243)
(-0.261799388,	0.080545092)
(-0.130899694,	0.041847166)
(0,	0)
(0.130899694,	0.041847166)
(0.261799388,	0.080545092)
(0.392699082,	0.116637243)
(0.523598776,	0.150571485)
(0.654498469,	0.182724878)
(0.785398163,	0.213422393)
(0.916297857,	0.242951614)
(1.047197551,	0.271574874)
(1.178097245,	0.299539977)
(1.308996939,	0.327090444)
(1.439896633,	0.354476234)
(1.570796327,	0.381966011)
(1.701696021,	0.40986231)
(1.832595715,	0.438521501)
(1.963495408,	0.468381315)
(2.094395102,	0.5)
(2.225294796,	0.534113162)
(2.35619449,	0.571716864)
(2.487094184,	0.614188119)
(2.617993878,	0.66345428)
(2.748893572,	0.722216615)
(2.879793266,	0.794217482)
(3.01069296,	0.884536508)
(3.141592654,	1)
};
    \addlegendentry{\small $|\hat{\beta}|$}   
    
\end{axis}

\end{tikzpicture}}
    \caption{}\label{AC_M_1<2}
  \end{subfigure} &
  \begin{subfigure}[t]{\linewidth}
    \centering \panelbox{\begin{tikzpicture}
\begin{axis}[
    height=5.2cm,
width=4.5cm,
    thick,
    xlabel={\textbf{\textit{k}L}},
    ylabel={\textbf{Modulus}},
    xmin=-3.1416, xmax=3.1416,
    ymin=0, ymax=1,
    every x tick/.style={color=black, thick},
    xtick={-3.1416,-1.5708,0,1.5708,3.1416},
    xticklabels={-$\pi$,-$\frac{\pi}{2}$,0,$\frac{\pi}{2}$,$\pi$},
    every y tick/.style={color=black, thick},
    ytick={0,0.25,0.5,0.75,1},
    yticklabels={$0$,$\frac{1}{4}$,$\frac{1}{2}$,$\frac{3}{4}$,$1$},
    legend style={at={(1.1,0.75)}, anchor=east}, 
    legend cell align={left},
]

\addplot[
    only marks,
    color=red,
    mark=*,
    mark size=1pt,
]
coordinates {
(-3.141592654,	1.22465E-16)
(-3.01069296,	0.060539567)
(-2.879793266,	0.109721837)
(-2.748893572,	0.147132568)
(-2.617993878,	0.17401665)
(-2.487094184,	0.192225024)
(-2.35619449,	0.203564468)
(-2.225294796,	0.209541504)
(-2.094395102,	0.211324865)
(-1.963495408,	0.209791783)
(-1.832595715,	0.205591887)
(-1.701696021,	0.199204418)
(-1.570796327,	0.190983006)
(-1.439896633,	0.181188842)
(-1.308996939,	0.170014717)
(-1.178097245,	0.157602333)
(-1.047197551,	0.144054819)
(-0.916297857,	0.129445862)
(-0.785398163,	0.113826487)
(-0.654498469,	0.09723019)
(-0.523598776,	0.079676979)
(-0.392699082,	0.061176695)
(-0.261799388,	0.041731935)
(-0.130899694,	0.021340843)
(0,	6.12323E-17)
(0.130899694,	0.021340843)
(0.261799388,	0.041731935)
(0.392699082,	0.061176695)
(0.523598776,	0.079676979)
(0.654498469,	0.09723019)
(0.785398163,	0.113826487)
(0.916297857,	0.129445862)
(1.047197551,	0.144054819)
(1.178097245,	0.157602333)
(1.308996939,	0.170014717)
(1.439896633,	0.181188842)
(1.570796327,	0.190983006)
(1.701696021,	0.199204418)
(1.832595715,	0.205591887)
(1.963495408,	0.209791783)
(2.094395102,	0.211324865)
(2.225294796,	0.209541504)
(2.35619449,	0.203564468)
(2.487094184,	0.192225024)
(2.617993878,	0.17401665)
(2.748893572,	0.147132568)
(2.879793266,	0.109721837)
(3.01069296,	0.060539567)
(3.141592654,	1.22465E-16)
};

\addplot[
    only marks,
    color=blue,
    mark=*,
    mark size=1pt,
]
coordinates {
(-3.141592654,	1)
(-3.01069296,	0.939460433)
(-2.879793266,	0.890278163)
(-2.748893572,	0.852867432)
(-2.617993878,	0.82598335)
(-2.487094184,	0.807774976)
(-2.35619449,	0.796435532)
(-2.225294796,	0.790458496)
(-2.094395102,	0.788675135)
(-1.963495408,	0.790208217)
(-1.832595715,	0.794408113)
(-1.701696021,	0.800795582)
(-1.570796327,	0.809016994)
(-1.439896633,	0.818811158)
(-1.308996939,	0.829985283)
(-1.178097245,	0.842397667)
(-1.047197551,	0.855945181)
(-0.916297857,	0.870554138)
(-0.785398163,	0.886173513)
(-0.654498469,	0.90276981)
(-0.523598776,	0.920323021)
(-0.392699082,	0.938823305)
(-0.261799388,	0.958268065)
(-0.130899694,	0.978659157)
(0,	1)
(0.130899694,	0.978659157)
(0.261799388,	0.958268065)
(0.392699082,	0.938823305)
(0.523598776,	0.920323021)
(0.654498469,	0.90276981)
(0.785398163,	0.886173513)
(0.916297857,	0.870554138)
(1.047197551,	0.855945181)
(1.178097245,	0.842397667)
(1.308996939,	0.829985283)
(1.439896633,	0.818811158)
(1.570796327,	0.809016994)
(1.701696021,	0.800795582)
(1.832595715,	0.794408113)
(1.963495408,	0.790208217)
(2.094395102,	0.788675135)
(2.225294796,	0.790458496)
(2.35619449,	0.796435532)
(2.487094184,	0.807774976)
(2.617993878,	0.82598335)
(2.748893572,	0.852867432)
(2.879793266,	0.890278163)
(3.01069296,	0.939460433)
(3.141592654,	1)
};

\end{axis}

\end{tikzpicture}}
    \caption{}\label{1O_M_1>2}
  \end{subfigure} &
  \begin{subfigure}[t]{\linewidth}
    \centering \hspace{-2.4em} \panelbox{\begin{tikzpicture}
\begin{axis}[
    height=5.2cm,
width=4.5cm,
    thick,
    xlabel={\textbf{\textit{k}L}},
    xmin=-3.1416, xmax=3.1416,
    ymin=0, ymax=1,
    every x tick/.style={color=black, thick},
    xtick={-3.1416,-1.5708,0,1.5708,3.1416},
    xticklabels={-$\pi$,-$\frac{\pi}{2}$,0,$\frac{\pi}{2}$,$\pi$},
    every y tick/.style={color=black, thick},
    ytick={0,0.25,0.5,0.75,1},
    yticklabels={$0$,$\frac{1}{4}$,$\frac{1}{2}$,$\frac{3}{4}$,$1$},
    legend style={at={(1.1,0.75)}, anchor=east}, 
    legend cell align={left},
]

\addplot[
    only marks,
    color=red,
    mark=*,
    mark size=1pt,
]
coordinates {
(-3.141592654,	1)
(-3.01069296,	0.884536508)
(-2.879793266,	0.794217482)
(-2.748893572,	0.722216615)
(-2.617993878,	0.66345428)
(-2.487094184,	0.614188119)
(-2.35619449,	0.571716864)
(-2.225294796,	0.534113162)
(-2.094395102,	0.5)
(-1.963495408,	0.468381315)
(-1.832595715,	0.438521501)
(-1.701696021,	0.40986231)
(-1.570796327,	0.381966011)
(-1.439896633,	0.354476234)
(-1.308996939,	0.327090444)
(-1.178097245,	0.299539977)
(-1.047197551,	0.271574874)
(-0.916297857,	0.242951614)
(-0.785398163,	0.213422393)
(-0.654498469,	0.182724878)
(-0.523598776,	0.150571485)
(-0.392699082,	0.116637243)
(-0.261799388,	0.080545092)
(-0.130899694,	0.041847166)
(0,	6.12323E-17)
(0.130899694,	0.041847166)
(0.261799388,	0.080545092)
(0.392699082,	0.116637243)
(0.523598776,	0.150571485)
(0.654498469,	0.182724878)
(0.785398163,	0.213422393)
(0.916297857,	0.242951614)
(1.047197551,	0.271574874)
(1.178097245,	0.299539977)
(1.308996939,	0.327090444)
(1.439896633,	0.354476234)
(1.570796327,	0.381966011)
(1.701696021,	0.40986231)
(1.832595715,	0.438521501)
(1.963495408,	0.468381315)
(2.094395102,	0.5)
(2.225294796,	0.534113162)
(2.35619449,	0.571716864)
(2.487094184,	0.614188119)
(2.617993878,	0.66345428)
(2.748893572,	0.722216615)
(2.879793266,	0.794217482)
(3.01069296,	0.884536508)
(3.141592654,	1)
};

\addplot[
    only marks,
    color=blue,
    mark=*,
    mark size=1pt,
]
coordinates {
(-3.141592654,	6.12323E-17)
(-3.01069296,	0.115463492)
(-2.879793266,	0.205782518)
(-2.748893572,	0.277783385)
(-2.617993878,	0.33654572)
(-2.487094184,	0.385811881)
(-2.35619449,	0.428283136)
(-2.225294796,	0.465886838)
(-2.094395102,	0.5)
(-1.963495408,	0.531618685)
(-1.832595715,	0.561478499)
(-1.701696021,	0.59013769)
(-1.570796327,	0.618033989)
(-1.439896633,	0.645523766)
(-1.308996939,	0.672909556)
(-1.178097245,	0.700460023)
(-1.047197551,	0.728425126)
(-0.916297857,	0.757048386)
(-0.785398163,	0.786577607)
(-0.654498469,	0.817275122)
(-0.523598776,	0.849428515)
(-0.392699082,	0.883362757)
(-0.261799388,	0.919454908)
(-0.130899694,	0.958152834)
(0,	1)
(0.130899694,	0.958152834)
(0.261799388,	0.919454908)
(0.392699082,	0.883362757)
(0.523598776,	0.849428515)
(0.654498469,	0.817275122)
(0.785398163,	0.786577607)
(0.916297857,	0.757048386)
(1.047197551,	0.728425126)
(1.178097245,	0.700460023)
(1.308996939,	0.672909556)
(1.439896633,	0.645523766)
(1.570796327,	0.618033989)
(1.701696021,	0.59013769)
(1.832595715,	0.561478499)
(1.963495408,	0.531618685)
(2.094395102,	0.5)
(2.225294796,	0.465886838)
(2.35619449,	0.428283136)
(2.487094184,	0.385811881)
(2.617993878,	0.33654572)
(2.748893572,	0.277783385)
(2.879793266,	0.205782518)
(3.01069296,	0.115463492)
(3.141592654,	1.83697E-16)
};

\end{axis}

\end{tikzpicture}}
    \caption{}\label{1O_M_1<2}
  \end{subfigure} \\
  \begin{subfigure}[t]{\linewidth}
    \centering \hspace{-1.4em} \panelbox{\begin{tikzpicture}
\begin{axis}[
    height=5.2cm,
width=4.5cm,
    thick,
    xlabel={\textbf{\textit{k}L}},
    ylabel={\textbf{Angle (rad)}},
    ylabel style={at={(axis description cs:-0.2,0.5)},anchor=south}, 
    xmin=-3.1416, xmax=3.1416,
    ymin=-3.1416, ymax=3.1416,
    every x tick/.style={color=black, thick},
    xtick={-3.1416,-1.5708,0,1.5708,3.1416},
    xticklabels={-$\pi$,-$\frac{\pi}{2}$,0,$\frac{\pi}{2}$,$\pi$},
    every y tick/.style={color=black, thick},
    ytick={-3.1416,-1.5708,0,1.5708,3.1416},
    yticklabels={-$\pi$,-$\frac{\pi}{2}$,0,$\frac{\pi}{2}$,$\pi$},
    every y tick/.style={color=black, thick},
]

\addplot[
  only marks,
    color=purple,
    mark=*,
    mark size=0.5pt,
]
coordinates {
(-3.141592654,	-1.570796327)
(-3.01069296,	-1.570796327)
(-2.879793266,	-1.570796327)
(-2.748893572,	-1.570796327)
(-2.617993878,	-1.570796327)
(-2.487094184,	-1.570796327)
(-2.35619449,	-1.570796327)
(-2.225294796,	-1.570796327)
(-2.094395102,	-1.570796327)
(-1.963495408,	-1.570796327)
(-1.832595715,	-1.570796327)
(-1.701696021,	-1.570796327)
(-1.570796327,	-1.570796327)
(-1.439896633,	-1.570796327)
(-1.308996939,	-1.570796327)
(-1.178097245,	-1.570796327)
(-1.047197551,	-1.570796327)
(-0.916297857,	-1.570796327)
(-0.785398163,	-1.570796327)
(-0.654498469,	-1.570796327)
(-0.523598776,	-1.570796327)
(-0.392699082,	-1.570796327)
(-0.261799388,	-1.570796327)
(-0.130899694,	-1.570796327)
(0,	1.570796327)
(0.130899694,	1.570796327)
(0.261799388,	1.570796327)
(0.392699082,	1.570796327)
(0.523598776,	1.570796327)
(0.654498469,	1.570796327)
(0.785398163,	1.570796327)
(0.916297857,	1.570796327)
(1.047197551,	1.570796327)
(1.178097245,	1.570796327)
(1.308996939,	1.570796327)
(1.439896633,	1.570796327)
(1.570796327,	1.570796327)
(1.701696021,	1.570796327)
(1.832595715,	1.570796327)
(1.963495408,	1.570796327)
(2.094395102,	1.570796327)
(2.225294796,	1.570796327)
(2.35619449,	1.570796327)
(2.487094184,	1.570796327)
(2.617993878,	1.570796327)
(2.748893572,	1.570796327)
(2.879793266,	1.570796327)
(3.01069296,	1.570796327)
(3.141592654,	1.570796327)
};
\end{axis}

\end{tikzpicture}}
    \caption{}\label{AC_P_1>2}
  \end{subfigure} &
  \begin{subfigure}[t]{\linewidth}
    \centering \hspace{-1em} \panelbox{\begin{tikzpicture}
\begin{axis}[
    height=5.2cm,
width=4.5cm,
    thick,
    xlabel={\textbf{\textit{k}L}},
    xmin=-3.1416, xmax=3.1416,
    ymin=-3.1416, ymax=3.1416,
    every x tick/.style={color=black, thick},
    xtick={-3.1416,-1.5708,0,1.5708,3.1416},
    xticklabels={-$\pi$,-$\frac{\pi}{2}$,0,$\frac{\pi}{2}$,$\pi$},
    every y tick/.style={color=black, thick},
    ytick={-3.1416,-1.5708,0,1.5708,3.1416},
    yticklabels={-$\pi$,-$\frac{\pi}{2}$,0,$\frac{\pi}{2}$,$\pi$},
    every y tick/.style={color=black, thick},
    legend style={at={(1.15,0)}, anchor=west}, 
    legend cell align={left},
]

\addplot[
  only marks,
    color=purple,
    mark=*,
    mark size=0.5pt,
]
coordinates {
(-3.141592654,	-1.570796327)
(-3.01069296,	-1.570796327)
(-2.879793266,	-1.570796327)
(-2.748893572,	-1.570796327)
(-2.617993878,	-1.570796327)
(-2.487094184,	-1.570796327)
(-2.35619449,	-1.570796327)
(-2.225294796,	-1.570796327)
(-2.094395102,	-1.570796327)
(-1.963495408,	-1.570796327)
(-1.832595715,	-1.570796327)
(-1.701696021,	-1.570796327)
(-1.570796327,	-1.570796327)
(-1.439896633,	-1.570796327)
(-1.308996939,	-1.570796327)
(-1.178097245,	-1.570796327)
(-1.047197551,	-1.570796327)
(-0.916297857,	-1.570796327)
(-0.785398163,	-1.570796327)
(-0.654498469,	-1.570796327)
(-0.523598776,	-1.570796327)
(-0.392699082,	-1.570796327)
(-0.261799388,	-1.570796327)
(-0.130899694,	-1.570796327)
(0,	1.570796327)
(0.130899694,	1.570796327)
(0.261799388,	1.570796327)
(0.392699082,	1.570796327)
(0.523598776,	1.570796327)
(0.654498469,	1.570796327)
(0.785398163,	1.570796327)
(0.916297857,	1.570796327)
(1.047197551,	1.570796327)
(1.178097245,	1.570796327)
(1.308996939,	1.570796327)
(1.439896633,	1.570796327)
(1.570796327,	1.570796327)
(1.701696021,	1.570796327)
(1.832595715,	1.570796327)
(1.963495408,	1.570796327)
(2.094395102,	1.570796327)
(2.225294796,	1.570796327)
(2.35619449,	1.570796327)
(2.487094184,	1.570796327)
(2.617993878,	1.570796327)
(2.748893572,	1.570796327)
(2.879793266,	1.570796327)
(3.01069296,	1.570796327)
(3.141592654,	1.570796327)
};

\addlegendentry{$\Delta\phi$}
\end{axis}

\end{tikzpicture}}
    \caption{}\label{AC_P_1<2}
  \end{subfigure} &
  \begin{subfigure}[t]{\linewidth}
    \centering \hspace{-1.4em} \panelbox{\begin{tikzpicture}
\begin{axis}[
    height=5.2cm,
width=4.5cm,
    thick,
    xlabel={\textbf{\textit{k}L}},
    ylabel={\textbf{Angle (rad)}},
    ylabel style={at={(axis description cs:-0.2,0.5)},anchor=south}, 
    xmin=-3.1416, xmax=3.1416,
    ymin=-3.1416, ymax=3.1416,
    every x tick/.style={color=black, thick},
    xtick={-3.1416,-1.5708,0,1.5708,3.1416},
    xticklabels={-$\pi$,-$\frac{\pi}{2}$,0,$\frac{\pi}{2}$,$\pi$},
    every y tick/.style={color=black, thick},
    ytick={-3.1416,-1.5708,0,1.5708,3.1416},
    yticklabels={-$\pi$,-$\frac{\pi}{2}$,0,$\frac{\pi}{2}$,$\pi$},
    every y tick/.style={color=black, thick},
]

\addplot[
  only marks,
    color=purple,
    mark=*,
    mark size=0.5pt,
]
coordinates {
(-3.141592654,	1.570796327)
(-3.01069296,	1.570796327)
(-2.879793266,	1.570796327)
(-2.748893572,	1.570796327)
(-2.617993878,	1.570796327)
(-2.487094184,	1.570796327)
(-2.35619449,	1.570796327)
(-2.225294796,	1.570796327)
(-2.094395102,	1.570796327)
(-1.963495408,	1.570796327)
(-1.832595715,	1.570796327)
(-1.701696021,	1.570796327)
(-1.570796327,	1.570796327)
(-1.439896633,	1.570796327)
(-1.308996939,	1.570796327)
(-1.178097245,	1.570796327)
(-1.047197551,	1.570796327)
(-0.916297857,	1.570796327)
(-0.785398163,	1.570796327)
(-0.654498469,	1.570796327)
(-0.523598776,	1.570796327)
(-0.392699082,	1.570796327)
(-0.261799388,	1.570796327)
(-0.130899694,	1.570796327)
(0,	1.570796327)
(0.130899694,	-1.570796327)
(0.261799388,	-1.570796327)
(0.392699082,	-1.570796327)
(0.523598776,	-1.570796327)
(0.654498469,	-1.570796327)
(0.785398163,	-1.570796327)
(0.916297857,	-1.570796327)
(1.047197551,	-1.570796327)
(1.178097245,	-1.570796327)
(1.308996939,	-1.570796327)
(1.439896633,	-1.570796327)
(1.570796327,	-1.570796327)
(1.701696021,	-1.570796327)
(1.832595715,	-1.570796327)
(1.963495408,	-1.570796327)
(2.094395102,	-1.570796327)
(2.225294796,	-1.570796327)
(2.35619449,	-1.570796327)
(2.487094184,	-1.570796327)
(2.617993878,	-1.570796327)
(2.748893572,	-1.570796327)
(2.879793266,	-1.570796327)
(3.01069296,	-1.570796327)
(3.141592654,	1.570796327)
};

\end{axis}

\end{tikzpicture}}
    \caption{}\label{1O_P_1>2}
  \end{subfigure} &
  \begin{subfigure}[t]{\linewidth}
    \centering \hspace{-3.1em} \panelbox{\begin{tikzpicture}
\begin{axis}[
    height=5.2cm,
width=4.5cm,
    thick,
    xlabel={\textbf{\textit{k}L}},
    xmin=-3.1416, xmax=3.1416,
    ymin=-3.1416, ymax=3.1416,
    every x tick/.style={color=black, thick},
    xtick={-3.1416,-1.5708,0,1.5708,3.1416},
    xticklabels={-$\pi$,-$\frac{\pi}{2}$,0,$\frac{\pi}{2}$,$\pi$},
    every y tick/.style={color=black, thick},
    ytick={-3.1416,-1.5708,0,1.5708,3.1416},
    yticklabels={-$\pi$,-$\frac{\pi}{2}$,0,$\frac{\pi}{2}$,$\pi$},
    every y tick/.style={color=black, thick},
]

\addplot[
  only marks,
    color=purple,
    mark=*,
    mark size=0.5pt,
]
coordinates {
(-3.141592654,	1.570796327)
(-3.01069296,	1.570796327)
(-2.879793266,	1.570796327)
(-2.748893572,	1.570796327)
(-2.617993878,	1.570796327)
(-2.487094184,	1.570796327)
(-2.35619449,	1.570796327)
(-2.225294796,	1.570796327)
(-2.094395102,	1.570796327)
(-1.963495408,	1.570796327)
(-1.832595715,	1.570796327)
(-1.701696021,	1.570796327)
(-1.570796327,	1.570796327)
(-1.439896633,	1.570796327)
(-1.308996939,	1.570796327)
(-1.178097245,	1.570796327)
(-1.047197551,	1.570796327)
(-0.916297857,	1.570796327)
(-0.785398163,	1.570796327)
(-0.654498469,	1.570796327)
(-0.523598776,	1.570796327)
(-0.392699082,	1.570796327)
(-0.261799388,	1.570796327)
(-0.130899694,	1.570796327)
(0,	1.570796327)
(0.130899694,	-1.570796327)
(0.261799388,	-1.570796327)
(0.392699082,	-1.570796327)
(0.523598776,	-1.570796327)
(0.654498469,	-1.570796327)
(0.785398163,	-1.570796327)
(0.916297857,	-1.570796327)
(1.047197551,	-1.570796327)
(1.178097245,	-1.570796327)
(1.308996939,	-1.570796327)
(1.439896633,	-1.570796327)
(1.570796327,	-1.570796327)
(1.701696021,	-1.570796327)
(1.832595715,	-1.570796327)
(1.963495408,	-1.570796327)
(2.094395102,	-1.570796327)
(2.225294796,	-1.570796327)
(2.35619449,	-1.570796327)
(2.487094184,	-1.570796327)
(2.617993878,	-1.570796327)
(2.748893572,	-1.570796327)
(2.879793266,	-1.570796327)
(3.01069296,	-1.570796327)
(3.141592654,	1.570796327)
};

\end{axis}

\end{tikzpicture}}
    \caption{}\label{1O_P_1<2}
  \end{subfigure} \\
\end{tabular}
\caption{Evolution of the modulus and phase of coefficient of the superposition of states at acoustic and optical branch at two different condition of stiffness of $\psi_1>\psi_2$ and $\psi_1<\psi_2$. \textbf{Top Panel:} Modulus of complex amplitudes $\hat\alpha$ and $\hat\beta$ of two mutually orthogonal states $\ket{E_1}$ and $\ket{E_2}$. $\hat\alpha$ is dominating at acoustic branch while $\beta$ is dominating at the optical branch at both stiffness conditions. \textbf{Bottom Panel:} Phase difference between the amplitude moduli $\Delta\phi=\arg(\hat\alpha)-\arg(\hat\beta)$, showing a locked $\pm\pi/2$ plateaus across the first Brillouin zone $kL \in [-\pi,\pi]$}
\label{Fig:Modulus_2Mass}
\end{figure}

Figure \ref{Fig:Modulus_2Mass} shows the normalized superposition coefficients of Eq.~(\ref{eq5}) across the first Brillouin zone $k L \in [-\pi, \pi]$. Panel (\subref{AC_M_1>2}) and (\subref{AC_M_1<2}) shows the moduli $|\hat\alpha|$ and $|\hat\beta|$ that weight the two mutually orthogonal eigenstates $\ket{E_1}$ and $\ket{E_2}$. For the illustrative stiffness ordering $(\psi_1, \psi_2) = (1, 1/2)$—and, more generally, whenever $\psi_1 > \psi_2$—the acoustic branch is dominated by the $\ket{E_1}$ component, i.e., $\lvert\hat\alpha\rvert > \lvert\hat\beta\rvert$; the situation inverts on the optical branch, for which $\lvert\hat\beta\rvert$ dominates (Fig.~\ref{1O_M_1>2}). Except at the high-symmetry points $k L = -\pi, 0, \pi$, both $|\hat\alpha|$ and $\lvert\hat\beta\rvert$ remain appreciable over substantial portions of the zone, indicating a mixed superposition rather than a pure-state selection. And, for the stiffness ordering $(\psi_1,\psi_2)=(2,1)$, which means, $\psi_1<\psi_2$, the acoustic branch is dominated by the $\ket{E_2}$ component, i.e., $\lvert\hat\alpha\rvert<\lvert\hat\beta\rvert$ (Fig.~\ref{AC_M_1<2}). Similar to before case, this phenomena flips at the optical branch (Fig.~\ref{1O_M_1<2}).

From Figs.~\ref{AC_P_1>2} and \ref{AC_P_1<2}, we see that the relative phase $\Delta \phi = \arg(\hat\alpha) - \arg(\hat\beta)$ locks to $-\pi/2$ or $+\pi/2$ and switches between those values when $k L$ advances by $\pi$ (i.e., upon crossing the zone midpoint). Consequently, $\Delta \phi$ is piecewise constant with a jump of $\pi$ at the inversion-symmetric point. However, the individual absolute phases evolve smoothly with $k L \in [-\pi, \pi]$. For the case $\psi_1 > \psi_2$ (Fig.~\ref{AC_P_1>2}), the phase of $\beta$ acquires two twists of $\pi$ after one closed traversal of the Brillouin zone, whereas the phase of $\hat\alpha$ changes monotonically without such a twist. This phenomena inverts on the optical branch (Fig.~\ref{1O_P_1>2}). On the contrary, for the case $\psi_1 < \psi_2$ (Fig.~\ref{AC_P_1<2} and \ref{1O_P_1<2}), both the phases of $\hat\alpha$ and $\hat\beta$ acquire a single twist of $\pi$ after one closed traversal of the Brillouin zone for the acoustic and optical branch. This behavior is consistent with the in-phase/out-of-phase character of acoustic/optical modes in diatomic chains and with inversion-symmetry-protected quantization of the Zak (Berry) phase to 0 or $\pi$ in one-dimensional lattices \cite{PhysRevLett.62.2747}.

The branch selection, amplitude symmetry, and Berry-phase results obtained above are fully consistent with established one-dimensional topological band theory and with our prior MD/SAAP analyses of periodic chains: for the two-mass cell, the Zak phase is $0$ when $\psi_1 > \psi_2$ and $\pi$ when $\psi_1 < \psi_2$ (up to the choice of origin). What is distinctive here is the explicit classical-superposition framework, in which the elastic state is written and manipulated as $\hat{\alpha}\ket{E_1} + \hat{\beta}\ket{E_2}$ and its evolution over $k$ traces well-defined loops on the Bloch sphere whose winding encodes the topology. This geometric–dynamical picture complements the usual bulk-invariant description by revealing how the same binary topology manifests in measurable amplitude–phase data and by tying the locked $\Delta\phi = \pm \pi/2$ plateaus and the $\lvert\hat{\alpha}\rvert/\lvert\hat{\beta}\rvert$ hierarchy to identifiable paths on $S^2$. In doing so, it connects directly with contemporary measurements of Zak phases in classical and quantum platforms and their role in predicting interface states in inversion-symmetric media \cite{RN252,PhysRevX.4.021017}. Moreover, framing the diatomic chain as a controllable two-level system makes the structure explicit and portable: the same machinery extends naturally to the symmetry-broken and time-modulated lattices treated later in the manuscript, where it distinguishes non-quantized geometric phases from origin conventions, links phase winding to mode hybridization, and provides the complex modal coefficients needed for Berry-connection evaluation \cite{PhysRevLett.82.2147}.

\subsubsection{Bloch-Sphere Parameterization}
\label{Bloch Sphere Time Independent}
Following the standard two-component spinor notation, we write
\begin{equation}
\boxed{ \hat{\alpha} = \cos \left( \tfrac{\theta}{2} \right), \quad 
        \hat{\beta} = e^{\mathrm{i} \varphi} \sin \left( \tfrac{\theta}{2} \right) }
        \label{eq8}
\end{equation}
so that the polar angle $\theta \in [0, \pi]$ encodes the relative weight of the two eigenstates while the azimuth $\varphi \in [-\pi, \pi)$ stores their relative phase. The north pole $(\theta=0)$ therefore represents the pure in-phase state $\ket{E_1}$, the south pole $(\theta = \pi)$ its out-of-phase companion $\ket{E_2}$, and any point on the sphere corresponds to a coherent, classical superposition
\begin{equation*}
|\psi(k)\rangle = \hat{\alpha}(k) \ket{E_1} + \hat{\beta}(k) \ket{E_2} \equiv \cos \left( \frac{\theta(k)}{2} \right) \ket{E_1} + e^{\mathrm{i} \varphi(k)} \sin \left( \frac{\theta(k)}{2} \right) \ket{E_2}.
\end{equation*}
To anchor the geometry, we evaluate $(\theta, \varphi)$ for three $k$-points (overall normalization is omitted for brevity):

Near a band extremum. For the state sampled at $k L = -1.57$, the amplitudes lead to
\begin{equation*}
\binom{\hat{\alpha}}{\hat{\beta}} = \binom{-0.0006 - 0.0001 \mathrm{i}}{-0.9497 - 0.3111 \mathrm{i}} = \left(0.0006 e^{-\mathrm{i} \pi}\right) \ket{E_1} + \left(0.9994 e^{-\mathrm{i} \pi}\right) \ket{E_2} = \cos \frac{\theta}{2} \ket{E_1} + e^{\mathrm{i} \varphi} \sin \frac{\theta}{2} \ket{E_2},
\end{equation*}
with the angles identified as $\theta = \pi$ and $\varphi = 0$. Geometrically the state sits at the south pole (pure $E_2$), as expected when one mode dominates.

Near the opposite extremum. For $k L = 0.0654$, we obtain
\begin{equation*}
\begin{split}
\binom{\hat{\alpha}}{\hat{\beta}}
&= \binom{-0.9857 - 0.1680 \mathrm{i}}{-8.6397 \times 10^{-5} - 1.9891 \times 10^{-5} \mathrm{i}}
=\left(0.9999 e^{-\mathrm{i} \pi}\right) \ket{E_1}
+ \left(8.8657 \times 10^{-5} e^{-\mathrm{i} \pi}\right) \ket{E_2} \\
&= \cos \frac{\theta}{2} \ket{E_1}
+ e^{\mathrm{i} \varphi} \sin \frac{\theta}{2} \ket{E_2}
\end{split}
\end{equation*}
with $\theta = 0$ and $\varphi = 0$. The state is at the north pole (pure $E_1$).

Nonseparable superposition At $k L = -0.4254$, the amplitudes yield
\begin{equation*}
\binom{\hat{\alpha}}{\hat{\beta}} = \binom{-0.5888 - 0.0386 \mathrm{i}}{0.0266 - 0.4091 \mathrm{i}} = \left(0.5091 e^{-\mathrm{i} \pi}\right) \ket{E_1} + \left(0.4099 e^{-\mathrm{i} \frac{\pi}{2}}\right) \ket{E_2} = \cos \frac{\theta}{2} \ket{E_1} + e^{\mathrm{i} \varphi} \sin \frac{\theta}{2} \ket{E_2},
\end{equation*}
with $\theta = \frac{\pi}{2}$ and $\varphi = \frac{\pi}{2}$. This corresponds to a balanced, classically nonseparable (i.e., phase-coherent) superposition with the state proportional to $\left(\lvert E_1 \rangle + \mathrm{i} \lvert E_2 \rangle \right)$ up to overall normalization. The geometry is now on the equator at azimuth $\varphi = \pi/2$.

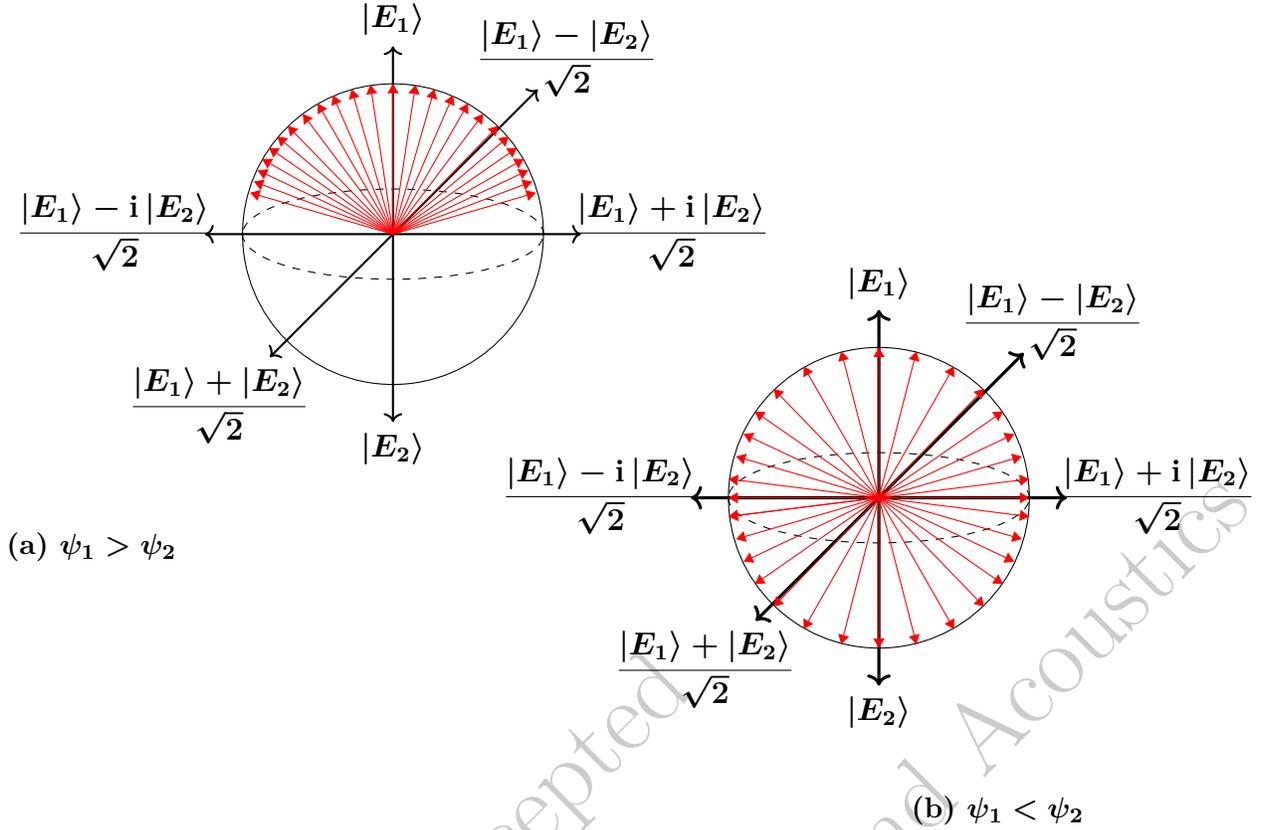
\begin{figure}[H]
    \setlength{\tabcolsep}{-0.02cm} 
    \renewcommand{\arraystretch}{0.2} 
    \begin{tabular}{p{6.5cm}p{7cm}}
        \begin{subfigure}{0.46\textwidth}
        \captionsetup{justification=raggedright, singlelinecheck=false}
        \begin{tikzpicture}[scale=1]
  \draw (0,0) circle (2);
  \draw[dashed] (-2,0) arc (180:360:2 and 0.6);
  \draw[dashed] (2,0) arc (0:180:2 and 0.6);
  \draw[thick,->] (0,0,0) -- (0,2.5,0);
  \draw[thick,->] (0,0,0) -- (0,-2.5,0);
  \draw[thick,->] (0,0,0) -- (2.5,0,0);
  \draw[thick,->] (0,0,0) -- (-2.5,0,0);
  \draw[thick,->] (0,0,0) -- (0,0,4.25);
  \draw[thick,->] (0,0,0) -- (0,0,-5);
  \node[anchor=south, font=\bfseries] at (0,2.5,0)  {$\boldsymbol{\ket{E_1}}$};
\node[anchor=north, font=\bfseries] at (0,-2.5,0) {$\boldsymbol{\ket{E_2}}$};

\node[anchor=mid,   font=\bfseries] at (3.7,0,0)
{$\boldsymbol{\dfrac{\ket{E_1}+\mathrm{i}\ket{E_2}}{\sqrt{2}}}$};
\node[anchor=mid,   font=\bfseries] at (-3.7,0,0)
{$\boldsymbol{\dfrac{\ket{E_1}-\mathrm{i}\ket{E_2}}{\sqrt{2}}}$};

\node[anchor=mid,   font=\bfseries] at (0,0,6)
{$\boldsymbol{\dfrac{\ket{E_1}+\ket{E_2}}{\sqrt{2}}}$};
\node[anchor=mid,   font=\bfseries] at (0,0,-6)
{$\boldsymbol{\dfrac{\ket{E_1}-\ket{E_2}}{\sqrt{2}}}$};

\draw[-Triangle,red] (0,0,0) -- (0,2,0);
\draw[-Triangle,red] (0,0,0) -- (-1.9,0.55,0);
\draw[-Triangle,red] (0,0,0) -- (-1.85,0.7,0);
\draw[-Triangle,red] (0,0,0) -- (-1.8,0.85,0);
\draw[-Triangle,red] (0,0,0) -- (-1.74,1,0);
\draw[-Triangle,red] (0,0,0) -- (-1.65,1.15,0);
\draw[-Triangle,red] (0,0,0) -- (-1.54,1.3,0);
\draw[-Triangle,red] (0,0,0) -- (-1.4,1.44,0);
\draw[-Triangle,red] (0,0,0) -- (-1.2,1.6,0);
\draw[-Triangle,red] (0,0,0) -- (-1,1.74,0);
\draw[-Triangle,red] (0,0,0) -- (-0.8,1.85,0);
\draw[-Triangle,red] (0,0,0) -- (-0.55,1.93,0);
\draw[-Triangle,red] (0,0,0) -- (-0.3,1.99,0);
\draw[-Triangle,red] (0,0,0) -- (1.9,0.55,0);
\draw[-Triangle,red] (0,0,0) -- (1.85,0.7,0);
\draw[-Triangle,red] (0,0,0) -- (1.8,0.85,0);
\draw[-Triangle,red] (0,0,0) -- (1.74,1,0);
\draw[-Triangle,red] (0,0,0) -- (1.65,1.15,0);
\draw[-Triangle,red] (0,0,0) -- (1.54,1.3,0);
\draw[-Triangle,red] (0,0,0) -- (1.4,1.44,0);
\draw[-Triangle,red] (0,0,0) -- (1.2,1.6,0);
\draw[-Triangle,red] (0,0,0) -- (1,1.74,0);
\draw[-Triangle,red] (0,0,0) -- (0.8,1.85,0);
\draw[-Triangle,red] (0,0,0) -- (0.55,1.93,0);
\draw[-Triangle,red] (0,0,0) -- (0.3,1.99,0);
\end{tikzpicture}
             \caption{ $\boldsymbol{\psi_1>\psi_2}$} 
            \label{Bloch_TI_1>2}
        \end{subfigure}
        &
        \\[-10em] 
        &
        \begin{subfigure}{0.46\textwidth}
        \captionsetup{justification=raggedleft, singlelinecheck=false}
\begin{tikzpicture}[scale=1]
height=2cm,
width=2cm,
  \draw (0,0) circle (2);
  \draw[dashed] (-2,0) arc (180:360:2 and 0.6);
  \draw[dashed] (2,0) arc (0:180:2 and 0.6);
  \draw[very thick,->] (0,0,0) -- (0,2.5,0);
  \draw[very thick,->] (0,0,0) -- (0,-2.5,0);
  \draw[very thick,->] (0,0,0) -- (2.5,0,0);
  \draw[very thick,->] (0,0,0) -- (-2.5,0,0);
  \draw[very thick,->] (0,0,0) -- (0,0,4.25);
  \draw[very thick,->] (0,0,0) -- (0,0,-5);
  \node[anchor=south, font=\bfseries] at (0,2.5,0)  {$\boldsymbol{\ket{E_1}}$};
\node[anchor=north, font=\bfseries] at (0,-2.5,0) {$\boldsymbol{\ket{E_2}}$};

\node[anchor=mid,   font=\bfseries] at (3.7,0,0)
  {$\boldsymbol{\dfrac{\ket{E_1}+\mathrm{i}\ket{E_2}}{\sqrt{2}}}$};
\node[anchor=mid,   font=\bfseries] at (-3.7,0,0)
  {$\boldsymbol{\dfrac{\ket{E_1}-\mathrm{i}\ket{E_2}}{\sqrt{2}}}$};

\node[anchor=mid,   font=\bfseries] at (0,0,6)
  {$\boldsymbol{\dfrac{\ket{E_1}+\ket{E_2}}{\sqrt{2}}}$};
\node[anchor=mid,   font=\bfseries] at (0,0,-6)
  {$\boldsymbol{\dfrac{\ket{E_1}-\ket{E_2}}{\sqrt{2}}}$};

\draw[-Triangle,red] (0,0,0) -- (0,-2,0);
\draw[-Triangle,red] (0,0,0) -- (-2,-0.25,0);
\draw[-Triangle,red] (0,0,0) -- (-1.9,-0.55,0);
\draw[-Triangle,red] (0,0,0) -- (-1.8,-0.85,0);
\draw[-Triangle,red] (0,0,0) -- (-1.65,-1.15,0);
\draw[-Triangle,red] (0,0,0) -- (-1.4,-1.45,0);
\draw[-Triangle,red] (0,0,0) -- (-1,-1.75,0);
\draw[-Triangle,red] (0,0,0) -- (-2,-0.25,0);
\draw[-Triangle,red] (0,0,0) -- (1,-1.75,0);
\draw[-Triangle,red] (0,0,0) -- (1.4,-1.45,0);
\draw[-Triangle,red] (0,0,0) -- (1.65,-1.15,0);
\draw[-Triangle,red] (0,0,0) -- (1.8,-0.85,0);
\draw[-Triangle,red] (0,0,0) -- (1.9,-0.55,0);
\draw[-Triangle,red] (0,0,0) -- (2,-0.25,0);
\draw[-Triangle,red] (0,0,0) -- (2,0.25,0);
\draw[-Triangle,red] (0,0,0) -- (-2,0.25,0);
\draw[-Triangle,red] (0,0,0) -- (-2,0,0);
\draw[-Triangle,red] (0,0,0) -- (-1.9,0.55,0);
\draw[-Triangle,red] (0,0,0) -- (-1.8,0.85,0);
\draw[-Triangle,red] (0,0,0) -- (-1.65,1.15,0);
\draw[-Triangle,red] (0,0,0) -- (-1.4,1.45,0);
\draw[-Triangle,red] (0,0,0) -- (-1,1.75,0);
\draw[-Triangle,red] (0,0,0) -- (1,1.75,0);
\draw[-Triangle,red] (0,0,0) -- (1.4,1.45,0);
\draw[-Triangle,red] (0,0,0) -- (1.65,1.15,0);
\draw[-Triangle,red] (0,0,0) -- (1.8,0.85,0);
\draw[-Triangle,red] (0,0,0) -- (1.9,0.55,0);
\draw[-Triangle,red] (0,0,0) -- (-0.5,-1.95,0);
\draw[-Triangle,red] (0,0,0) -- (0.5,-1.95,0);
\draw[-Triangle,red] (0,0,0) -- (0,2,0);
\draw[-Triangle,red] (0,0,0) -- (-0.5,1.95,0);
\draw[-Triangle,red] (0,0,0) -- (0.5,1.95,0);
\draw[-Triangle,red] (0,0,0) -- (2,0,0);
\end{tikzpicture}
        \caption{ $\boldsymbol{\psi_1<\psi_2}$}
            \label{Bloch_TI_1<2}
        \end{subfigure}
    \end{tabular}
    \caption{Bloch state demonstration of change of state between $\ket{E_1}$ and $\ket{E_2}$ at a two-mass system depicted in Hilbert Space at the acoustic branch with different stiffness matrices for the two cases \textbf{(a)} Stiffness ordering $\psi_1>\psi_2$, the trajectory of the Bloch states approaches the pure state $\ket{E_1}$ \textbf{(b)} Stiffness ordering $\psi_1<\psi_2$, exploring larger region of the Bloch sphere covering both pure state $\ket{E_1}$ and $\ket{E_2}$.} 
    \label{BlochState_TI}
\end{figure}

These points confirm that the lattice state moves smoothly between the poles and the equator as $k$ sweeps the Brillouin zone, exactly mirroring the interchange of the modulus hierarchy $|\hat{\alpha}| \gtrless |\hat{\beta}|$ identified in Fig.~\ref{Fig:Modulus_2Mass}. Using Eq.~(\ref{eq8}) at every discretized $k$, we reconstruct the full path $[\theta(k), \varphi(k)]$ on the sphere (Fig.~\ref{BlochState_TI}).

From Fig.~\ref{AC_M_1>2} $(\psi_1 > \psi_2)$ we have seen that the modulus hierarchy $\lvert\hat\alpha\rvert > \lvert\hat\beta\rvert$ holds over broad $k$-intervals, hence from Fig.~\ref{Bloch_TI_1>2} it is clear that the trajectory hugs the $\ket{E_1}$ pole: the system repeatedly approaches the pure state $\ket{E_1}$ whenever the real part of the masses dominates. $\hat{\alpha}$ dominance yields a state geometrically aligned with $\ket{E_1}$ because domination here refers to the modulus associated with the in-phase eigenvector. The corresponding polar angle remains in the "polar cap," $\theta \in (-\pi/2, +\pi/2)$, i.e., the path does not cross the equator except near isolated $k$ where $\lvert\hat\alpha\rvert \approx \lvert\hat\beta\rvert$. In contrast, Fig.~\ref{Bloch_TI_1<2} $(\psi_1 < \psi_2)$ does not bias the dynamics toward either $\ket{E_1}$ or $\ket{E_2}$: the trajectory makes a full longitude-like revolution ($\theta$ runs from $-\pi$ to $+\pi$) without ever settling at a perfectly pure $\ket{E_1}$ or $\ket{E_2}$ state (the path skirts the pole but does not terminate there).

In both stiffness orderings, the azimuth $\varphi$ is confined to a $\pi$-wide window, $\varphi \in [-\pi/2, +\pi/2]$, as dictated by the measured phase difference $\varphi = \arg (\hat\alpha) - \arg (\hat\beta)$. This constraint is visible in Figs. \ref{AC_P_1>2}--\subref{AC_P_1<2}: as $k$ passes through the zone center, the relative phase undergoes a discrete jump of magnitude $\pi$, sending $\varphi$ from $-\pi/2$ to $+\pi/2$ (or vice versa). That sign flip of the relative phase—equivalently, a reversal of the spinor's azimuth—encodes the usual $0/\pi$ Zak-phase bifurcation in inversion-symmetric 1D lattices \cite{RN253}.

Putting the two stiffness orderings side-by-side clarifies the accessible regions on $S^{2}$. For $\psi_1<\psi_2$ the trajectory explores a much larger fraction of the sphere: starting near $\ket{E_2}$, varying $k$ drives the state through equatorial mixtures $(\ket{E_1} + \mathrm{i} \ket{E_2})/\sqrt{2}$ and then toward $\ket{E_1}$ before a $\pi$ jump returns it to the opposite hemisphere; subsequent evolution repeats the pattern until the loop closes back near $\ket{E_2}$. For $\psi_1>\psi_2$ the loop is more localized: it begins at (or near) $\ket{E_1}$ and oscillates around the equator between $(\ket{E_1} - \mathrm{i} \ket{E_2})/\sqrt{2}$ without ever approaching the $\ket{E_2}$ pole. In short, $\psi_1 < \psi_2$ enables a broader sweep of superposed states within the same $k$-span, whereas $\psi_1 > \psi_2$ confines the dynamics to a single longitude family. The contrast highlights a central result of this paper: changing the spring-constant hierarchy toggles the accessible portion of the Bloch sphere and hence the class of elastic superpositions that can be realized. This tunability is the classical parallel to manipulating a quantum bit on its Bloch sphere. By linking a macroscopic design knob (the spring-stiffness ratio) to the solid angle swept on $S^{2}$, we provide an intuitive handle for engineering topological phase transitions in mechanical lattices.

In realistic elastic lattices, dissipation and manufacturing tolerances perturb the ideal lossless, perfectly periodic model. To assess their impact on the Bloch-sphere representation, we distinguish common-mode (uniform) damping from differential damping, and weak disorder from strong disorder. Our Bloch-sphere representation is constructed from the normalized complex coefficient vector obtained by projecting measured (or simulated) steady-state responses onto the orthonormal modal basis {$\ket{E_1},\ket{E_2}$}, so uniform amplitude decay factors do not affect the state direction on the sphere. Under weak, approximately uniform damping, the main effects are reduced response magnitude and signal-to-noise, together with a smooth frequency-dependent phase lag and modest deformation of the path; the normalized coefficient trajectory therefore remains close to the lossless result provided the band gap remains open and the symmetry protecting phase quantization is not effectively broken. In particular, the inversion-symmetry mechanism underpinning Zak-phase quantization is expressed in our framework as a locking of relative phases to piecewise-constant plateaus with discrete $\pi$ jumps at symmetry points, and this locking is expected to persist (up to small fluctuations) under weak, symmetry-preserving dissipation. By contrast, non-proportional or sublattice-asymmetric damping (for example, different loss on the two sublattices or frequency-dependent dissipation that unevenly weights the two basis components) introduces small but finite changes in the relative phase and can visibly deform the Bloch-sphere path. These deformations grow with damping contrast and are most pronounced near strong band hybridization, where an effective non-Hermitian description may require biorthogonal eigenvectors and can yield complex geometric phases. In this setting, strong damping or asymmetric loss can unlock the plateau behavior, leading to continuously drifting relative phases and de-quantized geometric phases, analogous to what is observed when inversion symmetry is explicitly broken.

Disorder introduces a distinct limitation: generic random disorder, including static disorder arising from small spatial fluctuations in masses and stiffnesses, breaks strict translational invariance, so Bloch wavenumber is no longer the natural path parameter and a strictly Bloch-parameterized loop $\Psi(k)$ is not expected in a finite disordered sample. Practically, one observes a broadened or noisy quasi-$k$ trajectory when coefficients are inferred from spatially resolved responses, and the Bloch-sphere trajectory becomes realization-dependent. Nevertheless, two regimes can be distinguished. For weak, symmetry-preserving disorder (for example, a periodic supercell that retains an inversion center), the geometric phase of isolated bands remains quantized unless the disorder closes a band gap, so the topological class is stable even though the sphere trajectory is perturbed. In this regime the geometric-phase signature may be computed using standard discretized overlap/Wilson-loop expressions or real-space formulations of 1D topological invariants developed for disordered systems \cite{prodan2016bulk}. For generic disorder that breaks the protecting symmetry, the quantization constraint is lifted and the relative phases can drift continuously, indicating a genuinely geometry-dependent (non-quantized) phase; strong disorder of this type can close the gap and drive a true topological transition, in which case neither quantized phase behavior nor edge-mode pinning is expected to persist. Finally, when damping or disorder prevents perfect loop closure in practice, the accumulated phase can be defined using open-path geometric-phase constructions, consistent with our space–time modulated setting where the geometric phase is accumulated along the actual trajectory rather than an idealized closed loop \cite{PhysRevLett.60.2339,PhysRevA.106.023513}.

\subsubsection{Quantum Analogue Logic Gates}

Having mapped the diatomic lattice to a two-level Hilbert space spanned by $\left(\ket{E_1}, \ket{E_2}\right)$, we represent the normalized state vector $(\hat{\alpha}, \hat{\beta})$ as a point on the Bloch sphere. In this representation, any norm-preserving transformation of $(\hat{\alpha}, \hat{\beta})$ corresponds to a rotation on the sphere, exactly as in a single-qubit system. Accordingly, logic gates in the present mechanical setting are realized as unitary maps that rotate the classical superposition state on $S^2$. Importantly, these gate-like operations represent geometric rotations of a classical spinor on the Bloch sphere and do not involve quantum entanglement and do not require wave function collapse or any other non-classical resource. The use of gate terminology reflects the mathematical isomorphism between the normalized two-component elastic state and the $SU(2)$ action on the Bloch sphere familiar from single-qubit control. This isomorphism provides a compact language for describing how band-resolved states evolve across the Brillouin zone or under space–time stiffness modulation, but the underlying physics remains fully deterministic and classical: the coefficients are reconstructed from measured displacements, and the “operations” are geometric rotations on $S^2$. 

Throughout, we adopt the computational basis $\ket{E_1}$, $\ket{E_2}$. Because phases and global signs depend on the ansatz and band labeling (Sec.~\ref{diatomic unit}), matrix representatives of elementary gates may differ by overall phases or sign conventions from conventional quantum logic gates; what matters physically is the induced rotation on the sphere and the relative phase between components, which are basis-invariant. This is also exploited in quantum mechanics, where different but unitarily equivalent conventions generate identical Bloch-sphere trajectories and Berry/Zak phases.

\noindent\textbf{Pauli–(Y)–type inversion about the (y)-axis.} At $kL = -\pi$, the complex amplitude coefficients take the form
$\begin{pmatrix}
	0 \\ -\mathrm{i}
\end{pmatrix}$
which corresponds to the out-of-phase state $\ket{E_2}$. Acting with the (conventionally phased) $y$–axis inversion,
\[
\begin{bmatrix}
	0 & \mathrm{i} \\ -\mathrm{i} & 0
\end{bmatrix}
\begin{pmatrix}
	0 \\ -\mathrm{i}
\end{pmatrix}
=
\begin{pmatrix}
	1 \\ 0
\end{pmatrix}
\]
advances the state along its Bloch-sphere trajectory to the north pole $\left|E_1\right\rangle$ (up to a global phase). In the canonical qubit convention this is a $\pi$–rotation about $\hat{y}$, i.e., $R_y(\pi)$, which flips $\ket{1} \mapsto \ket{0}$ with an overall phase. The two descriptions are equivalent modulo phase and reflect the same geometric half-turn on $S^2$ \cite{crooks2024quantum}.

\noindent\textbf{Pauli–(Z)–type phase flip about the (z)-axis.} Let us consider the equal-norm superposition at $kL = -\pi/2$,
\[
\frac{\ket{E_1} + \mathrm{i}\ket{E_2}}{\sqrt{2}},
\]
which is taken by a $z$–axis phase flip to
\[
\frac{\ket{E_1} - \mathrm{i}\ket{E_2}}{\sqrt{2}}.
\]
Within our basis convention this phase inversion is represented by
\[
\begin{bmatrix}
	0 & e^{\mathrm{i}\pi} \\ 1 & 0
\end{bmatrix}
\begin{pmatrix}
	-0.618\,\mathrm{i} \\ 0.382
\end{pmatrix}
=
\begin{pmatrix}
	0.382 \\ 0.618\,\mathrm{i}
\end{pmatrix},
\]
which embodies the $\left|E_2\right\rangle \mapsto -\left|E_2\right\rangle$ action (a $\pi$–rotation about $\hat{z}$) while respecting the observed complex amplitudes at those $k$–points. In the canonical computational basis this is the familiar $Z = \mathrm{diag}(1, -1)$, i.e., the phase-shift gate $P(\pi)$; the off-diagonal representative used here differs by a unitary change of basis and by phase convention but generates the same Bloch-sphere rotation.

\noindent\textbf{Hadamard followed by $S$ (quadrature) phase.} A particularly instructive sequence is the Hadamard transformation followed by a quadrature phase, written as
\[
H = \frac{1}{\sqrt{2}}
\begin{bmatrix}
	1 & 1 \\[3pt]
	1 & -1
\end{bmatrix},
\qquad
S =
\begin{bmatrix}
	0 & 1 \\[3pt]
	\mathrm{i} & 1
\end{bmatrix}.
\]
In our elastic-bit basis, $H$ rotates a pole state (e.g., $\ket{E_1}$) to an equatorial superposition, while the subsequent $S$ adjusts the relative quadrature between $\ket{E_1}$ and $\ket{E_2}$, steering the state azimuthally on the sphere. In the canonical qubit convention, $S$ is the pure phase $P(\pi/2) = \mathrm{diag}(1, \mathrm{i})$; the non-diagonal representative above captures the same azimuthal advance once the basis and band-dependent phases are accounted for. The net effect reproduces the equatorial quarter-turn standard in single-qubit control (Hadamard followed by a $\pi/2$ $z$–rotation).

The aforementioned algebra acquires a direct mechanical interpretation: in the static diatomic lattice, sweeping $k$ across the Brillouin zone produces great-circle trajectories whose parity encodes the $\{0, \pi\}$ Zak classes; under space–time modulation, sideband mixing enlarges the accessible region on $S^2$, enabling continuous paths between states such as $\left(\left|E_1\right\rangle \pm \mathrm{i}\left|E_2\right\rangle\right)/\sqrt{2}$ and their phase-flipped counterparts. This is precisely what is seen around hybridization points in Fig.~\ref{BlochState_TI}, where the $k$-parametric evolution of $(\hat{\alpha}, \hat{\beta})$ explores latitudes and longitudes rather than a single meridian; the transformation from $\left(\ket{E_1} + \mathrm{i}\ket{E_2}\right)/\sqrt{2}$ to $\left(\ket{E_1} - \mathrm{i}\ket{E_2}\right)/\sqrt{2}$ thereby realizes a $Z$-type gate continuously along the band. Because these gate-like maps are rotations on $S^2$, the enclosed solid angle controls the geometric phase accrued by the elastic state, in direct analogy with qubit holonomies. The present approach is distinctive in that it reconstructs the complex coefficients and manipulates the classical superposition state itself on the Bloch sphere—thereby demonstrating gate-level control without decoherence and without post-measurement statistics, a capability not found in earlier treatments of phononic analogues.

\subsubsection{Geometric Phase Formalism for the Two-Level (Bloch-Sphere) Manifold}

We next cast the geometric phase of the diatomic lattice into the canonical language of a two-level system. Any pure lattice state can be written on the Bloch sphere as
\begin{equation}
\ket{\psi} = \hat\alpha \ket{E_1} + \hat\beta \ket{E_2}; \quad \sqrt{\lvert\hat\alpha\rvert^2 + \lvert\hat\beta\rvert^2} = 1,
\end{equation}
so that its evolution is completely specified by the spherical angles $\theta(k) = 2 \arctan (\lvert\hat\beta / \hat\alpha\rvert)$ and $\varphi(k) = \arg(\hat\beta) - \arg(\hat\alpha)$. In the effective three-dimensional Hamiltonian space $\hat{H}(k)$ the adiabatic eigen-spinors are $\left\lvert +_k \right\rangle$ and $\left\lvert -_k \right\rangle$; choosing the usual gauge angle $\varepsilon$ to fix the overall phase, one obtains
\begin{equation}
\left\lvert +_k \right\rangle = e^{\mathrm{i} \varepsilon(\theta, \varphi)} \begin{pmatrix} \cos \frac{\theta(k)}{2} \\ e^{\mathrm{i} \varphi(k)} \sin \frac{\theta(k)}{2} \end{pmatrix}, \quad \text{with } \varepsilon = \frac{\pi}{2} \quad \text{(convenient Bloch gauge)}.
\end{equation}
For a closed trajectory $\zeta \subset \mathrm{BZ}$, the accumulated Zak phase is
\begin{equation}
\chi(\zeta) = \oint_{\zeta} A(k) \, d k, \quad A(k) = \mathrm{i} \left\langle +_k \mid \partial_k +_k \right\rangle,
\label{eq11}
\end{equation}
with the Berry vector (connection)
\begin{equation*}
\mathbf{B}_V(k) = \mathrm{i} \langle +_k \mid  \nabla_k \mid +_k \rangle.
\end{equation*}

On a discretized Brillouin zone—essential for numerical work—the manifestly gauge-invariant "overlap" formula reads \cite{RN252}:
\begin{equation}
\mathbf{B}_V(k_i) = \left( \cos \frac{\theta_i}{2}, e^{-\mathrm{i} \varphi_i} \sin \frac{\theta_i}{2} \right) \begin{pmatrix} \cos \frac{\theta_{i+1}}{2} \\ e^{\mathrm{i} \varphi_{i+1}} \sin \frac{\theta_{i+1}}{2} \end{pmatrix},
\label{eq12}
\end{equation}

\begin{equation}
    \chi_{\mathrm{disc}} = - \operatorname{Im} \left\{ \ln \left[ \prod_{i=1}^{N_c} \mathbf{B}_V(k_i) \right] \right\}.
    \label{eq13}
\end{equation}

Equations (\ref{eq11})--(\ref{eq13}) are the conventional geometric-phase expressions in both the continuum and discrete settings. In the continuum, the Berry connection reads $A(k) = \mathrm{i} \left\langle u_{k} \mid \partial_{k} u_{k} \right\rangle$ and the Zak (Berry) phase is its line integral over a closed loop in the Brillouin zone; on a discretized mesh the same object is evaluated as the imaginary part of the logarithm of the ordered product of nearest-neighbor overlaps. These are the forms we use in (\ref{eq11})--(\ref{eq13}), and they coincide with established treatments of Berry/Zak phases and their lattice discretizations \cite{PhysRevLett.82.2147,RevModPhys.82.1959,fukui2005chern}.

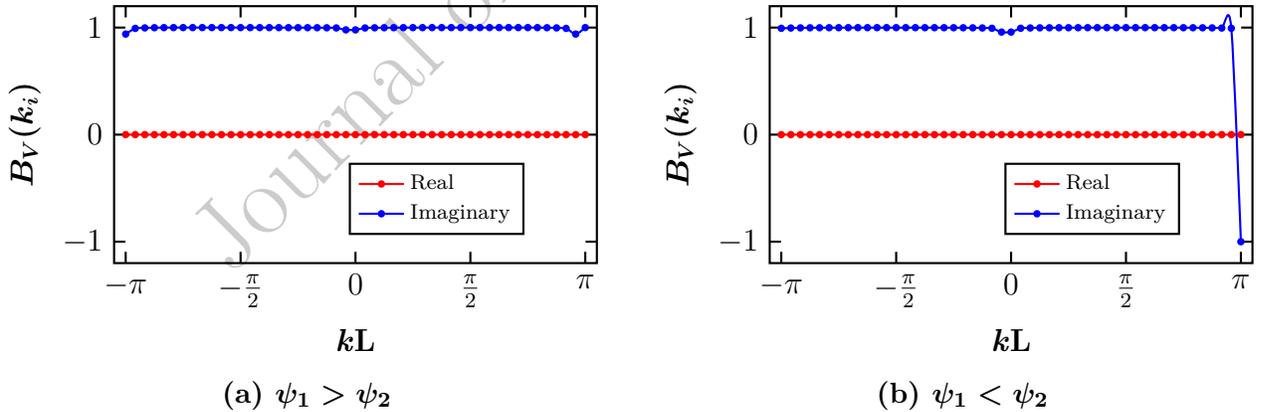
\begin{figure}[H]
     \centering
      \begin{subfigure}{0.49\textwidth}
         \centering
         \begin{tikzpicture}
\begin{axis}[
    height=5cm,
    width=8cm,
    thick,
    xlabel={\textbf{\textit{k}L}},
    ylabel={$\boldsymbol{B_V(k_i)}$},
    xmin=-3.3, xmax=3.3,
    ymin=-1.2, ymax=1.2,
    every x tick/.style={color=black, thick},
    xtick={-3.1416,-1.5708,0,1.5708,3.1416},
    xticklabels={$-\pi$,$-\frac{\pi}{2}$,0,$\frac{\pi}{2}$,$\pi$},
    every y tick/.style={color=black, thick},
    ytick={-1,0,1},
    legend style={at={(0.85,0.25)}, anchor=east}, 
    legend cell align={left},
]

\addplot[
    smooth,
    thick,
    color=red,
    mark=*,
    mark size=1pt,
]
coordinates {
(-3.141592654,	0)
(-3.01069296,	0)
(-2.879793266,	0)
(-2.748893572,	0)
(-2.617993878,	0)
(-2.487094184,	0)
(-2.35619449,	0)
(-2.225294796,	1.38778E-16)
(-2.094395102,	-1.38778E-16)
(-1.963495408,	0)
(-1.832595715,	0)
(-1.701696021,	0)
(-1.570796327,	0)
(-1.439896633,	0)
(-1.308996939,	0)
(-1.178097245,	0)
(-1.047197551,	0)
(-0.916297857,	0)
(-0.785398163,	0)
(-0.654498469,	0)
(-0.523598776,	0)
(-0.392699082,	0)
(-0.261799388,	0)
(-0.130899694,	0)
(0,	0)
(0.130899694,	0)
(0.261799388,	0)
(0.392699082,	0)
(0.523598776,	0)
(0.654498469,	0)
(0.785398163,	0)
(0.916297857,	0)
(1.047197551,	0)
(1.178097245,	0)
(1.308996939,	0)
(1.439896633,	0)
(1.570796327,	0)
(1.701696021,	0)
(1.832595715,	0)
(1.963495408,	-1.38778E-16)
(2.094395102,	1.38778E-16)
(2.225294796,	0)
(2.35619449,	0)
(2.487094184,	0)
(2.617993878,	0)
(2.748893572,	0)
(2.879793266,	0)
(3.01069296,	0)
(3.141592654,	0)

};

    \addlegendentry{\scriptsize Real}

\addplot[
    smooth,
    thick,
    color=blue,
    mark=*,
    mark size=1pt,
]
coordinates {
(-3.141592654,	0.939460433)
(-3.01069296,	0.99243317)
(-2.879793266,	0.997075083)
(-2.748893572,	0.998774715)
(-2.617993878,	0.999501499)
(-2.487094184,	0.999819694)
(-2.35619449,	0.999951779)
(-2.225294796,	0.999995777)
(-2.094395102,	0.999996881)
(-1.963495408,	0.999976307)
(-1.832595715,	0.999943928)
(-1.701696021,	0.999904015)
(-1.570796327,	0.999857887)
(-1.439896633,	0.999805088)
(-1.308996939,	0.999743814)
(-1.178097245,	0.999670835)
(-1.047197551,	0.999580941)
(-0.916297857,	0.999465624)
(-0.785398163,	0.999310181)
(-0.654498469,	0.999086815)
(-0.523598776,	0.998735619)
(-0.392699082,	0.998098185)
(-0.261799388,	0.996561559)
(-0.130899694,	0.978659157)
(0,	0.978659157)
(0.130899694,	0.996561559)
(0.261799388,	0.998098185)
(0.392699082,	0.998735619)
(0.523598776,	0.999086815)
(0.654498469,	0.999310181)
(0.785398163,	0.999465624)
(0.916297857,	0.999580941)
(1.047197551,	0.999670835)
(1.178097245,	0.999743814)
(1.308996939,	0.999805088)
(1.439896633,	0.999857887)
(1.570796327,	0.999904015)
(1.701696021,	0.999943928)
(1.832595715,	0.999976307)
(1.963495408,	0.999996881)
(2.094395102,	0.999995777)
(2.225294796,	0.999951779)
(2.35619449,	0.999819694)
(2.487094184,	0.999501499)
(2.617993878,	0.998774715)
(2.748893572,	0.997075083)
(2.879793266,	0.99243317)
(3.01069296,	0.939460433)
(3.141592654,	1)
};
    \addlegendentry{\scriptsize Imaginary}

\end{axis}

\end{tikzpicture}
         \caption{$\boldsymbol{\psi_1>\psi_2}$}
         \label{BV_1>2}
     \end{subfigure}
     \hfill
     \begin{subfigure}{0.49\textwidth}
         \centering
         \begin{tikzpicture}
\begin{axis}[
    height=5cm,
    width=8cm,
    thick,
    xlabel={\textbf{\textit{k}L}},
    ylabel={$\boldsymbol{B_V(k_i)}$},
    xmin=-3.3, xmax=3.3,
    ymin=-1.2, ymax=1.2,
    every x tick/.style={color=black, thick},
    xtick={-3.1416,-1.5708,0,1.5708,3.1416},
    xticklabels={$-\pi$,$-\frac{\pi}{2}$,0,$\frac{\pi}{2}$,$\pi$},
    every y tick/.style={color=black, thick},
    ytick={-1,0,1},
    legend style={at={(0.85,0.25)}, anchor=east}, 
    legend cell align={left},
]

\addplot[
    smooth,
    thick,
    color=red,
    mark=*,
    mark size=1pt,
]
coordinates {
(-3.141592654,	5.31165E-18)
(-3.01069296,	0)
(-2.879793266,	0)
(-2.748893572,	0)
(-2.617993878,	2.77556E-17)
(-2.487094184,	2.77556E-17)
(-2.35619449,	2.77556E-17)
(-2.225294796,	2.77556E-17)
(-2.094395102,	0)
(-1.963495408,	0)
(-1.832595715,	0)
(-1.701696021,	0)
(-1.570796327,	0)
(-1.439896633,	0)
(-1.308996939,	0)
(-1.178097245,	0)
(-1.047197551,	0)
(-0.916297857,	0)
(-0.785398163,	0)
(-0.654498469,	0)
(-0.523598776,	0)
(-0.392699082,	0)
(-0.261799388,	0)
(-0.130899694,	0)
(0,	0)
(0.130899694,	0)
(0.261799388,	0)
(0.392699082,	0)
(0.523598776,	0)
(0.654498469,	0)
(0.785398163,	0)
(0.916297857,	0)
(1.047197551,	0)
(1.178097245,	0)
(1.308996939,	0)
(1.439896633,	0)
(1.570796327,	0)
(1.701696021,	0)
(1.832595715,	0)
(1.963495408,	0)
(2.094395102,	2.77556E-17)
(2.225294796,	2.77556E-17)
(2.35619449,	2.77556E-17)
(2.487094184,	2.77556E-17)
(2.617993878,	0)
(2.748893572,	0)
(2.879793266,	0)
(3.01069296,	5.31165E-18)
(3.141592654,	-1.22465E-16)
};
    \addlegendentry{\scriptsize Real}

\addplot[
    smooth,
    thick,
    color=blue,
    mark=*,
    mark size=1pt,
]
coordinates {
(-3.141592654,	0.993311725)
(-3.01069296,	0.995812986)
(-2.879793266,	0.997246764)
(-2.748893572,	0.998093438)
(-2.617993878,	0.998604215)
(-2.487094184,	0.998918869)
(-2.35619449,	0.999116255)
(-2.225294796,	0.99924107)
(-2.094395102,	0.999318838)
(-1.963495408,	0.999364217)
(-1.832595715,	0.999385533)
(-1.701696021,	0.999387211)
(-1.570796327,	0.999371025)
(-1.439896633,	0.999336639)
(-1.308996939,	0.999281624)
(-1.178097245,	0.999200962)
(-1.047197551,	0.999085796)
(-0.916297857,	0.998920778)
(-0.785398163,	0.998678307)
(-0.654498469,	0.998304821)
(-0.523598776,	0.99768292)
(-0.392699082,	0.996498972)
(-0.261799388,	0.993533332)
(-0.130899694,	0.958152834)
(0,	0.958152834)
(0.130899694,	0.993533332)
(0.261799388,	0.996498972)
(0.392699082,	0.99768292)
(0.523598776,	0.998304821)
(0.654498469,	0.998678307)
(0.785398163,	0.998920778)
(0.916297857,	0.999085796)
(1.047197551,	0.999200962)
(1.178097245,	0.999281624)
(1.308996939,	0.999336639)
(1.439896633,	0.999371025)
(1.570796327,	0.999387211)
(1.701696021,	0.999385533)
(1.832595715,	0.999364217)
(1.963495408,	0.999318838)
(2.094395102,	0.99924107)
(2.225294796,	0.999116255)
(2.35619449,	0.998918869)
(2.487094184,	0.998604215)
(2.617993878,	0.998093438)
(2.748893572,	0.997246764)
(2.879793266,	0.995812986)
(3.01069296,	0.993311725)
(3.141592654,	-1)
};
    \addlegendentry{\scriptsize Imaginary}

\end{axis}

\end{tikzpicture}
         \caption{$\boldsymbol{\psi_1<\psi_2}$}
         \label{BV_1<2}
     \end{subfigure}
     \caption{Analytical result of the real and imaginary component of the Berry vector corresponding to the stiffness of (a) $\psi_1>\psi_2$ and (b) $\psi_1<\psi_2$.}
     \label{Fig3}
\end{figure}

With these definitions in hand, the unit-cell stiffness ordering fixes how the Bloch-sphere angles $\theta(k)$ and $\varphi(k)$ wind, and therefore how the geometric phase accumulates across the zone. The Berry vector extracted from Eq.~(\ref{eq12}) is purely imaginary in our data for both orderings (Figs. \ref{BV_1>2}--\subref{BV_1<2}), so the reported $B_V(k)$ traces are controlled by their imaginary part. When $\psi_{1} > \psi_{2}$ [Fig.~\ref{BV_1>2}], $B_V(k)$ is essentially flat in $k$, leading to a vanishing line integral and a Zak phase of $0$. When $\psi_{1} < \psi_{2}$ (Fig.~\ref{BV_1<2}), $B_V(k)$ changes sign once across the Brillouin zone, giving a net accumulation of $\pi$. This $0/\pi$ dichotomy is the consequence of inversion symmetry in diatomic (SSH-type) chains and matches direct measurements of the Zak phase in one-dimensional lattices \cite{PhysRevB.108.035403,RN252}.

The geometric content of Fig.~\ref{Fig3} can be read directly on the Bloch sphere. When the state dwells near a pole ($\theta \approx 0$ or $\pi$), neighboring $k$-points have almost purely real and positive overlaps, so $B_V(k)$ exhibits only small imaginary excursions and contributes negligibly to the total phase. By contrast, when the trajectory crosses the equator ($\theta \approx \pi/2$), the azimuth $\varphi$ winds rapidly; the overlaps acquire a large imaginary component, and the Zak phase picks up sharply. Varying the stiffness ratio (e.g., $\psi_{1}/\psi_{2} = 2$ versus $0.5$) shifts where $\theta(k)$ and $\varphi(k)$ vary most strongly, thereby relocating the $k$-intervals of enhanced Berry connection response. In other words, it is the twist of the state on the sphere---not its absolute location---that governs the geometric phase, in keeping with the general theory \cite{RevModPhys.82.1959}.

Seen through this lens, the Bloch sphere parameterization provides a compact and gauge-transparent narrative that is fully consistent with the spectral evaluation of the Berry/Zak phase via (\ref{eq11})--(\ref{eq13}). In the two-mass case, $\psi_{1} > \psi_{2}$ yields a trivial loop with total twist $\chi=0$, while $\psi_{1} < \psi_{2}$ yields a loop with a single twist $\chi = \pi$. This is precisely the classical-superposition analogue of the SSH chain's inversion-protected $0/\pi$ Zak-phase quantization: our diatomic lattice realizes the same binary topology, but the path lives on a classical Bloch sphere traced by the superposed modal amplitudes $(\alpha, \beta)$. The agreement between the Bloch sphere picture and the discretized-overlap evaluation in (\ref{eq12})--(\ref{eq13}) reflects the robustness of gauge-invariant lattice formulas on discrete $k$-meshes \cite{fukui2005chern}.

Finally, we emphasize that SSH-type quantization of the Zak phase is well known in electronic, photonic, and ultracold-atom settings; however, in our system the same topological content is accessed and visualized through strictly classical superpositions on a Bloch sphere, with the stiffness ordering playing the role of dimerization. This qubit-like geometric representation---paired with a manifestly gauge-invariant numerical evaluation via (\ref{eq11})--(\ref{eq13})---provides a physically transparent bridge between classical elasticity and band-topological concepts, and it underpins the later extensions to multi-mass cells and spatiotemporal modulation developed in this manuscript \cite{RN252,RevModPhys.82.1959}.

The results demonstrate that the Bloch sphere formulation provides direct access to the underlying Hilbert space structure of elastic lattices. By representing the intra-cell modal coefficients as a coherent state in a two level space and tracking their evolution through the Brillouin zone, the geometric mechanisms responsible for phase accumulation become visible and physically interpretable. The shape of the trajectory and the region of the sphere that is explored reveal whether the lattice is constrained to a limited family of configurations or if it can access more extensive superpositions analogous to qubit rotations. This viewpoint remains applicable even when inversion symmetry is relaxed or when modulation induces open paths that are not easily handled through traditional Berry phase evaluation. In this way, the Bloch sphere evolution links topological classification with controllable state dynamics and establishes the basis for the quantum analogous interpretation developed in the following sections, where symmetry breaking and time varying stiffness further expand the accessible portion of the Hilbert space.

\subsection{Triatomic Unit Cell}
\label{Three Mass Unit}
We now consider a unit cell composed of three identical point masses that interact through linear, nearest-neighbor springs of stiffnesses $\{\psi_{1}, \psi_{2}, \psi_{3}\}$. Denoting by $u_{n, N_i}(t)$ the displacement of mass $n=1,2,3$ in the $N_i$-th cell, the equations of motion read
\begin{equation}
\begin{aligned}
m \ddot{u}_{1, N_i}(t) &= \psi_{3} \bigl[u_{3, N_{i-1}}(t) - u_{1, N_i}(t)\bigr] - \psi_{1} \bigl[u_{1, N_i}(t) - u_{2, N_i}(t)\bigr], \\
m \ddot{u}_{2, N_i}(t) &= \psi_{1} \bigl[u_{1, N_i}(t) - u_{2, N_i}(t)\bigr] - \psi_{2} \bigl[u_{2, N_i}(t) - u_{3, N_i}(t)\bigr], \\
m \ddot{u}_{3, N_i}(t) &= \psi_{2} \bigl[u_{2, N_i}(t) - u_{3, N_i}(t)\bigr] - \psi_{3} \bigl[u_{3, N_{i}}(t) - u_{1, N_{i+1}}(t)\bigr].
\end{aligned}
\end{equation}
These are solved with the Bloch ansatz. $u_{n, N_i}(t) = A_n e^{\mathrm{i} k N_i L} e^{\mathrm{i} \omega t}$. As in Sec.~\ref{diatomic unit}, we examine how inversion symmetry of the spring pattern controls whether the Zak (Berry) phase is quantized ($0$ or $\pi$) or not. For a three-mass cell, inversion symmetry is the compact condition $\psi_i = \psi_{N_m - i}$ with $N_m = 3$; when it holds, the Zak phase of each isolated band is pinned to $0$ or $\pi \pmod{2\pi}$---a standard result we will exploit below. Breaking inversion, by contrast, typically de-quantizes the geometric phase while leaving the spectrum invariant under origin shifts \cite{PhysRevLett.127.147401,xiao2016coexistence}.
To analyze intra-cell motion, it is convenient to resolve the displacement vector into the orthonormal basis
\[
\ket{E_1} = \frac{1}{\sqrt{3}} \begin{bmatrix} 1 \\ 1 \\ 1 \end{bmatrix}, \quad
\ket{E_2} = \frac{1}{\sqrt{2}} \begin{bmatrix} 1 \\ 0 \\ -1 \end{bmatrix}, \quad
\ket{E_3} = \frac{1}{\sqrt{6}} \begin{bmatrix} 1 \\ -2 \\ 1 \end{bmatrix},
\]
so that
\begin{equation}
\begin{bmatrix}
u_{1, N_i}(t) \\
u_{2, N_i}(t) \\
u_{3, N_i}(t)
\end{bmatrix}
= 
\begin{bmatrix}
A_1 \\
A_2 \\
A_3
\end{bmatrix} e^{\mathrm{i} k N_i L} e^{i \omega t} \equiv (\alpha \ket{E_1} + \beta \ket{E_2} + \gamma \ket{E_3}) e^{i \omega t}.
\end{equation}

Here $A_n$ are the traveling-wave amplitudes of the physical masses, while $\alpha, \beta, \gamma$ are the complex modal coefficients in the internal basis $\{\ket{E_1}, \ket{E_2}, \ket{E_3}\}$. To compare trends across bands and parameter sets, we rescale the modal coefficients by the band-resolved physical amplitudes through
\begin{equation}
\begin{aligned}
\hat{\alpha} &= \frac{1}{\sqrt{(|A_1|)^2+(|A_2|)^2+(|A_3|)^2}} \left( \frac{1}{\sqrt{3}} A_1 + \frac{1}{\sqrt{2}} A_2 + \frac{1}{\sqrt{6}} A_3 \right) e^{\mathrm{i} k N_i L}, \\
\hat{\beta} &= \frac{1}{\sqrt{(|A_1|)^2+(|A_2|)^2+(|A_3|)^2}} \left( \frac{1}{\sqrt{3}} A_1 - \frac{2}{\sqrt{6}} A_3 \right) e^{\mathrm{i} k N_i L}, \\
\hat{\gamma} &= \frac{1}{\sqrt{(|A_1|)^2+(|A_2|)^2+(|A_3|)^2}} \left( \frac{1}{\sqrt{3}} A_1 - \frac{1}{\sqrt{2}} A_2 + \frac{1}{\sqrt{6}} A_3 \right) e^{\mathrm{i} k N_i L}
\end{aligned}   
\end{equation}
We study three representative stiffness sets:
\begin{table}[H]
    \centering
\begin{tabular}{lll}
(i) $\psi_1 = \psi_2,\ \psi_3 = 2\psi_1$; &
(ii) $\psi_1 = \psi_2,\ \psi_3 = \frac{1}{2}\psi_1$; &
(iii) $\psi_2 = \psi_3,\ \psi_1 = 2\psi_2$. 
\end{tabular}
\end{table}

Cases (i)--(ii) satisfy inversion symmetry and therefore must yield Zak phases equal to $0$ or $\pi \pmod{2\pi}$, while case (iii) explicitly breaks inversion and thus generically produces non-quantized geometric phases \cite{PhysRevLett.127.147401}. For each case, we compute the dispersion and the cell-wise complex amplitudes $(\hat\alpha, \hat\beta, \hat\gamma)$ across the Brillouin zone $k L \in [-\pi, \pi]$. The amplitude of the coefficients at different at representative stiffness are provided in the supplementary materials.

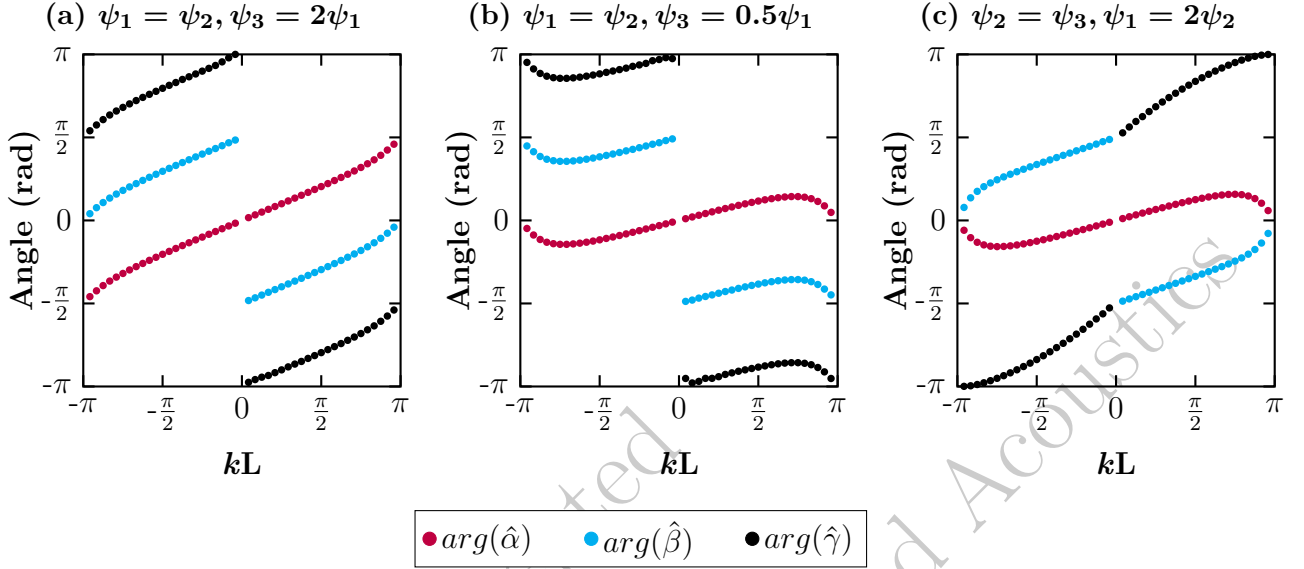
\begin{figure}[H]
\centering
\begin{subfigure}{0.32\textwidth}
  \centering
  \caption{$\boldsymbol{\psi_1=\psi_2, \psi_3=2\psi_1}$}\label{Phase_AC}
  \begin{tikzpicture}
\begin{axis}[
    height=6cm,
    width=5.8cm,
    thick,
    xlabel={\textbf{\textit{k}L}},
    ylabel={\textbf{Angle (rad)}},
    ylabel style={at={(axis description cs:-0.1,0.5)},anchor=south}, 
    xmin=-3.1416, xmax=3.1416,
    ymin=-3.1416, ymax=3.1416,
    every x tick/.style={color=black, thick},
    xtick={-3.1416,-1.5708,0,1.5708,3.1416},
    xticklabels={-$\pi$,-$\frac{\pi}{2}$,0,$\frac{\pi}{2}$,$\pi$},
    every y tick/.style={color=black, thick},
    ytick={-3.1416,-1.5708,0,1.5708,3.1416},
    yticklabels={-$\pi$,-$\frac{\pi}{2}$,0,$\frac{\pi}{2}$,$\pi$},
    every y tick/.style={color=black, thick},
    legend style={at={(1.15,0.3)}, anchor=east}, 
    legend cell align={left},
]

\addplot[
  only marks,
    color=purple,
    mark=*,
    mark size=1pt,
]
coordinates {	
(-3.01069296,	-1.441814814)
(-2.879793266,	-1.331713042)
(-2.748893572,	-1.232633931)
(-2.617993878,	-1.147299171)
(-2.487094184,	-1.072404437)
(-2.35619449,	-1.003097381)
(-2.225294796,	-0.936636374)
(-2.094395102,	-0.874013964)
(-1.963495408,	-0.813899974)
(-1.832595715,	-0.755066704)
(-1.701696021,	-0.697332457)
(-1.570796327,	-0.64081888)
(-1.439896633,	-0.585547067)
(-1.308996939,	-0.530599982)
(-1.178097245,	-0.47584956)
(-1.047197551,	-0.422255707)
(-0.916297857,	-0.368631354)
(-0.785398163,	-0.315586187)
(-0.654498469,	-0.262401355)
(-0.523598776,	-0.209724763)
(-0.392699082,	-0.156879862)
(-0.261799388,	-0.104738267)
(-0.130899694,	-0.052599252)
(0.130899694,	0.05262032)
(0.261799388,	0.105091711)
(0.392699082,	0.157564138)
(0.523598776,	0.209792263)
(0.654498469,	0.262825697)
(0.785398163,	0.315489047)
(0.916297857,	0.368884874)
(1.047197551,	0.422368354)
(1.178097245,	0.475997875)
(1.308996939,	0.53062554)
(1.439896633,	0.585447505)
(1.570796327,	0.640610803)
(1.701696021,	0.697079692)
(1.832595715,	0.754638366)
(1.963495408,	0.813276638)
(2.094395102,	0.873832154)
(2.225294796,	0.935970058)
(2.35619449,	1.002156496)
(2.487094184,	1.07215338)
(2.617993878,	1.146513503)
(2.748893572,	1.232128569)
(2.879793266,	1.331465351)
(3.01069296,	1.442019392)
};


\addplot[
  only marks,
    color=cyan,
    mark=*,
    mark size=1pt,
]
coordinates {
(-3.01069296,	0.127782764)
(-2.879793266,	0.241223472)
(-2.748893572,	0.338368261)
(-2.617993878,	0.422813904)
(-2.487094184,	0.498662565)
(-2.35619449,	0.568291375)
(-2.225294796,	0.634068534)
(-2.094395102,	0.696632546)
(-1.963495408,	0.757080415)
(-1.832595715,	0.816024957)
(-1.701696021,	0.873332185)
(-1.570796327,	0.929602704)
(-1.439896633,	0.985019325)
(-1.308996939,	1.04001388)
(-1.178097245,	1.094851508)
(-1.047197551,	1.148006859)
(-0.916297857,	1.201391414)
(-0.785398163,	1.255172541)
(-0.654498469,	1.308915311)
(-0.523598776,	1.360135064)
(-0.392699082,	1.413732923)
(-0.261799388,	1.469408067)
(-0.130899694,	1.52070286)
(0.130899694,	-1.515766397)
(0.261799388,	-1.464281858)
(0.392699082,	-1.412346942)
(0.523598776,	-1.360002853)
(0.654498469,	-1.308452728)
(0.785398163,	-1.255273962)
(0.916297857,	-1.20152622)
(1.047197551,	-1.148285275)
(1.178097245,	-1.0952509)
(1.308996939,	-1.040335813)
(1.439896633,	-0.985731842)
(1.570796327,	-0.930259574)
(1.701696021,	-0.873605986)
(1.832595715,	-0.816217209)
(1.963495408,	-0.757422351)
(2.094395102,	-0.696686854)
(2.225294796,	-0.634257114)
(2.35619449,	-0.568485607)
(2.487094184,	-0.498626661)
(2.617993878,	-0.422911983)
(2.748893572,	-0.338748539)
(2.879793266,	-0.241780827)
(3.01069296,	-0.127822531)

};


\addplot[
  only marks,
    color=black,
    mark=*,
    mark size=1pt,
]
coordinates {
(-3.01069296,	1.696847098)
(-2.879793266,	1.805794905)
(-2.748893572,	1.90746255)
(-2.617993878,	1.994527498)
(-2.487094184,	2.069959708)
(-2.35619449,	2.138679432)
(-2.225294796,	2.204866642)
(-2.094395102,	2.267691019)
(-1.963495408,	2.327557137)
(-1.832595715,	2.385637582)
(-1.701696021,	2.442773684)
(-1.570796327,	2.499787573)
(-1.439896633,	2.555770419)
(-1.308996939,	2.609200728)
(-1.178097245,	2.667266509)
(-1.047197551,	2.717731974)
(-0.916297857,	2.770978592)
(-0.785398163,	2.8243388)
(-0.654498469,	2.876843432)
(-0.523598776,	2.931810875)
(-0.392699082,	2.999556301)
(-0.261799388,	3.076214335)
(-0.130899694,	3.141260857)
(0.130899694,	-3.062864795)
(0.261799388,	-3.010932718)
(0.392699082,	-2.965792263)
(0.523598776,	-2.93336729)
(0.654498469,	-2.880196985)
(0.785398163,	-2.827240742)
(0.916297857,	-2.770774224)
(1.047197551,	-2.717732823)
(1.178097245,	-2.667939932)
(1.308996939,	-2.610460458)
(1.439896633,	-2.557234846)
(1.570796327,	-2.501045695)
(1.701696021,	-2.444377946)
(1.832595715,	-2.387476632)
(1.963495408,	-2.329649483)
(2.094395102,	-2.268078144)
(2.225294796,	-2.206173323)
(2.35619449,	-2.140149286)
(2.487094184,	-2.068756493)
(2.617993878,	-1.99487183)
(2.748893572,	-1.908119778)
(2.879793266,	-1.809032956)
(3.01069296,	-1.694566933)
};

\end{axis}

\end{tikzpicture}
\end{subfigure}\hfill
\begin{subfigure}{0.32\textwidth}
  \centering
  \caption{$\boldsymbol{\psi_1=\psi_2, \psi_3=0.5\psi_1}$}\label{Phase_1O}
  \begin{tikzpicture}
\begin{axis}[
    height=6cm,
    width=5.8cm,
    thick,
    xlabel={\textbf{\textit{k}L}},
    ylabel={\textbf{Angle (rad)}},
    ylabel style={at={(axis description cs:-0.1,0.5)},anchor=south}, 
    xmin=-3.1416, xmax=3.1416,
    ymin=-3.1416, ymax=3.1416,
    every x tick/.style={color=black, thick},
    xtick={-3.1416,-1.5708,0,1.5708,3.1416},
    xticklabels={-$\pi$,-$\frac{\pi}{2}$,0,$\frac{\pi}{2}$,$\pi$},
    every y tick/.style={color=black, thick},
    ytick={-3.1416,-1.5708,0,1.5708,3.1416},
    yticklabels={-$\pi$,-$\frac{\pi}{2}$,0,$\frac{\pi}{2}$,$\pi$},
    every y tick/.style={color=black, thick},
    legend style={at={(1.15,0.3)}, anchor=east}, 
    legend cell align={left},
]

\addplot[
  only marks,
    color=purple,
    mark=*,
    mark size=1pt,
]
coordinates {	
(-3.01069296,	-0.151725515)
(-2.879793266,	-0.275718382)
(-2.748893572,	-0.360663749)
(-2.617993878,	-0.412794891)
(-2.487094184,	-0.439951117)
(-2.35619449,	-0.451980057)
(-2.225294796,	-0.45241956)
(-2.094395102,	-0.444011126)
(-1.963495408,	-0.430125931)
(-1.832595715,	-0.412180332)
(-1.701696021,	-0.390776658)
(-1.570796327,	-0.36705801)
(-1.439896633,	-0.340665596)
(-1.308996939,	-0.313601709)
(-1.178097245,	-0.28490566)
(-1.047197551,	-0.254990517)
(-0.916297857,	-0.224627399)
(-0.785398163,	-0.193491077)
(-0.654498469,	-0.161944343)
(-0.523598776,	-0.12987625)
(-0.392699082,	-0.097701454)
(-0.261799388,	-0.065349627)
(-0.130899694,	-0.032598041)
(0.130899694,	0.032726791)
(0.261799388,	0.065682034)
(0.392699082,	0.097882108)
(0.523598776,	0.130278856)
(0.654498469,	0.162266387)
(0.785398163,	0.193858334)
(0.916297857,	0.225052571)
(1.047197551,	0.255131072)
(1.178097245,	0.28496165)
(1.308996939,	0.313548228)
(1.439896633,	0.340666478)
(1.570796327,	0.36672516)
(1.701696021,	0.39036627)
(1.832595715,	0.411647649)
(1.963495408,	0.429994859)
(2.094395102,	0.443339328)
(2.225294796,	0.45128209)
(2.35619449,	0.451036647)
(2.487094184,	0.439530002)
(2.617993878,	0.411320913)
(2.748893572,	0.359083878)
(2.879793266,	0.275701909)
(3.01069296,	0.151838447)
};


\addplot[
  only marks,
    color=cyan,
    mark=*,
    mark size=1pt,
]
coordinates {
(-3.01069296,	1.408978218)
(-2.879793266,	1.299743112)
(-2.748893572,	1.210150503)
(-2.617993878,	1.157562241)
(-2.487094184,	1.129953531)
(-2.35619449,	1.118175582)
(-2.225294796,	1.117962115)
(-2.094395102,	1.125833719)
(-1.963495408,	1.140414901)
(-1.832595715,	1.157982649)
(-1.701696021,	1.179896341)
(-1.570796327,	1.203353056)
(-1.439896633,	1.229896955)
(-1.308996939,	1.256335924)
(-1.178097245,	1.285312438)
(-1.047197551,	1.316073352)
(-0.916297857,	1.344332838)
(-0.785398163,	1.376061764)
(-0.654498469,	1.407912029)
(-0.523598776,	1.441328123)
(-0.392699082,	1.470301737)
(-0.261799388,	1.510697978)
(-0.130899694,	1.542851329)
(0.130899694,	-1.531827639)
(0.261799388,	-1.506835325)
(0.392699082,	-1.46983716)
(0.523598776,	-1.438911159)
(0.654498469,	-1.409089281)
(0.785398163,	-1.37746595)
(0.916297857,	-1.345225687)
(1.047197551,	-1.316893976)
(1.178097245,	-1.285813935)
(1.308996939,	-1.257326126)
(1.439896633,	-1.230174544)
(1.570796327,	-1.204647618)
(1.701696021,	-1.180438663)
(1.832595715,	-1.159037502)
(1.963495408,	-1.140661892)
(2.094395102,	-1.126997901)
(2.225294796,	-1.119671481)
(2.35619449,	-1.11883609)
(2.487094184,	-1.129999012)
(2.617993878,	-1.159131087)
(2.748893572,	-1.212017963)
(2.879793266,	-1.299050888)
(3.01069296,	-1.408205482)
};


\addplot[
  only marks,
    color=black,
    mark=*,
    mark size=1pt,
]
coordinates {
(-3.01069296,	2.989660091)
(-2.879793266,	2.866223777)
(-2.748893572,	2.781224776)
(-2.617993878,	2.728729037)
(-2.487094184,	2.700896761)
(-2.35619449,	2.688663371)
(-2.225294796,	2.688320156)
(-2.094395102,	2.697353954)
(-1.963495408,	2.710596044)
(-1.832595715,	2.72935982)
(-1.701696021,	2.750441546)
(-1.570796327,	2.772891916)
(-1.439896633,	2.801277736)
(-1.308996939,	2.826822362)
(-1.178097245,	2.853435436)
(-1.047197551,	2.881099837)
(-0.916297857,	2.91519279)
(-0.785398163,	2.941811675)
(-0.654498469,	2.970922455)
(-0.523598776,	3.022251169)
(-0.392699082,	3.044180235)
(-0.261799388,	3.079160874)
(-0.130899694,	3.061503196)
(0.130899694,	-2.989740432)
(0.261799388,	-3.07453927)
(0.392699082,	-3.045049222)
(0.523598776,	-2.988037271)
(0.654498469,	-2.987754436)
(0.785398163,	-2.95633957)
(0.916297857,	-2.913490701)
(1.047197551,	-2.883917356)
(1.178097245,	-2.853448858)
(1.308996939,	-2.82724424)
(1.439896633,	-2.801287096)
(1.570796327,	-2.775353781)
(1.701696021,	-2.752535903)
(1.832595715,	-2.730493882)
(1.963495408,	-2.712614727)
(2.094395102,	-2.699547142)
(2.225294796,	-2.691882615)
(2.35619449,	-2.691456542)
(2.487094184,	-2.702900034)
(2.617993878,	-2.730271937)
(2.748893572,	-2.782440057)
(2.879793266,	-2.86516787)
(3.01069296,	-2.990210951)
};

\end{axis}

\end{tikzpicture}
\end{subfigure}\hfill
\begin{subfigure}{0.32\textwidth}
  \centering
  \caption{$\boldsymbol{\psi_2=\psi_3, \psi_1=2\psi_2}$}\label{Phase_2O}
  \begin{tikzpicture}
\begin{axis}[
    height=6cm,
    width=5.8cm,
    thick,
    xlabel={\textbf{\textit{k}L}},
    ylabel={\textbf{Angle (rad)}},
    ylabel style={at={(axis description cs:-0.1,0.5)},anchor=south}, 
    xmin=-3.1416, xmax=3.1416,
    ymin=-3.1416, ymax=3.1416,
    every x tick/.style={color=black, thick},
    xtick={-3.1416,-1.5708,0,1.5708,3.1416},
    xticklabels={-$\pi$,-$\frac{\pi}{2}$,0,$\frac{\pi}{2}$,$\pi$},
    every y tick/.style={color=black, thick},
    ytick={-3.1416,-1.5708,0,1.5708,3.1416},
    yticklabels={-$\pi$,-$\frac{\pi}{2}$,0,$\frac{\pi}{2}$,$\pi$},
    every y tick/.style={color=black, thick},
    legend style={at={(1.15,0.3)}, anchor=east}, 
    legend cell align={left},
]

\addplot[
  only marks,
    color=purple,
    mark=*,
    mark size=1pt,
]
coordinates {	
(-3.01069296,	-0.18595477)
(-2.879793266,	-0.32621977)
(-2.748893572,	-0.413450026)
(-2.617993878,	-0.462768061)
(-2.487094184,	-0.487429088)
(-2.35619449,	-0.495639391)
(-2.225294796,	-0.492013343)
(-2.094395102,	-0.480938796)
(-1.963495408,	-0.464521192)
(-1.832595715,	-0.443763061)
(-1.701696021,	-0.419753975)
(-1.570796327,	-0.393363156)
(-1.439896633,	-0.36532498)
(-1.308996939,	-0.335614548)
(-1.178097245,	-0.304381236)
(-1.047197551,	-0.272791549)
(-0.916297857,	-0.240059203)
(-0.785398163,	-0.206853072)
(-0.654498469,	-0.172927544)
(-0.523598776,	-0.138821737)
(-0.392699082,	-0.104097776)
(-0.261799388,	-0.069735498)
(-0.130899694,	-0.035081052)
(0.130899694,	0.035045795)
(0.261799388,	0.069908944)
(0.392699082,	0.104694286)
(0.523598776,	0.138880113)
(0.654498469,	0.173207155)
(0.785398163,	0.20676267)
(0.916297857,	0.2402653)
(1.047197551,	0.272855094)
(1.178097245,	0.30449505)
(1.308996939,	0.335654158)
(1.439896633,	0.365231798)
(1.570796327,	0.393202398)
(1.701696021,	0.419548739)
(1.832595715,	0.443526486)
(1.963495408,	0.464153103)
(2.094395102,	0.480834531)
(2.225294796,	0.491574797)
(2.35619449,	0.495070946)
(2.487094184,	0.48739499)
(2.617993878,	0.462347472)
(2.748893572,	0.413275485)
(2.879793266,	0.325870799)
(3.01069296,	0.186027187)
};


\addplot[
  only marks,
    color=cyan,
    mark=*,
    mark size=1pt,
]
coordinates {
(-3.01069296,	0.245902551)
(-2.879793266,	0.427665654)
(-2.748893572,	0.550114604)
(-2.617993878,	0.638556383)
(-2.487094184,	0.709139522)
(-2.35619449,	0.769118219)
(-2.225294796,	0.823325458)
(-2.094395102,	0.873688747)
(-1.963495408,	0.921735566)
(-1.832595715,	0.968410994)
(-1.701696021,	1.01374064)
(-1.570796327,	1.058161693)
(-1.439896633,	1.101824143)
(-1.308996939,	1.145283534)
(-1.178097245,	1.189062581)
(-1.047197551,	1.231256656)
(-0.916297857,	1.27373103)
(-0.785398163,	1.317035359)
(-0.654498469,	1.36036407)
(-0.523598776,	1.401195267)
(-0.392699082,	1.444095683)
(-0.261799388,	1.489949797)
(-0.130899694,	1.531823732)
(0.130899694,	-1.527068602)
(0.261799388,	-1.485304692)
(0.392699082,	-1.442961765)
(0.523598776,	-1.400984679)
(0.654498469,	-1.359858098)
(0.785398163,	-1.317151213)
(0.916297857,	-1.273997192)
(1.047197551,	-1.231554311)
(1.178097245,	-1.189726596)
(1.308996939,	-1.145887403)
(1.439896633,	-1.102543943)
(1.570796327,	-1.058799399)
(1.701696021,	-1.014058589)
(1.832595715,	-0.96897588)
(1.963495408,	-0.922365469)
(2.094395102,	-0.873791217)
(2.225294796,	-0.823407084)
(2.35619449,	-0.769405176)
(2.487094184,	-0.709229578)
(2.617993878,	-0.638602352)
(2.748893572,	-0.550453093)
(2.879793266,	-0.427838431)
(3.01069296,	-0.245766564)
};


\addplot[
  only marks,
    color=black,
    mark=*,
    mark size=1pt,
]
coordinates {
(-3.01069296,	-3.139026277)
(-2.879793266,	-3.128514338)
(-2.748893572,	-3.104283301)
(-2.617993878,	-3.071596502)
(-2.487094184,	-3.031847174)
(-2.35619449,	-2.987310409)
(-2.225294796,	-2.937380559)
(-2.094395102,	-2.883374227)
(-1.963495408,	-2.825858588)
(-1.832595715,	-2.764318102)
(-1.701696021,	-2.699704356)
(-1.570796327,	-2.632027262)
(-1.439896633,	-2.559930641)
(-1.308996939,	-2.484744725)
(-1.178097245,	-2.406001651)
(-1.047197551,	-2.324747479)
(-0.916297857,	-2.238651212)
(-0.785398163,	-2.151433226)
(-0.654498469,	-2.059524518)
(-0.523598776,	-1.962830021)
(-0.392699082,	-1.863490646)
(-0.261799388,	-1.76748779)
(-0.130899694,	-1.655320665)
(0.130899694,	1.657222213)
(0.261799388,	1.768018068)
(0.392699082,	1.872243197)
(0.523598776,	1.964156826)
(0.654498469,	2.059800308)
(0.785398163,	2.150986501)
(0.916297857,	2.240292622)
(1.047197551,	2.324584693)
(1.178097245,	2.406290423)
(1.308996939,	2.485142097)
(1.439896633,	2.558990382)
(1.570796327,	2.631124916)
(1.701696021,	2.699097093)
(1.832595715,	2.764323774)
(1.963495408,	2.82562575)
(2.094395102,	2.883369231)
(2.225294796,	2.936672506)
(2.35619449,	2.987332822)
(2.487094184,	3.032856701)
(2.617993878,	3.071849011)
(2.748893572,	3.104234664)
(2.879793266,	3.126980726)
(3.01069296,	3.139886103)
};

\end{axis}

\end{tikzpicture}
\end{subfigure}

\vspace{-0.5em} 

\begin{tikzpicture}
\begin{axis}[
  hide axis,
  xmin=0,xmax=1,ymin=0,ymax=1,
  legend columns=3,
  legend cell align=left,
  legend style={
    /tikz/every even column/.append style={column sep=1.5em}
  }
]
  \addlegendimage{only marks, mark=*, mark size=2.5pt, purple}
  \addlegendentry{$arg(\hat\alpha)$}
  \addlegendimage{only marks, mark=*, mark size=2.5pt, cyan}
  \addlegendentry{$arg(\hat\beta)$}
  \addlegendimage{only marks, mark=*, mark size=2.5pt, black}
  \addlegendentry{$arg(\hat\gamma)$}
\end{axis}
\end{tikzpicture}

\caption{Absolute phases of the complex coefficients at the acoustic branch at the set parameters (i) $\psi_1=\psi_2, \psi_3=2\psi_1$, (ii) $\psi_1=\psi_2, \psi_3=0.5\psi_1$, and (iii) $\psi_2=\psi_3, \psi_1=2\psi_2$.}
\label{Fig:3MassPhase}
\end{figure}

When inversion symmetry is present, the dominant modal components exhibit rigid phase locking: the pairwise differences remain fixed within each half zone,
\begin{equation*}
\arg(\hat\alpha)-\arg(\hat\beta)\simeq -\frac{\pi}{2}, \qquad
\arg(\hat\beta)-\arg(\hat\gamma)\simeq +\frac{\pi}{2},
\end{equation*}
up to the expected $\pi$ flips at $kL = 0,\pm\pi$. Correspondingly, the absolute phases $\arg(\hat\alpha)$, $\arg(\hat\beta)$, and $\arg(\hat\gamma)$ accumulate a single net $\pi$ twist in the $\pi$-Zak case (Fig.~\ref{Phase_AC}) and show either zero or two twists in the $0$-Zak case (Fig.~\ref{Phase_1O}), matching the usual Zak classification of inversion-symmetric 1D lattices \cite{PhysRevLett.127.147401,xiao2016coexistence,longhi2018probing}. When inversion is broken, these relations unlock: the pairwise phase differences vary smoothly with $k$, and the absolute phases develop mixed slopes rather than coherent twist patterns (Fig.~\ref{Phase_2O}), reflecting the transition to a non-quantized, geometry-dependent Berry phase \cite{xiao2016coexistence,PhysRevResearch.4.023120}.

Figure~\ref{Fig:3MassPhase} show that the three-mass unit cell generates a normalized modal state $\alpha\ket{E_1}+\beta\ket{E_2}+\gamma\ket{E_3}$ (up to phase and normalization). Unlike the diatomic spinor, this three-component state cannot be represented on a single $S^2$: pure three-level states reside in the eight-dimensional $SU(3)$ Bloch set \cite{KIMURA2003339}, and only their pairwise projections $\{\ket{E_1},\ket{E_2}\}$, $\{\ket{E_2},\ket{E_3}\}$, and $\{\ket{E_1},\ket{E_3}\}$ map to standard Bloch spheres \cite{eltschka2021shape}. Under inversion, the fixed phase differences ensure that each projection traces a simple latitude–longitude arc whose total azimuthal winding encodes the $0/\pi$ Zak parity \cite{RN253}. When inversion is broken, these projected paths deform and lose quantization, consistent with the unpinned phase evolution in Fig.~\ref{Phase_2O}.

Thus, the three-mass cell generalizes the diatomic Bloch-sphere framework by producing three mutually consistent two-level projections whose windings capture the Zak class, while simultaneously admitting an $SU(3)$ geometric interpretation familiar from quantum qutrits \cite{RevModPhys.82.1959,KIMURA2003339}. Beyond serving as a higher-dimensional analogue, the three-mass unit cell is important because it represents the minimal mechanical system in which genuinely three-component superpositions, multi-angle phase relations, and $SU(3)$-type geometric behavior can emerge. These features cannot be realized in two-degree-of-freedom lattices and open the door to classical implementations of richer mode-coupling, multi-parameter Berry phases, and higher-dimensional state control. In this sense, the three-mass architecture provides a compact mechanical platform for exploring geometrical and topological effects that go beyond the two-level (Bloch-sphere) setting.

\section{Lattice with Time-Dependent Stiffness}
\label{Time Dependent Stiffness}
\subsection{Two-Mass System}

To extend the static analysis, we examine the same diatomic elastic lattice when the two masses are coupled by springs whose stiffness is modulated sinusoidally in both space and time.

To compute the dispersion and to extract complex amplitudes and phases for each branch, we use a two-step MD $\to$ SAAP workflow (see Supplementary Material, similar to \cite{10.1121/1.5114911}): first determine $\omega_j(k)$ from random-phase excitation; then re-excite each band with traveling-wave initial conditions to extract complex amplitudes and phases, and finally project onto $\{\ket{E_1}, \ket{E_2}\}$ to obtain $\alpha(k), \beta(k)$.
In selecting the spatiotemporal modulation parameters, two physical constraints determine the admissible regime. First, the modulation velocity must remain below the intrinsic elastic wave speed of the medium \cite{1444496}. If $V_m=\omega_m/k_m$ exceeds $c_0$, the modulation outruns the propagating wave and cannot maintain phase synchrony, suppressing coherent frequency conversion and producing nonphysical interactions. Thus, $V_m < c_0$, $c_0=\sqrt{E/\rho}$. Using material properties of Ge Se chalcogenide glass ($GeSe_4$) with $E=13.8~GPa$, $\rho=4361~kg/m^3$, yields a sound speed $c_0 \approx 1780~m/s$ \cite{PhysRevLett.92.245501}. All modulation velocities in this study, including the representative $V_m=350~m/s$, lie safely below this limit, ensuring stable subsonic modulation–wave interaction. Second, the stiffness modulation must remain positive throughout the cycle. This requires $\Delta\psi/\psi_0<1$, which guarantees physically realizable springs at all times. The modulation amplitudes used here satisfy this condition. Together, these constraints ensure that the observed hybridization, sideband formation, and geometric-phase trends arise from physically valid, stable interactions.

For each wave number, we analyze two modulation patterns, denoted $\psi_n^{+}$ and $\psi_n^{-}$ (definitions below). The modulation of the force constant adopts the form introduced earlier in Eq.~(\ref{eq1}) and implemented previously in SAAP-based studies of spatiotemporal lattices, where two MD passes with carefully chosen initial conditions are required to resolve both the frequencies and the phases of the traveling modes. 
The modal–decomposition procedure used in this work follows the two–step Spectral Analysis of Amplitudes and Phases (SAAP) workflow introduced in our previous study \cite{10.1121/1.5114911,HASAN2019114843}. In the first step, molecular dynamics (MD) is performed with broadband initial conditions to resolve the dispersion $\omega_j(k)$ across the Brillouin zone. In the second step, a separate MD pass is carried out using traveling–wave initial conditions that selectively excite a single $(k,j)$ branch, enabling mode–resolved phase extraction.
Mathematically, the two-step procedure is required because the quantities of interest depend on both the eigenfrequencies $\omega_j(k)$ and the associated complex amplitudes $A_{n,j}(k)$, which cannot be obtained simultaneously from a single MD realization. The first MD pass supplies only the spectral support $\{\omega_j(k)\}$ through the temporal Fourier transform of $u_{n,Ni}(t)$, since the response contains a mixed superposition of all modes at that $k$-value. To extract the correct complex phase for a specific branch, the Bloch component of interest must be isolated, and its frequency must already be known. This dependency makes the dispersion from step one an essential prerequisite for the second step. With $\omega_j(k)$ determined, the mode can then be selectively excited using traveling–wave initial conditions $u_{n,Ni}(0)=\cos(kN_iL)$ and $\dot{u}_{n,Ni}(0)=\omega_j(k)\sin(kN_iL)$, corresponding to the real part of a single Bloch state $\mathrm{e}^{i(kN_iL-\omega_j t)}$. Under these controlled conditions, the projection:
\begin{equation*}
A_{n,j}(k)=\frac{1}{\tau_0}\int_0^{t_0} u_{n,Ni}(t)\,
\mathrm{e}^{\,i(kN_iL-\omega_j t)}\,dt
\end{equation*}
returns a unique, gauge-consistent complex coefficient for that mode (detail in supplementary materials) \cite{10.1121/1.5114911,HASAN2019114843}. Therefore, resolving $\omega_j(k)$ first and then determining $A_{n,j}(k)$ in a second MD pass is mathematically necessary for obtaining the modal amplitudes and phases in a mode-pure and physically meaningful manner.

In particular, the second MD pass must employ traveling-wave initial conditions to avoid arbitrary phase offsets---an essential distinction between SAAP and eigenvector-only approaches. Specifically,
\begin{equation}
\psi_n(t) = \psi_0 \pm \Delta \psi \sin \left( k_m x_n - \omega_m t \right),
\end{equation}
where $\psi_0$ is the baseline stiffness, $\Delta \psi$ the modulation amplitude, and $k_m$ and $\omega_m$ the modulation wave number and angular frequency, respectively. Introducing the spatial period $\lambda_m = 2\pi / k_m = L = N_m a$ (with inter-mass spacing $a$), we may equivalently write
\begin{equation}
\psi_n(t) = \psi_0 \pm \Delta \psi \sin \left( \frac{2\pi (x_n - V_m t)}{L} \right).
\end{equation}

We label $\psi_n^{+}$ for the ``$+$'' and $\psi_n^{-}$ for the ``$-$'' choice. These two cases are related by an origin shift of half a cell (from the first mass to mid-cell).

Consequently, they share the same band structure but differ by Berry-phase conventions tied to origin choice---an invariance of the spectrum under unitary re-labelings that simultaneously allows different Berry phases under translations of the real-space origin \cite{10.1121/1.5114911}.

\begin{figure}[H]
    \centering
    \input{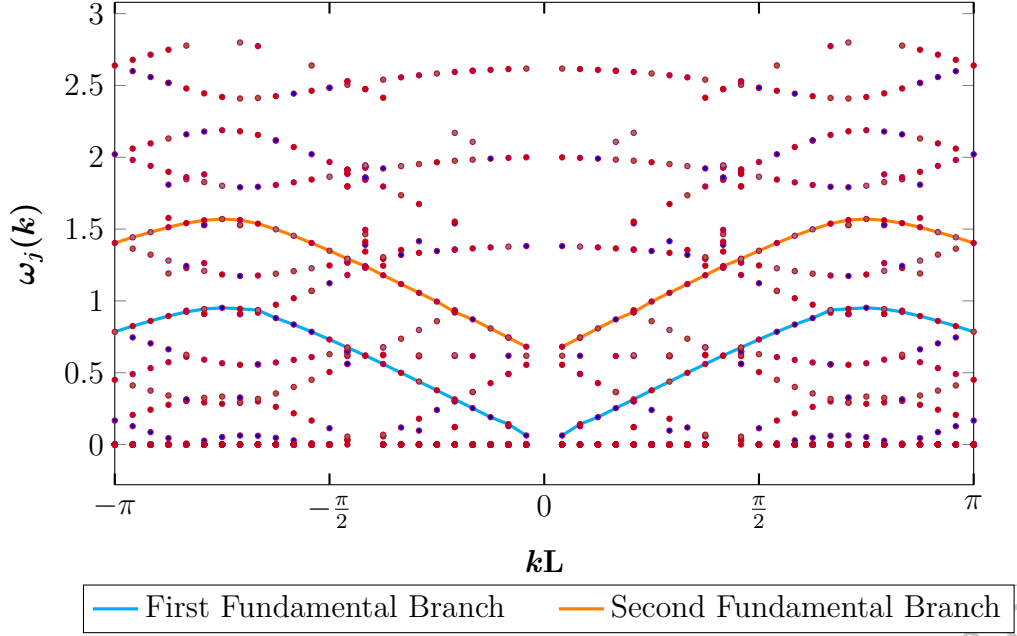}
    \newline
    \begin{tikzpicture}
\begin{axis}[
  hide axis,
  xmin=0,xmax=1,ymin=0,ymax=1,
  legend columns=3,
  legend cell align=left,
  legend style={
    /tikz/every even column/.append style={column sep=1.5em}
  }
]
  \addlegendimage{smooth, very thick, cyan}
  \addlegendentry{First Fundamental Branch}
  \addlegendimage{smooth, very thick, orange}
  \addlegendentry{Second Fundamental Branch}
\end{axis}
\end{tikzpicture}
    \caption{Elastic wave band structure for variable periodical sinusoidal modulation of stiffness at $V_m=350m/s$. The blue line defines the first fundamental branch (FFB), and the orange line represents the second fundamental branch (SFB).}
    \label{fig:Spatiotemporal}
    
\end{figure}

Figure~\ref{fig:Spatiotemporal} shows the dispersion of the space-time-modulated superlattice (as also reported in our prior work on \cite{HASAN2019114843}). The moving modulation generates directional hybridization (one-sided) gaps and Doppler-shifted replicas of the static branches---canonical signatures of non-reciprocal phononic media with space-time periodicity and Willis-type effective descriptions. The placement and multiplicity of gaps depend on the modulation speed and harmonic content, consistent with general theories and with beam-lattice implementations of spatiotemporal elasticity \cite{doi:10.1098/rspa.2017.0188,Trainiti_2016,NASSAR201797}.

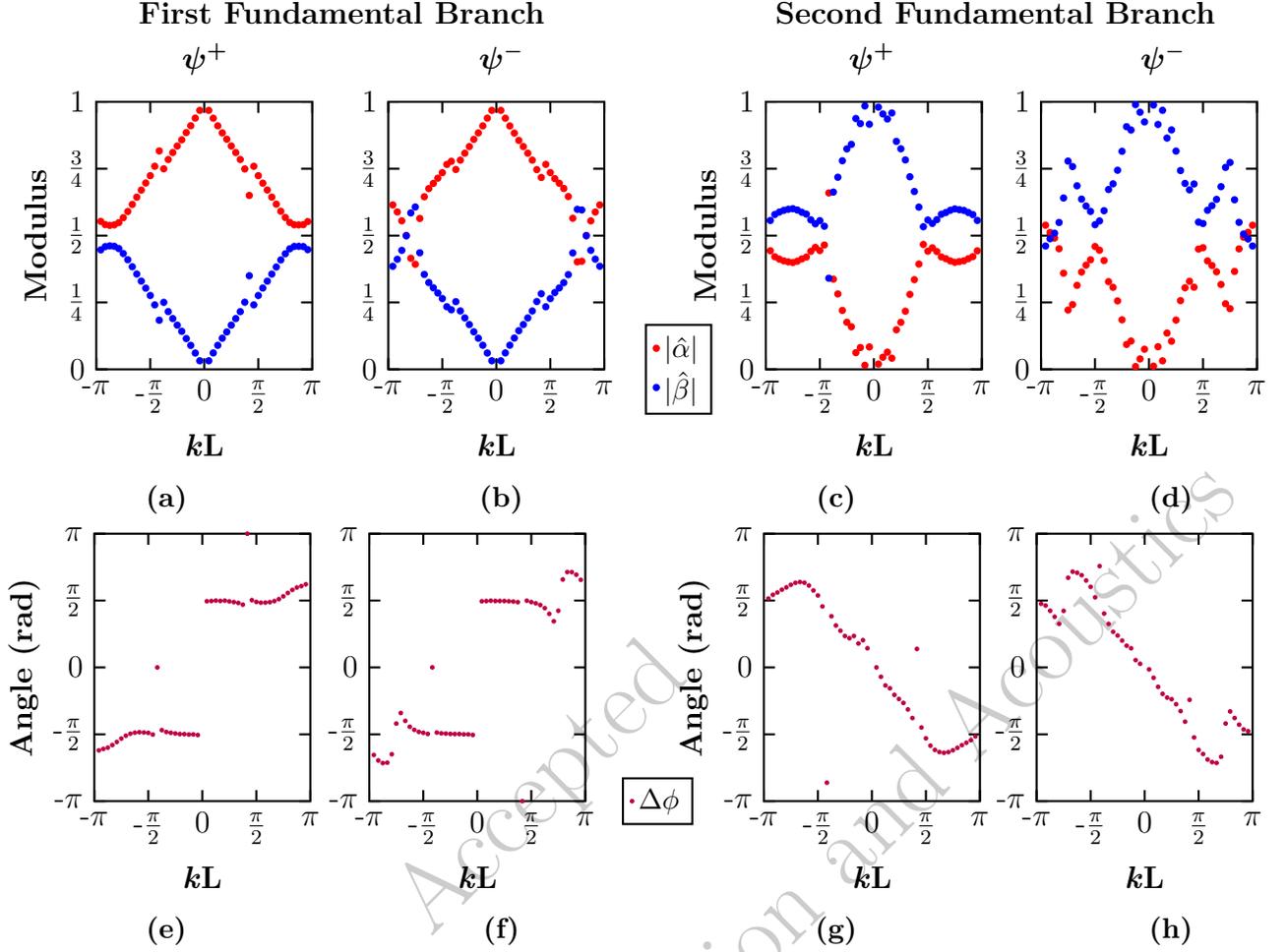
\begin{figure}[H]
\centering
\setlength{\tabcolsep}{6pt}
\renewcommand{\arraystretch}{1.2}

\begin{tabular}{@{}p{.24\textwidth} p{.24\textwidth} p{.24\textwidth} p{.24\textwidth}@{}}
  \multicolumn{2}{c}{\textbf{First Fundamental Branch}} &
  \multicolumn{2}{c}{\textbf{Second Fundamental Branch}} \\
  \multicolumn{1}{c}{ \hspace{2em}$\boldsymbol{\psi^+}$} &
  \multicolumn{1}{c}{ $\boldsymbol{\psi^-}$} &
  \multicolumn{1}{c}{\hspace{2em} $\boldsymbol{\psi^+}$} &
  \multicolumn{1}{c}{ $\boldsymbol{\psi^-}$} \\
  \begin{subfigure}[t]{\linewidth}
    \centering \panelbox{\begin{tikzpicture}
\begin{axis}[
height=5.2cm,
width=4.5cm,
    thick,
    xlabel={\textbf{\textit{k}L}},
    ylabel={\textbf{Modulus}},
    xmin=-3.1416, xmax=3.1416,
    ymin=0, ymax=1,
    every x tick/.style={color=black, thick},
    xtick={-3.1416,-1.5708,0,1.5708,3.1416},
    xticklabels={-$\pi$,-$\frac{\pi}{2}$,0,$\frac{\pi}{2}$,$\pi$},
    every y tick/.style={color=black, thick},
    ytick={0,0.25,0.5,0.75,1},
    yticklabels={$0$,$\frac{1}{4}$,$\frac{1}{2}$,$\frac{3}{4}$,$1$},
    legend style={at={(1.15,1.25)}, anchor=east}, 
    legend cell align={left},
]

\addplot[
    only marks,
    color=red,
    mark=*,
    mark size=1pt,
]
coordinates {
(-3.01069296,0.553236961)
(-2.879793266,0.541010922)
(-2.748893572,0.538306594)
(-2.617993878,0.540052236)
(-2.487094184,0.548382109)
(-2.356194490,0.566984540)
(-2.225294796,0.590155048)
(-2.094395102,0.616546691)
(-1.963495408,0.643632295)
(-1.832595715,0.670161585)
(-1.701696021,0.696588188)
(-1.570796327,0.723983351)
(-1.439896633,0.759143486)
(-1.308996939,0.816480271)
(-1.178097245,0.749602778)
(-1.047197551,0.784302743)
(-0.916297857,0.810414835)
(-0.785398163,0.834886323)
(-0.654498469,0.857980090)
(-0.523598776,0.885040358)
(-0.392699082,0.910558734)
(-0.261799388,0.939123534)
(-0.130899694,0.968061835)
(0.130899694,0.968115937)
(0.261799388,0.938739860)
(0.392699082,0.910309810)
(0.523598776,0.883861710)
(0.654498469,0.860262727)
(0.785398163,0.834923544)
(0.916297857,0.810381888)
(1.047197551,0.784398211)
(1.178097245,0.749103467)
(1.308996939,0.650152275)
(1.439896633,0.759045771)
(1.570796327,0.723991118)
(1.701696021,0.695945641)
(1.832595715,0.670398110)
(1.963495408,0.643837861)
(2.094395102,0.617406950)
(2.225294796,0.590964519)
(2.356194490,0.567900354)
(2.487094184,0.549299855)
(2.617993878,0.540477291)
(2.748893572,0.540168581)
(2.879793266,0.540947614)
(3.010692960,0.552460331)
};

\addplot[
    only marks,
    color=blue,
    mark=*,
    mark size=1pt,
]
coordinates {
(-3.01069296,0.446763039)
(-2.879793266,0.458989078)
(-2.748893572,0.461693406)
(-2.617993878,0.459947764)
(-2.487094184,0.451617891)
(-2.356194490,0.433015460)
(-2.225294796,0.409844952)
(-2.094395102,0.383453309)
(-1.963495408,0.356367705)
(-1.832595715,0.329838415)
(-1.701696021,0.303411812)
(-1.570796327,0.276016649)
(-1.439896633,0.240856514)
(-1.308996939,0.183519729)
(-1.178097245,0.250397222)
(-1.047197551,0.215697257)
(-0.916297857,0.189585165)
(-0.785398163,0.165113677)
(-0.654498469,0.142019910)
(-0.523598776,0.114959642)
(-0.392699082,0.089441266)
(-0.261799388,0.060876466)
(-0.130899694,0.031938165)
(0.130899694,0.031884063)
(0.261799388,0.061260140)
(0.392699082,0.089690190)
(0.523598776,0.116138290)
(0.654498469,0.139737273)
(0.785398163,0.165076456)
(0.916297857,0.189618112)
(1.047197551,0.215601789)
(1.178097245,0.250896533)
(1.308996939,0.349847725)
(1.439896633,0.240954229)
(1.570796327,0.276008882)
(1.701696021,0.304054359)
(1.832595715,0.329601890)
(1.963495408,0.356162139)
(2.094395102,0.382593050)
(2.225294796,0.409035481)
(2.356194490,0.432099646)
(2.487094184,0.450700145)
(2.617993878,0.459522709)
(2.748893572,0.459831419)
(2.879793266,0.459052386)
(3.010692960,0.447539669)
};
       
\end{axis}

\end{tikzpicture}}
    \caption{}\label{M_FFB_psi+}
  \end{subfigure} &
  \begin{subfigure}[t]{\linewidth}
    \centering \panelbox{\begin{tikzpicture}
\begin{axis}[
    height=5.2cm,
width=4.5cm,
    thick,
    xlabel={\textbf{\textit{k}L}},
    xmin=-3.1416, xmax=3.1416,
    ymin=0, ymax=1,
    every x tick/.style={color=black, thick},
    xtick={-3.1416,-1.5708,0,1.5708,3.1416},
    xticklabels={-$\pi$,-$\frac{\pi}{2}$,0,$\frac{\pi}{2}$,$\pi$},
    every y tick/.style={color=black, thick},
    ytick={0,0.25,0.5,0.75,1},
    yticklabels={$0$,$\frac{1}{4}$,$\frac{1}{2}$,$\frac{3}{4}$,$1$},
    legend style={at={(1.5,0)}, anchor=east}, 
    legend cell align={left},
]

\addplot[
    only marks,
    color=red,
    mark=*,
    mark size=1pt,
]
coordinates {
(-3.01069296,0.614658407)
(-2.879793266,0.588420471)
(-2.748893572,0.555202568)
(-2.617993878,0.500910969)
(-2.487094184,0.415219835)
(-2.356194490,0.392951502)
(-2.225294796,0.564165160)
(-2.094395102,0.645255255)
(-1.963495408,0.675698779)
(-1.832595715,0.695089412)
(-1.701696021,0.714692381)
(-1.570796327,0.735364313)
(-1.439896633,0.765998211)
(-1.308996939,0.777602538)
(-1.178097245,0.747058843)
(-1.047197551,0.781835818)
(-0.916297857,0.807947640)
(-0.785398163,0.832738966)
(-0.654498469,0.859703221)
(-0.523598776,0.883811075)
(-0.392699082,0.910020842)
(-0.261799388,0.939039208)
(-0.130899694,0.968039466)
(0.130899694,0.968076699)
(0.261799388,0.938657920)
(0.392699082,0.909837845)
(0.523598776,0.883136072)
(0.654498469,0.855700297)
(0.785398163,0.832761953)
(0.916297857,0.807884273)
(1.047197551,0.781954097)
(1.178097245,0.746579588)
(1.308996939,0.717330451)
(1.439896633,0.766210254)
(1.570796327,0.735490132)
(1.701696021,0.713103907)
(1.832595715,0.695648112)
(1.963495408,0.676114937)
(2.094395102,0.646087054)
(2.225294796,0.564895951)
(2.356194490,0.401816844)
(2.487094184,0.404128708)
(2.617993878,0.500541076)
(2.748893572,0.555037329)
(2.879793266,0.588099443)
(3.010692960,0.614626954)
};
    \addlegendentry{\small $|\hat\alpha|$}

\addplot[
    only marks,
    color=blue,
    mark=*,
    mark size=1pt,
]
coordinates {
(-3.01069296,0.385341593)
(-2.879793266,0.411579529)
(-2.748893572,0.444797432)
(-2.617993878,0.499089031)
(-2.487094184,0.584780165)
(-2.356194490,0.607048498)
(-2.225294796,0.435834840)
(-2.094395102,0.354744745)
(-1.963495408,0.324301221)
(-1.832595715,0.304910588)
(-1.701696021,0.285307619)
(-1.570796327,0.264635687)
(-1.439896633,0.234001789)
(-1.308996939,0.222397462)
(-1.178097245,0.252941157)
(-1.047197551,0.218164182)
(-0.916297857,0.192052360)
(-0.785398163,0.167261034)
(-0.654498469,0.140296779)
(-0.523598776,0.116188925)
(-0.392699082,0.089979158)
(-0.261799388,0.060960792)
(-0.130899694,0.031960534)
(0.130899694,0.031923301)
(0.261799388,0.061342080)
(0.392699082,0.090162155)
(0.523598776,0.116863928)
(0.654498469,0.144299703)
(0.785398163,0.167238047)
(0.916297857,0.192115727)
(1.047197551,0.218045903)
(1.178097245,0.253420412)
(1.308996939,0.282669549)
(1.439896633,0.233789746)
(1.570796327,0.264509868)
(1.701696021,0.286896093)
(1.832595715,0.304351888)
(1.963495408,0.323885063)
(2.094395102,0.353912946)
(2.225294796,0.435104049)
(2.356194490,0.598183156)
(2.487094184,0.595871292)
(2.617993878,0.499458924)
(2.748893572,0.444962671)
(2.879793266,0.411900557)
(3.010692960,0.385373046)
};
\addlegendentry{\small $|\hat\beta|$}

\end{axis}

\end{tikzpicture}}
    \caption{}\label{M_FFB_psi-}
  \end{subfigure} &
  \begin{subfigure}[t]{\linewidth}
    \centering \panelbox{\begin{tikzpicture}
\begin{axis}[
    height=5.2cm,
width=4.5cm,
    thick,
    xlabel={\textbf{\textit{k}L}},
    ylabel={\textbf{Modulus}},
    xmin=-3.1416, xmax=3.1416,
    ymin=0, ymax=1,
    every x tick/.style={color=black, thick},
    xtick={-3.1416,-1.5708,0,1.5708,3.1416},
    xticklabels={-$\pi$,-$\frac{\pi}{2}$,0,$\frac{\pi}{2}$,$\pi$},
    every y tick/.style={color=black, thick},
    ytick={0,0.25,0.5,0.75,1},
    yticklabels={$0$,$\frac{1}{4}$,$\frac{1}{2}$,$\frac{3}{4}$,$1$},
    legend style={at={(1.15,1.25)}, anchor=east}, 
    legend cell align={left},
]

\addplot[
    only marks,
    color=red,
    mark=*,
    mark size=1pt,
]
coordinates {
(-3.01069296,0.44335681)
(-2.879793266,0.420930321)
(-2.748893572,0.412004647)
(-2.617993878,0.405491098)
(-2.487094184,0.401553093)
(-2.35619449,0.399458849)
(-2.225294796,0.404387304)
(-2.094395102,0.41281103)
(-1.963495408,0.41966035)
(-1.832595715,0.441232742)
(-1.701696021,0.456043783)
(-1.570796327,0.44417915)
(-1.439896633,0.464855519)
(-1.308996939,0.659442138)
(-1.178097245,0.337294993)
(-1.047197551,0.281820277)
(-0.916297857,0.220176615)
(-0.785398163,0.175941943)
(-0.654498469,0.159046495)
(-0.523598776,0.062373153)
(-0.392699082,0.0813001)
(-0.261799388,0.015186238)
(-0.130899694,0.083671135)
(0.130899694,0.019820642)
(0.261799388,0.045250908)
(0.392699082,0.064159934)
(0.523598776,0.04161127)
(0.654498469,0.147895361)
(0.785398163,0.175559505)
(0.916297857,0.217138861)
(1.047197551,0.28009543)
(1.178097245,0.335109791)
(1.308996939,0.393540255)
(1.439896633,0.466879146)
(1.570796327,0.441731118)
(1.701696021,0.454468341)
(1.832595715,0.44039842)
(1.963495408,0.420238605)
(2.094395102,0.413680956)
(2.225294796,0.405270173)
(2.35619449,0.400235428)
(2.487094184,0.403047743)
(2.617993878,0.405802485)
(2.748893572,0.413562921)
(2.879793266,0.421484961)
(3.01069296,0.443179372)
};

\addplot[
    only marks,
    color=blue,
    mark=*,
    mark size=1pt,
]
coordinates {
(-3.01069296,0.55664319)
(-2.879793266,0.579069679)
(-2.748893572,0.587995353)
(-2.617993878,0.594508902)
(-2.487094184,0.598446907)
(-2.35619449,0.600541151)
(-2.225294796,0.595612696)
(-2.094395102,0.58718897)
(-1.963495408,0.58033965)
(-1.832595715,0.558767258)
(-1.701696021,0.543956217)
(-1.570796327,0.55582085)
(-1.439896633,0.535144481)
(-1.308996939,0.340557862)
(-1.178097245,0.662705007)
(-1.047197551,0.718179723)
(-0.916297857,0.779823385)
(-0.785398163,0.824058057)
(-0.654498469,0.840953505)
(-0.523598776,0.937626847)
(-0.392699082,0.9186999)
(-0.261799388,0.984813762)
(-0.130899694,0.916328865)
(0.130899694,0.980179358)
(0.261799388,0.954749092)
(0.392699082,0.935840066)
(0.523598776,0.95838873)
(0.654498469,0.852104639)
(0.785398163,0.824440495)
(0.916297857,0.782861139)
(1.047197551,0.71990457)
(1.178097245,0.664890209)
(1.308996939,0.606459745)
(1.439896633,0.533120854)
(1.570796327,0.558268882)
(1.701696021,0.545531659)
(1.832595715,0.55960158)
(1.963495408,0.579761395)
(2.094395102,0.586319044)
(2.225294796,0.594729827)
(2.35619449,0.599764572)
(2.487094184,0.596952257)
(2.617993878,0.594197515)
(2.748893572,0.586437079)
(2.879793266,0.578515039)
(3.01069296,0.556820628)
};

\end{axis}

\end{tikzpicture}}
    \caption{}\label{M_SFB_psi+}
  \end{subfigure} &
  \begin{subfigure}[t]{\linewidth}
    \centering \hspace{-2.4em} \panelbox{\begin{tikzpicture}
\begin{axis}[
    height=5.2cm,
width=4.5cm,
    thick,
    xlabel={\textbf{\textit{k}L}},
    xmin=-3.1416, xmax=3.1416,
    ymin=0, ymax=1,
    every x tick/.style={color=black, thick},
    xtick={-3.1416,-1.5708,0,1.5708,3.1416},
    xticklabels={-$\pi$,-$\frac{\pi}{2}$,0,$\frac{\pi}{2}$,$\pi$},
    every y tick/.style={color=black, thick},
    ytick={0,0.25,0.5,0.75,1},
    yticklabels={$0$,$\frac{1}{4}$,$\frac{1}{2}$,$\frac{3}{4}$,$1$},
    legend style={at={(1.15,1.25)}, anchor=east}, 
    legend cell align={left},
]

\addplot[
    only marks,
    color=red,
    mark=*,
    mark size=1pt,
]
coordinates {
(-3.01069296,0.538931569)
(-2.879793266,0.511735142)
(-2.748893572,0.490865935)
(-2.617993878,0.45086456)
(-2.487094184,0.359451448)
(-2.35619449,0.221635629)
(-2.225294796,0.242006145)
(-2.094395102,0.313897172)
(-1.963495408,0.363716773)
(-1.832595715,0.389469833)
(-1.701696021,0.409557372)
(-1.570796327,0.459319336)
(-1.439896633,0.445211551)
(-1.308996939,0.406848708)
(-1.178097245,0.327756811)
(-1.047197551,0.30647622)
(-0.916297857,0.255756106)
(-0.785398163,0.185041837)
(-0.654498469,0.092879102)
(-0.523598776,0.105449708)
(-0.392699082,0.010473097)
(-0.261799388,0.038949007)
(-0.130899694,0.075111066)
(0.130899694,0.012008383)
(0.261799388,0.084499015)
(0.392699082,0.031698439)
(0.523598776,0.135452895)
(0.654498469,0.105023984)
(0.785398163,0.184133113)
(0.916297857,0.256876982)
(1.047197551,0.304952337)
(1.178097245,0.329035284)
(1.308996939,0.307034115)
(1.439896633,0.449538034)
(1.570796327,0.454801143)
(1.701696021,0.405256902)
(1.832595715,0.388080766)
(1.963495408,0.364631148)
(2.094395102,0.31708375)
(2.225294796,0.245066142)
(2.35619449,0.226740073)
(2.487094184,0.366047961)
(2.617993878,0.450371078)
(2.748893572,0.493998636)
(2.879793266,0.512438822)
(3.01069296,0.538866349)
};


\addplot[
    only marks,
    color=blue,
    mark=*,
    mark size=1pt,
]
coordinates {
(-3.01069296,0.461068431)
(-2.879793266,0.488264858)
(-2.748893572,0.509134065)
(-2.617993878,0.54913544)
(-2.487094184,0.640548552)
(-2.35619449,0.778364371)
(-2.225294796,0.757993855)
(-2.094395102,0.686102828)
(-1.963495408,0.636283227)
(-1.832595715,0.610530167)
(-1.701696021,0.590442628)
(-1.570796327,0.540680664)
(-1.439896633,0.554788449)
(-1.308996939,0.593151292)
(-1.178097245,0.672243189)
(-1.047197551,0.69352378)
(-0.916297857,0.744243894)
(-0.785398163,0.814958163)
(-0.654498469,0.907120898)
(-0.523598776,0.894550292)
(-0.392699082,0.989526903)
(-0.261799388,0.961050993)
(-0.130899694,0.924888934)
(0.130899694,0.987991617)
(0.261799388,0.915500985)
(0.392699082,0.968301561)
(0.523598776,0.864547105)
(0.654498469,0.894976016)
(0.785398163,0.815866887)
(0.916297857,0.743123018)
(1.047197551,0.695047663)
(1.178097245,0.670964716)
(1.308996939,0.692965885)
(1.439896633,0.550461966)
(1.570796327,0.545198857)
(1.701696021,0.594743098)
(1.832595715,0.611919234)
(1.963495408,0.635368852)
(2.094395102,0.68291625)
(2.225294796,0.754933858)
(2.35619449,0.773259927)
(2.487094184,0.633952039)
(2.617993878,0.549628922)
(2.748893572,0.506001364)
(2.879793266,0.487561178)
(3.01069296,0.461133651)
};


\end{axis}

\end{tikzpicture}}
    \caption{}\label{M_SFB_psi-}
  \end{subfigure} \\
  \begin{subfigure}[t]{\linewidth}
    \centering \hspace{-1.4em} \panelbox{\begin{tikzpicture}
\begin{axis}[
    height=5.2cm,
width=4.5cm,
    thick,
    xlabel={\textbf{\textit{k}L}},
    ylabel={\textbf{Angle (rad)}},
    ylabel style={at={(axis description cs:-0.2,0.5)},anchor=south}, 
    xmin=-3.1416, xmax=3.1416,
    ymin=-3.1416, ymax=3.1416,
    every x tick/.style={color=black, thick},
    xtick={-3.1416,-1.5708,0,1.5708,3.1416},
    xticklabels={-$\pi$,-$\frac{\pi}{2}$,0,$\frac{\pi}{2}$,$\pi$},
    every y tick/.style={color=black, thick},
    ytick={-3.1416,-1.5708,0,1.5708,3.1416},
    yticklabels={-$\pi$,-$\frac{\pi}{2}$,0,$\frac{\pi}{2}$,$\pi$},
    every y tick/.style={color=black, thick},
    legend style={at={(1.15,0.3)}, anchor=east}, 
    legend cell align={left},
]

\addplot[
  only marks,
    color=purple,
    mark=*,
    mark size=0.5pt,
]
coordinates {
(-3.01069296,-1.948300399)
(-2.879793266,-1.913264629)
(-2.748893572,-1.883716449)
(-2.617993878,-1.822167870)
(-2.487094184,-1.750348271)
(-2.356194490,-1.670600739)
(-2.225294796,-1.603817082)
(-2.094395102,-1.556987201)
(-1.963495408,-1.532463461)
(-1.832595715,-1.521036400)
(-1.701696021,-1.527696018)
(-1.570796327,-1.536549228)
(-1.439896633,-1.576346193)
(-1.308996939,-0.000476594)
(-1.178097245,-1.471885626)
(-1.047197551,-1.515484520)
(-0.916297857,-1.535179789)
(-0.785398163,-1.551134729)
(-0.654498469,-1.564995617)
(-0.523598776,-1.565960595)
(-0.392699082,-1.575391351)
(-0.261799388,-1.573719393)
(-0.130899694,-1.587317839)
(0.130899694,1.555442917)
(0.261799388,1.560686216)
(0.392699082,1.570453569)
(0.523598776,1.562376965)
(0.654498469,1.569543400)
(0.785398163,1.551229043)
(0.916297857,1.535744097)
(1.047197551,1.515763993)
(1.178097245,1.474856481)
(1.308996939,3.141116047)
(1.439896633,1.581543738)
(1.570796327,1.538861829)
(1.701696021,1.521708892)
(1.832595715,1.523227299)
(1.963495408,1.533641286)
(2.094395102,1.558999104)
(2.225294796,1.604668692)
(2.356194490,1.670894410)
(2.487094184,1.749887696)
(2.617993878,1.820863111)
(2.748893572,1.877924164)
(2.879793266,1.910119186)
(3.010692960,1.953520296)
};
\end{axis}

\end{tikzpicture}}
    \caption{}\label{P_FFB_psi+}
  \end{subfigure} &
  \begin{subfigure}[t]{\linewidth}
    \centering \hspace{-2.4em} \panelbox{\begin{tikzpicture}
\begin{axis}[
    height=5.2cm,
width=4.5cm,
    thick,
    xlabel={\textbf{\textit{k}L}},
    xmin=-3.1416, xmax=3.1416,
    ymin=-3.1416, ymax=3.1416,
    every x tick/.style={color=black, thick},
    xtick={-3.1416,-1.5708,0,1.5708,3.1416},
    xticklabels={-$\pi$,-$\frac{\pi}{2}$,0,$\frac{\pi}{2}$,$\pi$},
    every y tick/.style={color=black, thick},
    ytick={-3.1416,-1.5708,0,1.5708,3.1416},
    yticklabels={-$\pi$,-$\frac{\pi}{2}$,0,$\frac{\pi}{2}$,$\pi$},
    every y tick/.style={color=black, thick},
    legend style={at={(1.5,0)}, anchor=east}, 
    legend cell align={left},
]

\addplot[
  only marks,
    color=purple,
    mark=*,
    mark size=0.5pt,
]
coordinates {
(-3.01069296,-2.056570478)
(-2.879793266,-2.184889584)
(-2.748893572,-2.246160174)
(-2.617993878,-2.236919695)
(-2.487094184,-2.039007967)
(-2.356194490,-1.318565324)
(-2.225294796,-1.070327025)
(-2.094395102,-1.255065359)
(-1.963495408,-1.393689849)
(-1.832595715,-1.465426710)
(-1.701696021,-1.514079309)
(-1.570796327,-1.535248270)
(-1.439896633,-1.561451151)
(-1.308996939,0.000477883)
(-1.178097245,-1.529256518)
(-1.047197551,-1.547411773)
(-0.916297857,-1.553582276)
(-0.785398163,-1.560709163)
(-0.654498469,-1.563802401)
(-0.523598776,-1.562870780)
(-0.392699082,-1.571418322)
(-0.261799388,-1.571987660)
(-0.130899694,-1.585633426)
(0.130899694,1.554557481)
(0.261799388,1.559596564)
(0.392699082,1.567979421)
(0.523598776,1.561539176)
(0.654498469,1.561340977)
(0.785398163,1.560919688)
(0.916297857,1.554457622)
(1.047197551,1.548030904)
(1.178097245,1.533978704)
(1.308996939,-3.141114761)
(1.439896633,1.567711086)
(1.570796327,1.535911909)
(1.701696021,1.506439850)
(1.832595715,1.466341141)
(1.963495408,1.393879064)
(2.094395102,1.259742667)
(2.225294796,1.084130781)
(2.356194490,1.331618555)
(2.487094184,2.066763713)
(2.617993878,2.240799179)
(2.748893572,2.239351388)
(2.879793266,2.179108224)
(3.010692960,2.058707887)
};
 \addlegendentry{\small $\Delta\phi$}   
\end{axis}

\end{tikzpicture}}
    \caption{}\label{P_FFB_psi-}
  \end{subfigure} &
  \begin{subfigure}[t]{\linewidth}
    \centering \hspace{-1.4em} \panelbox{\begin{tikzpicture}
\begin{axis}[
    height=5.2cm,
width=4.5cm,
    thick,
    xlabel={\textbf{\textit{k}L}},
    ylabel={\textbf{Angle (rad)}},
    ylabel style={at={(axis description cs:-0.2,0.5)},anchor=south}, 
    xmin=-3.1416, xmax=3.1416,
    ymin=-3.1416, ymax=3.1416,
    every x tick/.style={color=black, thick},
    xtick={-3.1416,-1.5708,0,1.5708,3.1416},
    xticklabels={-$\pi$,-$\frac{\pi}{2}$,0,$\frac{\pi}{2}$,$\pi$},
    every y tick/.style={color=black, thick},
    ytick={-3.1416,-1.5708,0,1.5708,3.1416},
    yticklabels={-$\pi$,-$\frac{\pi}{2}$,0,$\frac{\pi}{2}$,$\pi$},
    every y tick/.style={color=black, thick},
    legend style={at={(1.15,1.25)}, anchor=east}, 
    legend cell align={left},
]

\addplot[
  only marks,
    color=purple,
    mark=*,
    mark size=0.5pt,
]
coordinates {
(-3.01069296,1.619290309)
(-2.879793266,1.707111756)
(-2.748893572,1.760533385)
(-2.617993878,1.826632211)
(-2.487094184,1.881789025)
(-2.35619449,1.942443566)
(-2.225294796,1.987303075)
(-2.094395102,2.007099515)
(-1.963495408,1.990600769)
(-1.832595715,1.924862774)
(-1.701696021,1.808921589)
(-1.570796327,1.690994269)
(-1.439896633,1.433528593)
(-1.308996939,-2.707355189)
(-1.178097245,1.202803702)
(-1.047197551,0.985615484)
(-0.916297857,0.862133898)
(-0.785398163,0.737654402)
(-0.654498469,0.686657998)
(-0.523598776,0.738737574)
(-0.392699082,0.562312337)
(-0.261799388,0.64)
(-0.130899694,0.45)
(0.130899694,0)
(0.261799388,-0.22)
(0.392699082,-0.423439229)
(0.523598776,-0.49)
(0.654498469,-0.644752963)
(0.785398163,-0.74136219)
(0.916297857,-0.833809695)
(1.047197551,-0.986620344)
(1.178097245,-1.188355598)
(1.308996939,0.434237502)
(1.439896633,-1.410790158)
(1.570796327,-1.665125465)
(1.701696021,-1.837538498)
(1.832595715,-1.937058518)
(1.963495408,-1.986007508)
(2.094395102,-2.005447933)
(2.225294796,-1.988479573)
(2.35619449,-1.945467422)
(2.487094184,-1.888078778)
(2.617993878,-1.826012259)
(2.748893572,-1.768096863)
(2.879793266,-1.709401718)
(3.01069296,-1.618998741)
};

\end{axis}

\end{tikzpicture}}
    \caption{}\label{P_SFB_psi+}
  \end{subfigure} &
  \begin{subfigure}[t]{\linewidth}
    \centering \hspace{-3.1em} \panelbox{\begin{tikzpicture}
\begin{axis}[
    height=5.2cm,
width=4.5cm,
    thick,
    xlabel={\textbf{\textit{k}L}},
    xmin=-3.1416, xmax=3.1416,
    ymin=-3.1416, ymax=3.1416,
    every x tick/.style={color=black, thick},
    xtick={-3.1416,-1.5708,0,1.5708,3.1416},
    xticklabels={-$\pi$,-$\frac{\pi}{2}$,0,$\frac{\pi}{2}$,$\pi$},
    every y tick/.style={color=black, thick},
    ytick={-3.1416,-1.5708,0,1.5708,3.1416},
    yticklabels={-$\pi$,-$\frac{\pi}{2}$,0,$\frac{\pi}{2}$,$\pi$},
    every y tick/.style={color=black, thick},
    legend style={at={(1.15,1.25)}, anchor=east}, 
    legend cell align={left},
]

\addplot[
  only marks,
    color=purple,
    mark=*,
    mark size=0.5pt,
]
coordinates {
(-3.01069296,1.496119034)
(-2.879793266,1.45393799)
(-2.748893572,1.33106734)
(-2.617993878,1.189947807)
(-2.487094184,1.022855882)
(-2.35619449,1.328478779)
(-2.225294796,2.107340723)
(-2.094395102,2.249531061)
(-1.963495408,2.224459993)
(-1.832595715,2.166720065)
(-1.701696021,2.042965683)
(-1.570796327,1.893215073)
(-1.439896633,1.646551872)
(-1.308996939,2.380576213)
(-1.178097245,1.265917565)
(-1.047197551,1.022488133)
(-0.916297857,0.836699449)
(-0.785398163,0.748962271)
(-0.654498469,0.630622701)
(-0.523598776,0.509792476)
(-0.392699082,0.450012003)
(-0.261799388,0.17149293)
(-0.130899694,0.085)
(0.130899694,-0.046)
(0.261799388,-0.243788283)
(0.392699082,-0.456617997)
(0.523598776,-0.621155643)
(0.654498469,-0.708546597)
(0.785398163,-0.746345921)
(0.916297857,-0.863856578)
(1.047197551,-1.026215095)
(1.178097245,-1.274435435)
(1.308996939,-0.761016491)
(1.439896633,-1.64411498)
(1.570796327,-1.937839098)
(1.701696021,-2.033822784)
(1.832595715,-2.163335447)
(1.963495408,-2.223792708)
(2.094395102,-2.243216677)
(2.225294796,-2.09833942)
(2.35619449,-1.317303228)
(2.487094184,-1.033952186)
(2.617993878,-1.188149578)
(2.748893572,-1.344012247)
(2.879793266,-1.455841475)
(3.01069296,-1.496455745)
};
\end{axis}

\end{tikzpicture}}
    \caption{}\label{P_SFB_psi-}
  \end{subfigure} \\
\end{tabular}
\caption{Amplitude and Phase modulation of the two-mass system for the time-dependent stiffness of the harmonic spring for the stiffness of $\psi^+$ and $\psi^-$ at the first (FFB) and second (SFB) fundamental branch. \textbf{Top Panel:} Modulus of complex amplitudes $|\hat\alpha|$ and $|\hat\beta|$. \textbf{Bottom Panel:} Phase difference between the amplitude moduli $\Delta\phi=\arg(\hat\alpha)-\arg(\hat\beta)$.}
\label{Fig:MP_2Mass_TD}
\end{figure}

From the second set of MD simulations with traveling-wave initial conditions, we extract the complex superposition coefficients in the two-state in-cell basis $(\hat{\alpha}, \hat{\beta})$ exactly as in the static case. Relative to the time-independent lattice, the modulation sharpens the $k$-dependence of both moduli and phases (Fig.~\ref{Fig:MP_2Mass_TD}). Most notably, the relative phase $\Delta \phi = \arg (\hat{\alpha}) - \arg (\hat{\beta})$ is no longer piecewise constant: it winds across the Brillouin zone as carrier harmonics mix with Floquet sidebands, whereas in the static lattice inversion symmetry pins $\Delta \phi$ to $\{-\pi/2, +\pi/2\}$ on intervals separated by $\pi$-jumps at high-symmetry points. The modulus hierarchy on the FFB largely mirrors the static case---$|\hat{\alpha}| > |\hat{\beta}|$ over wide $k$-intervals---yet the $\psi^{+}$ and $\psi^{-}$ patterns differ by origin-dependent flips that track the $0/\pi$ swapping induced by unit-cell translations. Under temporal modulation, this swapping persists but the geometric phase becomes open-path: it is accumulated along the actual FFB trajectory in $k$ and does not require a closed loop. This is precisely the Samuel-Bhandari generalization of the Berry phase to non-cyclic evolution \cite{PhysRevLett.60.2339}.

On the SFB, the coefficient roles invert relative to the FFB: $|\hat{\beta}|$ becomes dominant over extended $k$-ranges and $\Delta \phi(k)$ exhibits larger excursions with alternating increases and decreases across the hybridization regions (Fig.~\ref{Fig:MP_2Mass_TD}). This behavior is expected because the SFB approaches---and in places crosses---additional hybridization points where Brillouin harmonics intersect the folded static bands, producing sharp kinks or discrete jumps in both amplitude and phase. The emergence and placement of these discontinuities with increasing modulation velocity are consistent with our prior perturbative estimates and with published analyses of spatiotemporally modulated elastic lattices \cite{Trainiti_2016}.

Taken together, these results lead to three conclusions germane to the manuscript's central thesis. First, the moving stiffness modulation preserves branch identification (FFB vs.\ SFB) and the equivalence of $\psi^{+}$ and $\psi^{-}$ up to an origin shift, but it qualitatively alters the evolution of $\Delta \phi(k)$: the piecewise-constant behavior of the static chain is replaced by a modulation-induced winding (Fig.~\ref{Fig:MP_2Mass_TD}). Second, the Berry-phase convention associated with origin choice (0 for $\psi^{+}$, $\pi$ for $\psi^{-}$) endures under temporal modulation, now interpreted as an open-path geometric phase---precisely the scenario anticipated by non-cyclic geometric-phase theory and verified numerically in our earlier work on \emph{Geometric-phase invariance}. Third, the Bloch-sphere trajectories expand from nearly planar arcs to fully three-dimensional loops, reflecting multi-harmonic coupling and enabling classical superpositions that operate like qubit rotations on the Bloch sphere. This geometric, coefficient-level description is the distinctive contribution of the present manuscript compared with prior studies of non-reciprocal elastic media: here, the band eigenstates are explicitly cast and measured as classical superpositions whose trajectories on the Bloch sphere make the corresponding Berry/Zak phases---closed or open---directly visible and origin-aware. That synthesis dovetails with foundational treatments of space-time modulation in phononic crystals and Willis materials, and it clarifies how directional gaps and Doppler-shifted replicas coexist with robust geometric-phase assignments in time-dependent lattices \cite{doi:10.1098/rspa.2017.0188,Trainiti_2016}.

\subsection{Bloch-Sphere Picture With Time-Dependent Stiffness}

\begin{figure}[H]
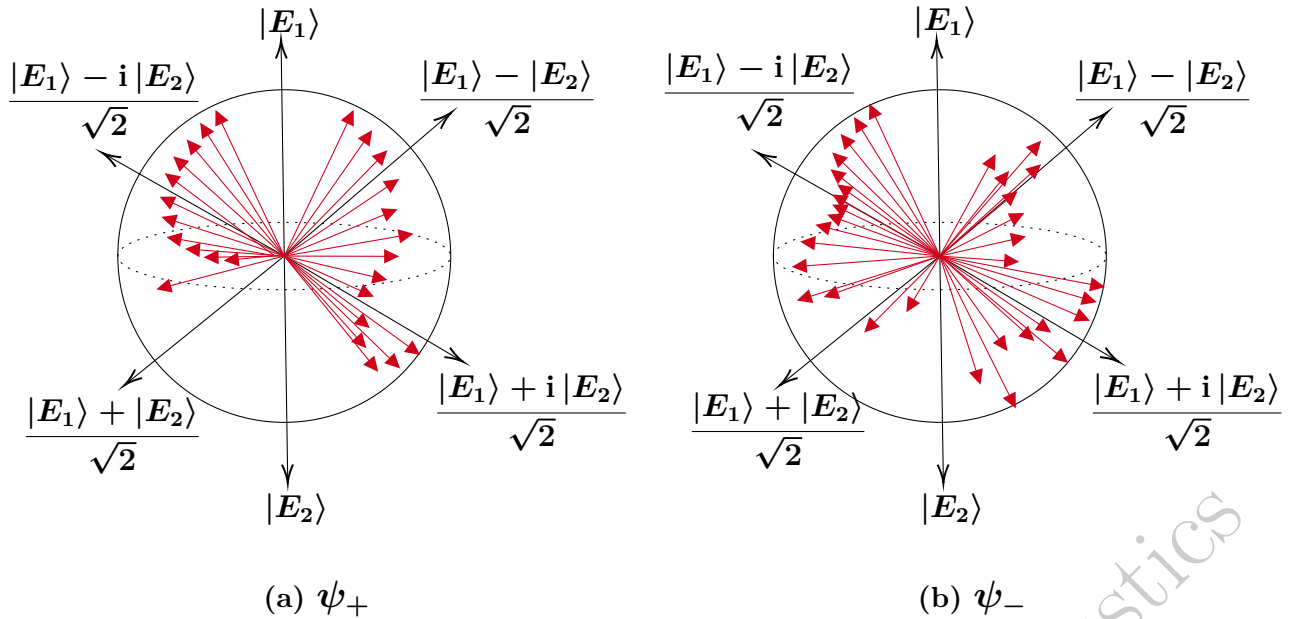

     \centering
      \begin{subfigure}{0.49\textwidth}
         \centering
         \input{Figure_Bloch_TD/Figure_Bloch_psi_plus}
         \caption{\large $\boldsymbol{\psi_+}$}
         \label{Bloch_psi+}
     \end{subfigure}
     \hfill
     \begin{subfigure}{0.49\textwidth}
         \centering
         \input{Figure_Bloch_TD/Figure_Bloch_psi_minus}
         \caption{\large $\boldsymbol{\psi_-}$}
         \label{Bloch_psi-}
     \end{subfigure}
     \caption{Bloch state demonstration of change of state between $|E_1\rangle$ and $|E_2\rangle$ at a two-mass system depicted in Hilbert Space with different stiffness matrices for the two cases (a) $\psi^+$ (b) $\psi^-$ at the FFB.} 
     \label{BlochState_TD}
\end{figure}

The Bloch-sphere representation remains a powerful geometric lens for the two-mass cell. With time-independent stiffness, the state vector traces an approximate circle (one ``plane'' on the sphere) as $k$ sweeps the zone. Under space-time modulation, the extra quadratures generated by sideband mixing allow the state to explore a much larger portion of the sphere---sampling both longitudes and latitudes rather than remaining confined to a single great-circle arc (Fig.~\ref{BlochState_TD}). Physically, near hybridization points the instantaneous diatomic state admits admixtures that effectively span the full two-state manifold, and the accumulated geometric phase follows the non-cyclic, geodesically closed prescription \cite{PhysRevLett.60.2339}.

Figure \ref{BlochState_TD} represents the Bloch state in the spatiotemporal modulation's FFB. Unlike the previous case, where the stiffness is independent of time and one direction, Bloch states are explored. We can analyze much larger regions of the Bloch states by switching to time-dependent stiffness of the lattices. The modulus hierarchy $\lvert\hat\alpha\rvert>\lvert\hat\beta\rvert$ on the $k$ band indicates that the states are approaching the pure state $|E_1\rangle$. However, the random change in relative phase ($\arg(\hat\alpha)-\arg(\hat\beta)$) indicates a state approach that deviates from the straight path as before. 

The Bloch-state representation further enables the realization of quantum-analogous gate operations. For example, as illustrated in Fig.~\ref{Bloch_psi-}, the state $(\ket{E_1} \ +\mathrm{i}\ket{E_2})/\sqrt{2}$ can be transformed into $(\ket{E_1} \ -\mathrm{i}\ket{E_2})/\sqrt{2}$ by an operation analogous to the Pauli-$Z$ gate, which effectively maps $\ket{E_2}\rightarrow -\,\ket{E_2}$. In the elastic lattice, this transformation can be traced continuously through the $k$-space sweep of the band structure, demonstrating that state manipulation on the Bloch sphere has a direct counterpart in the underlying mechanical system.

Whereas earlier work established non-reciprocity and hybridization in space-time modulated lattices, the present treatment reconstructs the classical superposition state itself---via SAAP---from MD data and represents its $k$-parametric evolution on the Bloch sphere. This provides a physically transparent, gauge-clean route to Berry/Zak phases (including their origin dependence) in both static and time-modulated settings, thereby demonstrating a classical analogue of qubit control that to our knowledge has not been articulated in this coefficient-level, Bloch-sphere form for diatomic elastic chains.

\section{Concluding Remarks}
\label{Conclusion}
We have developed and validated a compact, superposition-based description of one-dimensional elastic lattices that makes their band topology visible and computable directly from amplitude–phase data. By recasting intra-cell vibrations into normalized combinations of orthogonal eigenstates and tracking the corresponding complex coefficients across the Brillouin zone, we obtained a two-level (spinor) representation for diatomic chains in which the relative phase between the in-phase and out-of-phase modes is locked by inversion symmetry, yielding the expected quantization of the Zak phase to $0$ or $\pi$. Extending this framework to triatomic cells demonstrated how breaking inversion symmetry de-quantizes the individual band phases while preserving global band-sum constraints, clarifying which features are symmetry protected and which are gauge dependent.

Within this static two-level setting, we further operationalized the geometry of the modal coefficients by interpreting norm-preserving transformations of $(\hat{\alpha},\hat{\beta})$ as $\mathrm{SU}(2)$ rotations on the Bloch sphere. This viewpoint connects classical lattice dynamics with qubit kinematics: phase-locked states correspond to meridian-tracing trajectories, while symmetry-enforced $\pi$-jumps appear as discrete azimuthal inversions. These gate-like interpretations provide a basis-aware language for state manipulation and recast the Bloch-sphere trajectories as controllable rotations rather than merely descriptive curves.

A second principal outcome arises when the stiffness is driven by a moving spatiotemporal modulation. In this case the static, piecewise-constant phase behavior is replaced by modulation-induced winding across $k$, reflecting hybridization between carrier harmonics and Floquet sidebands. The associated geometric phase becomes the open-path Samuel–Bhandari form, accumulated along the actual Floquet trajectory. On the Bloch sphere, the resulting paths expand from nearly planar arcs to genuinely three-dimensional loops, giving a classical analogue of qubit-like rotations driven by space–time coupling while preserving the branch structure and the origin-dependent $0/\pi$ exchange familiar from the static lattice.

The Bloch-sphere construction developed in this work is intrinsically a two-component, one-dimensional descriptor: a diatomic unit cell yields an $SU(2)$-type normalized state ($\hat\alpha(k),\hat\beta(k)$) that is parameterized by a single Bloch wavenumber $k$, and the associated topological quantity is a one-dimensional Berry (Zak) phase accumulated as $k$ traverses the 1D Brillouin zone. Accordingly, the Bloch-sphere trajectories presented here provide a geometric visualization of inversion-protected phase locking and quantized $0/\pi$ Zak response in 1D lattices. 

Extending this coefficient-trajectory viewpoint to genuinely two-dimensional invariants, such as Chern numbers, is fundamentally more demanding and is not achieved by a single Bloch-sphere trajectory. A Chern number depends on the Berry curvature $\Omega(k_x,k_y)$ defined over a 2D Brillouin zone and requires a curvature integral over ($k_x,k_y$), which cannot be inferred from a single 1D loop. A practical 2D extension would therefore require (i) reconstructing the relevant Bloch eigenvectors on a 2D grid in reciprocal space (or in a 2D synthetic parameter space), (ii) addressing gauge continuity across the grid or using gauge-invariant discretized curvature formulas, and (iii) generalizing beyond an $SU(2)$ visualization when more than two internal modal components participate. We view the present amplitude-phase state reconstruction as a useful building block toward such measurements, but we emphasize that extracting 2D topological indices constitutes a separate technical development.

Taken together, these results distinguish the present manuscript from prior amplitude–phase or eigenvector-based treatments. Rather than inferring Berry/Zak phases from overlap integrals, we construct the physical state itself—band by band—and follow its evolution on the Bloch sphere, making symmetry-enforced quantization transparent in diatomic lattices and clarifying de-quantization mechanisms in triatomic ones. The same construction remains valid under temporal modulation, producing well-defined open-path geometric phases without gauge ambiguity. Furthermore, by articulating a gate-level interpretation, the framework not only visualizes but prescribes $\mathrm{SU}(2)$-equivalent rotations in a classical medium, linking plateaus, jumps, and windings in the modal phase directly to controllable state transformations. Collectively, these contributions advance earlier approaches by providing a unified, Hilbert-space-based viewpoint applicable to both static and time-modulated lattices and directly connected to experimental observables.

\newpage
\paragraph{Acknowledgements}

M.A.H. acknowledges partial support from NSF Grant No. 2204382. This work was also supported by the Science and Technology Center New Frontiers of Sound (NewFoS) through NSF Grant No. 2242925.

\paragraph{Author Contributions}

M.A.H. conceived the idea of the research. All authors analyzed the findings and contributed to the scientific discussion and the manuscript’s writing.

\paragraph{Competing Interests}

The authors declare no competing interests.

\paragraph{Data Availability Statement}

The data that support our findings of the present study are available from the corresponding author upon reasonable request.

\newpage
\printbibliography

\newpage
\appendix
\phantomsection
\pdfbookmark[1]{Supplementary Materials}{supp}
\section*{Supplementary Materials}

\setcounter{figure}{0}
\setcounter{equation}{0}
\setcounter{subsection}{0}
\setcounter{subsubsection}{0}

\renewcommand{\thefigure}{S\arabic{figure}}
\renewcommand{\theequation}{s\arabic{equation}}
\renewcommand{\thesubsection}{S.\arabic{subsection}}
\renewcommand{\thesubsubsection}{\thesubsection.\arabic{subsubsection}}

\subsection{Traveling Wave Solution in the Unit Cell}
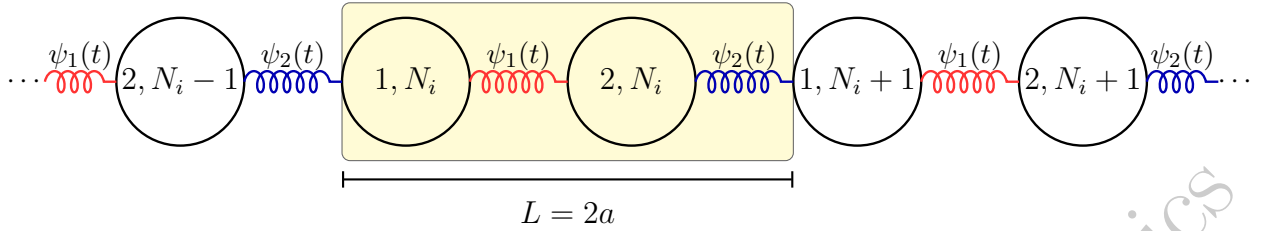
\begin{figure}[H]
    \centering
    \begin{tikzpicture}[
  line width=0.9pt,
  spring/.style   ={decorate,decoration={coil,aspect=0.6,segment length=5pt,amplitude=4pt}},
  springKone/.style={spring, draw=red!80},          
  springKtwo/.style={spring, draw=blue!70!black},   
  klabel/.style   ={font=\small, inner sep=1pt}, 
  dotlbl/.style   ={font=\small},
]

\def\R{0.85cm}       
\def\gap{3cm}      

\coordinate (mA) at ($(-2*\gap,0)$);
\coordinate (mB) at ($(-1*\gap,0)$);
\coordinate (mC) at ($( 0*\gap,0)$);
\coordinate (mD) at ($( 1*\gap,0)$);
\coordinate (mE) at ($( 2*\gap,0)$);

\draw (mA) circle (\R);
\draw (mB) circle (\R);
\draw (mC) circle (\R);
\draw (mD) circle (\R);
\draw (mE) circle (\R);

\node at (mA) {$2,N_i-1$};
\node at (mB) {$1,N_i$};
\node at (mC) {$2, N_i$};
\node at (mD) {$1,N_i+1$};
\node at (mE) {$2,N_i+1$};

\draw[springKone] ($(mA)+(-1.8cm,0)$) -- node[klabel,above=3pt] {$\psi_1(t)$} ($(mA)+(-\R,0)$);
\node[dotlbl] at ($(mA)+(-2.05cm,0)$) {$\cdots$};

\draw[springKtwo] ($(mA)+(\R,0)$) -- node[klabel,above=3pt] {$\psi_2(t)$} ($(mB)+(-\R,0)$);

\draw[springKone] ($(mB)+(\R,0)$) -- node[klabel,above=3pt] {$\psi_1(t)$} ($(mC)+(-\R,0)$);

\draw[springKtwo] ($(mC)+(\R,0)$) -- node[klabel,above=3pt] {$\psi_2(t)$} ($(mD)+(-\R,0)$);

\draw[springKone] ($(mD)+(\R,0)$) -- node[klabel,above=3pt] {$\psi_1(t)$} ($(mE)+(-\R,0)$);

\draw[springKtwo] ($(mE)+(\R,0)$) -- node[klabel,above=3pt] {$\psi_2(t)$} ($(mE)+(1.8cm,0)$);
\node[dotlbl] at  ($(mE)+(2.05cm,0)$) {$\cdots$};

\begin{scope}[on background layer]
  \def\pady{1.05cm}   
  \def\padx{0.0cm}    

  \coordinate (xL) at ($(mB)+(-\R,0)$);
  \coordinate (xR) at ($(mD)+(-\R,0)$);

  \draw[fill=yellow!20, draw=black!60, rounded corners=3pt]
    ($(xL)+(-\padx, \pady)$) rectangle ($(xR)+(\padx, -\pady)$);
\end{scope}

\draw[|-|, thick]
  ($(mB)+(-\R,-1.3cm)$) -- node[below=4pt, font=\itshape] {$L=2a$} ($(mD)+(-\R,-1.3cm)$);

\end{tikzpicture}
    \caption{Schematic of a one-dimensional diatomic mass–spring lattice consisting of two identical masses per unit cell and alternating spring constants. Each unit cell of length $L=2a$ contains two masses, $(1,N_i)$ and $(2,N_i)$, coupled by springs with stiffness $\psi_1(t)$ and $\psi_2(t)$. $a$ is the inter-mass spacing. The shaded region highlights a representative unit cell of the periodic structure.}
    \label{Mass-Spring System}
\end{figure}

We first consider a system of two identical masses per unit cell with spring constants modulation for which the analytical solutions of the eigen-problem are available and can be compared with numerical results obtained via SAAP. Figure~\ref{Mass-Spring System} represents the schematic of the 1D elastic lattices.

\subsubsection{Two-Mass Unit Cell}

The traveling wave solution of the Eq.~(\ref{eq1}) using the compact ansatz of Eq.~(\ref{eq2}) realizes that the amplitude $A_n$ are periodical in reciprocal space. That is also in contrast to the Ansatz of the general form $u_n=A_n'e^{\mathrm{i}kx_n}e^{\mathrm{i}kN_iL}e^{\mathrm{i}\omega t}$, where $A_n'e^{\mathrm{i}kx_n}$ is periodic in reciprocal space. We note that both ansatz are related by a unitary transformation that yields the same spectrum. Plugging in the Eq.~(\ref{eq2}) in Eq.~(\ref{eq1}) gives us the equation of motion as,

\begin{equation*}
\begin{aligned}
    (-m\omega^2+\psi_1+\psi_2)A_1-(\psi_1+\psi_2e^{-\mathrm{i}kL})A_2 & =0\\
    -(\psi_1+\psi_2e^{+\mathrm{i}kL})A_1+(-m\omega^2+\psi_1+\psi_2)A_2 & =0
\end{aligned}
\end{equation*}

\noindent These form a set of linear equations in the amplitudes $A_1$ and $A_2$. These can be rearranged in matrix form:

\begin{equation}
    \begin{pmatrix}
        \gamma & -\tau^*\\
        -\tau & \gamma
    \end{pmatrix}
    \begin{pmatrix}
        A_1 \\ A_2
    \end{pmatrix}
    =
    \begin{pmatrix}
        0 \\ 0
    \end{pmatrix}
    \label{eq:amp matrix}
\end{equation}

Here, we have defined $\gamma=-m\omega^2+\psi_1+\psi_2$ and $\tau=\psi_1+\psi_2e^{\mathrm{i}kL}$, where $\tau^*$ denotes the complex conjugate of $\tau$. The nontrivial solutions lead to:

\begin{equation*}
    \gamma^2-\tau\tau^*=0\quad\Rightarrow\quad\gamma=\pm\sqrt{\tau\tau^*}
\end{equation*}

\noindent Nontrivial solutions require the determinant to vanish:
\begin{equation*}
\begin{aligned}
&\big(\psi_1+\psi_2 - m\omega^2\big)^2
  -\big(\psi_1+\psi_2 e^{-\mathrm{i}kL}\big)\big(\psi_1+\psi_2 e^{\mathrm{i}kL}\big)=0\\[4pt]
\Rightarrow\quad & {\big[(\psi_1+\psi_2)^2 - 2(\psi_1+\psi_2)\,m\omega^2 + m^2\omega^4\big]}
 \;-\;
 {\Big[\psi_1^2+\psi_1\psi_2\!\big(e^{\mathrm{i}kL}+e^{-\mathrm{i}kL}\big)+\psi_2^2\Big]}=0\\[4pt]
\Rightarrow\quad & m^2\omega^4 - 2(\psi_1+\psi_2)m\omega^2
  +(\psi_1+\psi_2)^2
  -\psi_1^2-\psi_2^2-\psi_1\psi_2\!\big(e^{\mathrm{i}kL}+e^{-\mathrm{i}kL}\big)=0\\[4pt]
\Rightarrow\quad & m^2\omega^4 - 2(\psi_1+\psi_2)m\omega^2
  +(\psi_1+\psi_2)^2
  -\psi_1^2-\psi_2^2-2\psi_1\psi_2\cos(kL)=0\\[4pt]
\Rightarrow\quad & m^2\omega^4 - 2(\psi_1+\psi_2)m\omega^2
  +\big[\psi_1^2+2\psi_1\psi_2+\psi_2^2-\psi_1^2-\psi_2^2-2\psi_1\psi_2\cos(kL)\big]=0\\[4pt]
\Rightarrow\quad & m^2\omega^4 - 2(\psi_1+\psi_2)m\omega^2
  +2\psi_1\psi_2\big(1-\cos kL\big) =0\\[4pt]
\Rightarrow\quad & m^2\omega^4 - 2(\psi_1+\psi_2)m\omega^2
+4\psi_1\psi_2\sin^2\!\Big(\frac{kL}{2}\Big)=0,\qquad \left[(1-\cos{kL})=2\sin^2\left(\frac{kL}{2}\right)\right]
\end{aligned}
\end{equation*}

\noindent which is a quadratic equation for $\omega^2$. Solving,
\begin{equation*}
\omega^2_{\pm}
=\frac{\psi_1+\psi_2}{m}
\left[
1 \pm \sqrt{\,1-\frac{4\psi_1\psi_2}{(\psi_1+\psi_2)^2}
\sin^2\!\Big(\frac{kL}{2}\Big)}
\;\right],
\end{equation*}
giving the acoustic $(+)$ and optical $(-)$ branches.

\begin{figure}[H]
    \centering
    \begin{tikzpicture}
  \begin{axis}[
  height=6cm,
  width=12cm,
    xlabel={\textbf{\textit{k}L}},
    ylabel={$\boldsymbol{\omega_j(k)}$},
    xmin=-3.141592654, xmax=3.141592654,
    ymin=0, ymax=2.5,
    every x tick/.style={color=black, thick},
    xtick={-3.1416,-1.5708,0,1.5708,3.1416},
    xticklabels={$-\pi$,$-\frac{\pi}{2}$,0,$\frac{\pi}{2}$,$\pi$},
    every y tick/.style={color=black, thick},
    ytick={0,0.5,1,1.5,2,2.5},
    legend style={at={(0.98,0.02)},anchor=south east}, %
  ]

\addplot[
    only marks,
    color=red,
    mark=*,
    mark size=1.5pt,
]
coordinates {
(-3.141592654,	1.414213562)
(-3.00790786,	1.407945815)
(-2.874223066,	1.389711593)
(-2.740538272,	1.36096265)
(-2.606853479,	1.323496735)
(-2.473168685,	1.279018243)
(-2.339483891,	1.22893933)
(-2.205799097,	1.174355585)
(-2.072114303,	1.116095972)
(-1.93842951,	1.054787238)
(-1.804744716,	0.9909097)
(-1.671059922,	0.924839541)
(-1.537375128,	0.856879145)
(-1.403690335,	0.787278396)
(-1.270005541,	0.716249541)
(-1.136320747,	0.643977611)
(-1.002635953,	0.570627785)
(-0.86895116,	0.496350649)
(-0.735266366,	0.421286008)
(-0.601581572,	0.345565685)
(-0.467896778,	0.26931561)
(-0.334211984,	0.192657423)
(-0.200527191,	0.115709722)
(-0.066842397,	0.038589081)
(0.066842397,	0.038589081)
(0.200527191,	0.115709722)
(0.334211984,	0.192657423)
(0.467896778,	0.26931561)
(0.601581572,	0.345565685)
(0.735266366,	0.421286008)
(0.86895116,	0.496350649)
(1.002635953,	0.570627785)
(1.136320747,	0.643977611)
(1.270005541,	0.716249541)
(1.403690335,	0.787278396)
(1.537375128,	0.856879145)
(1.671059922,	0.924839541)
(1.804744716,	0.9909097)
(1.93842951,	1.054787238)
(2.072114303,	1.116095972)
(2.205799097,	1.174355585)
(2.339483891,	1.22893933)
(2.473168685,	1.279018243)
(2.606853479,	1.323496735)
(2.740538272,	1.36096265)
(2.874223066,	1.389711593)
(3.00790786,	1.407945815)
(3.141592654,	1.414213562)
};
\addlegendentry{\scriptsize Acoustic}

\addplot[
    only marks,
    color=blue,
    mark=*,
    mark size=1.5pt,
]
coordinates {
(-3.141592654,	2)
(-3.00790786,	2.004417267)
(-2.874223066,	2.0171023)
(-2.740538272,	2.036610091)
(-2.606853479,	2.061154141)
(-2.473168685,	2.089045795)
(-2.339483891,	2.118893136)
(-2.205799097,	2.149625307)
(-2.072114303,	2.180442565)
(-1.93842951,	2.210751881)
(-1.804744716,	2.24011115)
(-1.671059922,	2.268186902)
(-1.537375128,	2.294723977)
(-1.403690335,	2.319524246)
(-1.270005541,	2.34243177)
(-1.136320747,	2.363322415)
(-1.002635953,	2.382096541)
(-0.86895116,	2.398673807)
(-0.735266366,	2.412989453)
(-0.601581572,	2.42499162)
(-0.467896778,	2.434639419)
(-0.334211984,	2.441901537)
(-0.200527191,	2.446755251)
(-0.066842397,	2.449185759)
(0.066842397,	2.449185759)
(0.200527191,	2.446755251)
(0.334211984,	2.441901537)
(0.467896778,	2.434639419)
(0.601581572,	2.42499162)
(0.735266366,	2.412989453)
(0.86895116,	2.398673807)
(1.002635953,	2.382096541)
(1.136320747,	2.363322415)
(1.270005541,	2.34243177)
(1.403690335,	2.319524246)
(1.537375128,	2.294723977)
(1.671059922,	2.268186902)
(1.804744716,	2.24011115)
(1.93842951,	2.210751881)
(2.072114303,	2.180442565)
(2.205799097,	2.149625307)
(2.339483891,	2.118893136)
(2.473168685,	2.089045795)
(2.606853479,	2.061154141)
(2.740538272,	2.036610091)
(2.874223066,	2.0171023)
(3.00790786,	2.004417267)
(3.141592654,	2)
};
\addlegendentry{\scriptsize Optical}

\end{axis}
\end{tikzpicture}
    \caption{Dispersion relation of the diatomic lattice in the first Brillouin zone ($-\pi \leq kL \leq \pi$). The lower curve (red) is the acoustic branch, which tends to zero frequency at $k=0$, while the upper curve (blue) is the optical branch, which starts at finite frequency. A band gap exists between the two branches when $\psi_1\neq\psi_2$, reflecting the difference in stiffness.}
    \label{fig:band dispersion}
\end{figure}
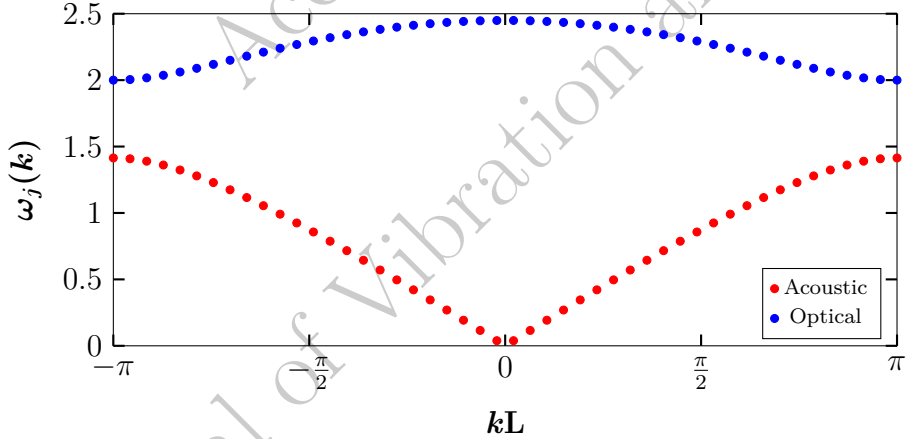

\paragraph{Acoustic branch $(+)$:} 

At small wavenumber $k$, the frequency tends to zero ($\omega \to 0$). This corresponds to long-wavelength oscillations where all masses move in phase, similar to a uniform translation. The branch increases approximately linearly with $k$ near the origin, mimicking sound waves, which is why it is called the acoustic branch.

\paragraph{Optical branch $(-)$:} 

At $k=0$, this branch begins at a finite cutoff frequency 
\[
\omega = \sqrt{\tfrac{2(\psi_1+\psi_2)}{m}}.
\]
This corresponds to out-of-phase oscillations between the two sublattice masses. The frequency remains finite even for long wavelengths, which is why it is termed the optical branch.

Figure \ref{fig:band dispersion} displays both branches across the first Brillouin zone ($-\pi \leq kL \leq \pi$). The periodicity in reciprocal space is visible, as the dispersion repeats with period $2\pi$. The band gap between the acoustic and optical branches is set by the difference in coupling constants $\psi_1$ and $\psi_2$. If $\psi_1=\psi_2$, the square-root factor becomes unity and the gap closes, producing continuous bands that meet at the Brillouin zone boundary. For unequal stiffness, a finite frequency gap opens, which is evident in Fig. \ref{fig:band dispersion}. In this specific plot, the dispersion was computed for a lattice of 48 unit cells, discretizing the Brillouin zone into 48 allowed $k$-points. The acoustic branch appears in the lower half, while the optical branch appears in the upper half, with both symmetric about $k=0$.

\subsubsection{Correlation of the Amplitudes}
Taking the complex conjugate amplitudes of Eq.~(\ref{eq:amp matrix}), we get,
\begin{equation}
\begin{pmatrix}
\gamma & -\tau \\
-\tau^{*} & \gamma
\end{pmatrix}
\begin{pmatrix}
A_1^{*} \\
A_2^{*}
\end{pmatrix}
=
\begin{pmatrix}
0 \\
0
\end{pmatrix}
\;\;\Rightarrow\;\;
\begin{pmatrix}
\gamma A_1^{*} - \tau A_2^{*} \\
-\tau^{*} A_1^{*} + \gamma A_2^{*}
\end{pmatrix}
=
\begin{pmatrix}
0 \\
0
\end{pmatrix}
\;\;\Rightarrow\;\;
\begin{pmatrix}
-\tau A_2^{*} + \gamma A_1^{*} \\
\gamma A_2^{*} - \tau^{*} A_1^{*}
\end{pmatrix}
=
\begin{pmatrix}
0 \\
0
\end{pmatrix}
\label{eq:amp matrix conj}
\end{equation}

\paragraph{Acoustic Branch:}

Comparing Eq.~(\ref{eq:amp matrix}) and (\ref{eq:amp matrix conj}) gives us the corelation of the amplitudes as
\begin{equation*}
    A_1=A_2^*\qquad A_2=A_1^*
\end{equation*}
From Eq.~(\ref{eq:amp matrix}), we can write the equation as,
\[
\begin{pmatrix}
\gamma & -\tau^{*} \\
-\tau & \gamma
\end{pmatrix}
\begin{pmatrix}
A_1 \\
A_2
\end{pmatrix}
=
\begin{pmatrix}
0 \\
0
\end{pmatrix}
\]

\[
\begin{aligned}
    &\gamma A_1 - \tau^{*} A_2 = 0 \\
    \Rightarrow\;& \gamma A_1 = \tau^{*} A_2 \;\;\Rightarrow\;\; (\gamma A_1)^2 = (\tau^{*} A_2)^2 \\
    \Rightarrow\;& \gamma^2 (A_1)^2 = (\tau^{*})^2 (A_2)^2 \;\;\Rightarrow\;\; (\tau \tau^{*})(A_1)^2 = (\tau^{*}\tau)(A_2)^2 \\
    \Rightarrow\;& \frac{(A_1)^2}{(A_2)^2} = \frac{\tau^{*}}{\tau} \;\;\Rightarrow\;\; \frac{A_1}{A_2} = \pm \frac{\sqrt{\tau^{*}}}{\sqrt{\tau}}
\end{aligned}
\]
Hence,
\[
\begin{pmatrix}
A_1 \\
A_2
\end{pmatrix}
\propto
\begin{pmatrix}
\sqrt{\tau^{*}} \\
\pm \sqrt{\tau}
\end{pmatrix}
\]

\paragraph{Optical branch:}

The nontrivial condition is \(\det=0\Rightarrow \gamma^2-|\tau|^2=0\), so
\(\gamma=\pm|\tau|\).
At \(k=0\) the \emph{optical} mode is out of phase; this corresponds to
\[
\ \gamma=-\,|\tau|\ 
\]

\noindent From the first row of Eq.~(\ref{eq:amp matrix}),
\[
\gamma A_1=\tau^{*}A_2 \quad\Rightarrow\quad
\frac{A_2}{A_1}=\frac{\gamma}{\tau^{*}}=
-\,\frac{|\tau|}{\tau^{*}}
= -\,e^{\mathrm{i}\theta},\qquad \theta\equiv \arg\tau 
\]
Hence an eigenvector for the optical branch is
\begin{equation*}
\label{eq:opt-vec-basic}
\begin{pmatrix}A_1\\[2pt] A_2\end{pmatrix}
\propto
\begin{pmatrix}1\\[2pt]-\,e^{\mathrm{i}\theta}\end{pmatrix}
=
e^{-\mathrm{i}\theta/2}
\begin{pmatrix}e^{-\mathrm{i}\theta/2}\\[2pt]-\,e^{+\mathrm{i}\theta/2}\end{pmatrix}.
\end{equation*}

\noindent Choose a consistent branch of the complex square root so that
\[
\sqrt{\tau}= \sqrt{|\tau|}\,e^{\mathrm{i}\theta/2},\qquad
\sqrt{\tau^{*}}= \sqrt{|\tau|}\,e^{-\mathrm{i}\theta/2}.
\]

\noindent since
\(
\sqrt{\tau^{*}}:\!-\sqrt{\tau}= e^{-\mathrm{i}\theta/2}:\!-e^{+\mathrm{i}\theta/2}
\):
\[
\begin{pmatrix}A_1\\ A_2\end{pmatrix}
\propto
\begin{pmatrix}\sqrt{e^{-\mathrm{i}\pi}\,\tau}\\[2pt]\pm\sqrt{e^{+\mathrm{i}\pi}\,\tau^{*}}\end{pmatrix}.
\]

\subsection{Three Mass Unit Cell}
The diatomic chain represents the simplest topological elastic system, but its two–level modal structure confines the Bloch state to a single phase relationship between the basis vectors. A triatomic unit cell relaxes this restriction by introducing a third independent modal component, enabling the state to exhibit nontrivial phase cycling across multiple pairs of eigenvectors. In this case, the Bloch mode traces a path in a higher–dimensional Hilbert space, where both amplitude dominance and multi–phase locking determine the band topology. This configuration forms the minimal classical platform capable of exhibiting qutrit–like state evolution and $SU(3)$–type geometric structure. To highlight these effects, Fig.~\ref{Fig:Modulation_3Mass} shows how the normalized coefficients $(\hat{\alpha},\hat{\beta},\hat{\gamma})$ vary over the Brillouin zone for different stiffness orderings that either preserve or break inversion symmetry.

\tikzset{every picture/.style={baseline=(current bounding box.south)}}
\captionsetup[subfigure]{skip=2pt}

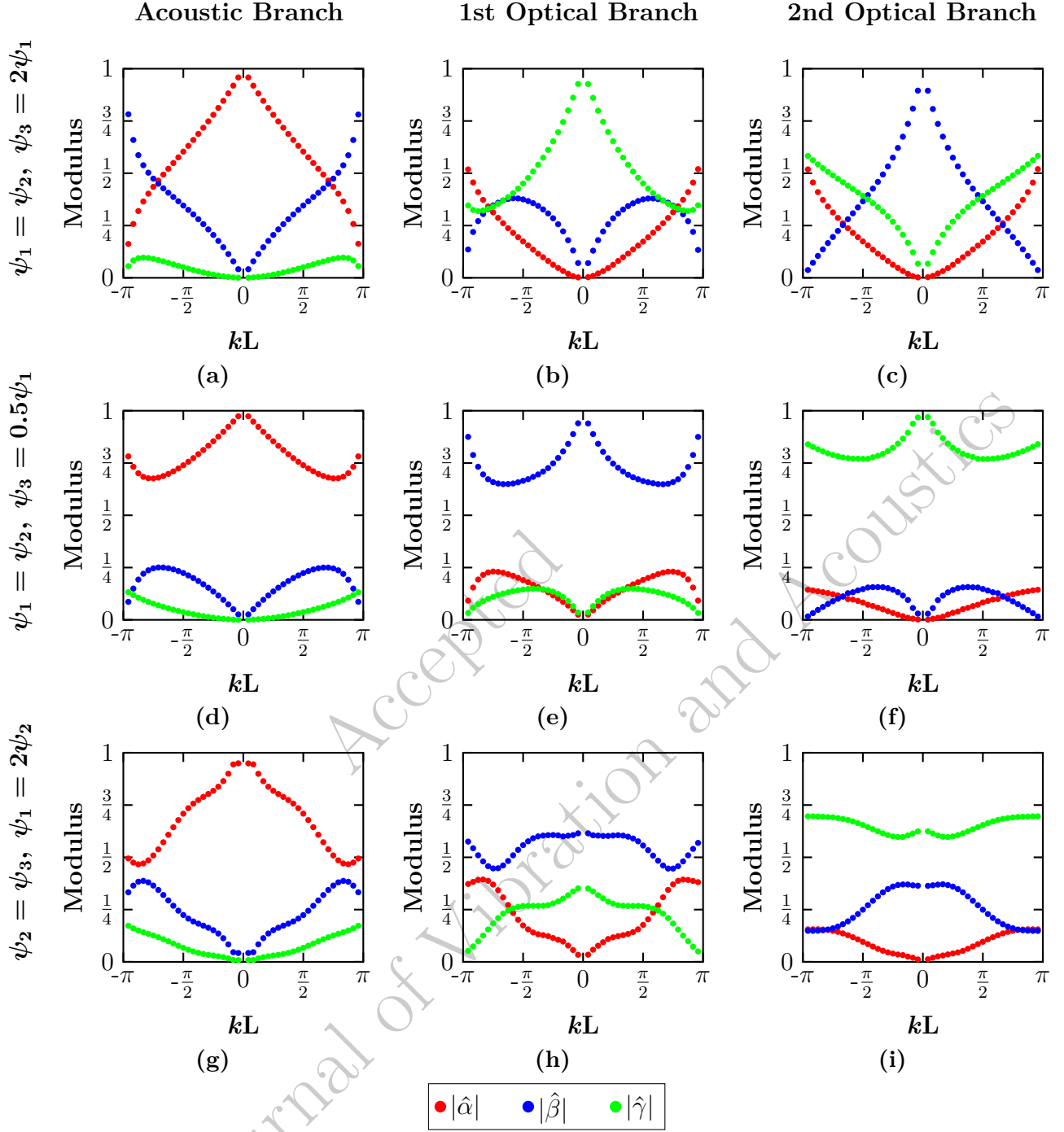
\begin{figure}[H]
\centering
\setlength{\tabcolsep}{6pt}
\renewcommand{\arraystretch}{1.2}

\begin{tabular}{@{}m{0.02\textwidth}  
                p{0.3\textwidth} p{0.3\textwidth} p{0.3\textwidth}@{}}
   & \multicolumn{1}{c}{\textbf{\hspace{1.8em}Acoustic Branch}}
   & \multicolumn{1}{c}{\textbf{\hspace{1.8em}1st Optical Branch}}
   & \multicolumn{1}{c}{\textbf{\hspace{1.8em}2nd Optical Branch}} \\[0.8em]

   \rotatebox{90}{\makebox[0pt][l]{\hspace{3.2em}$\boldsymbol{\psi_1=\psi_2,\ \psi_3=2\psi_1}$}} &
     \begin{subfigure}[t]{\linewidth}\centering
       \begin{tikzpicture}
\begin{axis}[
    height=5cm,
    width=5.5cm,
    thick,
    xlabel={\textbf{\textit{k}L}},
    ylabel={\textbf{Modulus}},
    xmin=-3.1416, xmax=3.1416,
    ymin=0, ymax=1,
    every x tick/.style={color=black, thick},
    xtick={-3.1416,-1.5708,0,1.5708,3.1416},
    xticklabels={-$\pi$,-$\frac{\pi}{2}$,0,$\frac{\pi}{2}$,$\pi$},
    every y tick/.style={color=black, thick},
    ytick={0,0.25,0.5,0.75,1},
    yticklabels={$0$,$\frac{1}{4}$,$\frac{1}{2}$,$\frac{3}{4}$,$1$},
    legend style={at={(2.5,0.3)}, anchor=east}, 
    legend cell align={left},
]

\addplot[
    only marks,
    color=red,
    mark=*,
    mark size=1pt,
]
coordinates {
(-3.01,0.162)
(-2.879793266,0.257466259)
(-2.748893572,0.319327357)
(-2.617993878,0.365057522)
(-2.487094184,0.402342148)
(-2.35619449,0.434961892)
(-2.225294796,0.465234121)
(-2.094395102,0.493787323)
(-1.963495408,0.521471132)
(-1.832595715,0.54887419)
(-1.701696021,0.576149246)
(-1.570796327,0.603510845)
(-1.439896633,0.631110694)
(-1.308996939,0.659351661)
(-1.178097245,0.688294639)
(-1.047197551,0.717710969)
(-0.916297857,0.748231611)
(-0.785398163,0.779658257)
(-0.654498469,0.812463721)
(-0.523598776,0.846405666)
(-0.392699082,0.882185649)
(-0.261799388,0.919198778)
(-0.130899694,0.95825549)
(0.130899694,0.958247914)
(0.261799388,0.918948382)
(0.392699082,0.881739519)
(0.523598776,0.846363215)
(0.654498469,0.812213432)
(0.785398163,0.779692295)
(0.916297857,0.74809946)
(1.047197551,0.71768178)
(1.178097245,0.688228351)
(1.308996939,0.659321914)
(1.439896633,0.631224014)
(1.570796327,0.603663815)
(1.701696021,0.576234007)
(1.832595715,0.548951758)
(1.963495408,0.521627037)
(2.094395102,0.493826837)
(2.225294796,0.46537952)
(2.35619449,0.435134605)
(2.487094184,0.402391031)
(2.617993878,0.365192167)
(2.748893572,0.319587755)
(2.879793266,0.257743857)
(3.01069296,0.162781722)
};

\addplot[
    only marks,
    color=blue,
    mark=*,
    mark size=1pt,
]
coordinates {
(-3.01,0.782)
(-2.879793266,0.659282182)
(-2.748893572,0.586699487)
(-2.617993878,0.538347880)
(-2.487094184,0.502807267)
(-2.356194490,0.474440443)
(-2.225294796,0.449891625)
(-2.094395102,0.427754415)
(-1.963495408,0.406906009)
(-1.832595715,0.386511474)
(-1.701696021,0.366198697)
(-1.570796327,0.345718314)
(-1.439896633,0.324915280)
(-1.308996939,0.303163169)
(-1.178097245,0.280324256)
(-1.047197551,0.256676609)
(-0.916297857,0.231527035)
(-0.785398163,0.204926687)
(-0.654498469,0.176414103)
(-0.523598776,0.146224476)
(-0.392699082,0.113590888)
(-0.261799388,0.078793712)
(-0.130899694,0.041186887)
(0.130899694,0.041188211)
(0.261799388,0.079000600)
(0.392699082,0.113960933)
(0.523598776,0.146275568)
(0.654498469,0.176687617)
(0.785398163,0.204893031)
(0.916297857,0.231616093)
(1.047197551,0.256697674)
(1.178097245,0.280310527)
(1.308996939,0.303101753)
(1.439896633,0.324813445)
(1.570796327,0.345584786)
(1.701696021,0.366100738)
(1.832595715,0.386311295)
(1.963495408,0.406659287)
(2.094395102,0.427697265)
(2.225294796,0.449756676)
(2.356194490,0.474185994)
(2.487094184,0.502643097)
(2.617993878,0.538050208)
(2.748893572,0.586228876)
(2.879793266,0.659147551)
(3.010692960,0.780903330)
};

    
\addplot[
    only marks,
    color=green,
    mark=*,
    mark size=1pt,
]
coordinates {
(-3.01,0.0558)
(-2.879793266,0.083251559)
(-2.748893572,0.093973156)
(-2.617993878,0.096594598)
(-2.487094184,0.094850585)
(-2.356194490,0.090597665)
(-2.225294796,0.084874254)
(-2.094395102,0.078458262)
(-1.963495408,0.071622860)
(-1.832595715,0.064614336)
(-1.701696021,0.057652056)
(-1.570796327,0.050770841)
(-1.439896633,0.043974026)
(-1.308996939,0.037485170)
(-1.178097245,0.031381105)
(-1.047197551,0.025612423)
(-0.916297857,0.020241354)
(-0.785398163,0.015415056)
(-0.654498469,0.011122176)
(-0.523598776,0.007369857)
(-0.392699082,0.004223463)
(-0.261799388,0.002007510)
(-0.130899694,0.000557623)
(0.130899694,0.000563875)
(0.261799388,0.002051018)
(0.392699082,0.004299548)
(0.523598776,0.007361217)
(0.654498469,0.011098951)
(0.785398163,0.015414674)
(0.916297857,0.020284447)
(1.047197551,0.025620546)
(1.178097245,0.031461122)
(1.308996939,0.037576333)
(1.439896633,0.043962540)
(1.570796327,0.050751399)
(1.701696021,0.057665256)
(1.832595715,0.064736947)
(1.963495408,0.071713677)
(2.094395102,0.078475898)
(2.225294796,0.084863804)
(2.356194490,0.090679401)
(2.487094184,0.094965872)
(2.617993878,0.096757625)
(2.748893572,0.094183370)
(2.879793266,0.083108592)
(3.010692960,0.056314948)
};
    
\end{axis}

\end{tikzpicture}\caption{}\label{M_AC_s1s2_2s3}
     \end{subfigure} &
     \begin{subfigure}[t]{\linewidth}\centering
       \begin{tikzpicture}
\begin{axis}[
    height=5cm,
    width=5.5cm,
    thick,
    xlabel={\textbf{\textit{k}L}},
    ylabel={\textbf{Modulus}},
    xmin=-3.1416, xmax=3.1416,
    ymin=0, ymax=1,
    every x tick/.style={color=black, thick},
    xtick={-3.1416,-1.5708,0,1.5708,3.1416},
    xticklabels={-$\pi$,-$\frac{\pi}{2}$,0,$\frac{\pi}{2}$,$\pi$},
    every y tick/.style={color=black, thick},
    ytick={0,0.25,0.5,0.75,1},
    yticklabels={$0$,$\frac{1}{4}$,$\frac{1}{2}$,$\frac{3}{4}$,$1$},
    legend style={at={(1.05,1.25)}, anchor=east}, 
    legend cell align={left},
]

\addplot[
    only marks,
    color=red,
    mark=*,
    mark size=1pt,
]
coordinates {
(-3.01069296,0.518435312)
(-2.879793266,0.456240319)
(-2.748893572,0.409950395)
(-2.617993878,0.372800626)
(-2.487094184,0.341303099)
(-2.35619449,0.312764697)
(-2.225294796,0.286582761)
(-2.094395102,0.26100386)
(-1.963495408,0.237178381)
(-1.832595715,0.214954902)
(-1.701696021,0.194072236)
(-1.570796327,0.174016053)
(-1.439896633,0.155078118)
(-1.308996939,0.134621908)
(-1.178097245,0.115426561)
(-1.047197551,0.098387822)
(-0.916297857,0.080429028)
(-0.785398163,0.062606896)
(-0.654498469,0.047334596)
(-0.523598776,0.032678209)
(-0.392699082,0.019433005)
(-0.261799388,0.007619764)
(-0.130899694,0.002066028)
(0.130899694,0.002951772)
(0.261799388,0.00942597)
(0.392699082,0.020839331)
(0.523598776,0.032196894)
(0.654498469,0.046409416)
(0.785398163,0.062446654)
(0.916297857,0.080414508)
(1.047197551,0.098290635)
(1.178097245,0.115913724)
(1.308996939,0.135649208)
(1.439896633,0.153951425)
(1.570796327,0.173065472)
(1.701696021,0.193728633)
(1.832595715,0.215619299)
(1.963495408,0.237780414)
(2.094395102,0.261106542)
(2.225294796,0.286285111)
(2.35619449,0.313374802)
(2.487094184,0.342124088)
(2.617993878,0.373421614)
(2.748893572,0.409990249)
(2.879793266,0.455673925)
(3.01069296,0.519784115)
};

\addplot[
    only marks,
    color=blue,
    mark=*,
    mark size=1pt,
]
coordinates {
(-3.01069296,0.135600557)
(-2.879793266,0.219200307)
(-2.748893572,0.271197554)
(-2.617993878,0.305014123)
(-2.487094184,0.328342258)
(-2.35619449,0.346139922)
(-2.225294796,0.359117979)
(-2.094395102,0.36988082)
(-1.963495408,0.377195634)
(-1.832595715,0.380348236)
(-1.701696021,0.37948414)
(-1.570796327,0.374997362)
(-1.439896633,0.367068047)
(-1.308996939,0.357053633)
(-1.178097245,0.343892634)
(-1.047197551,0.327088034)
(-0.916297857,0.307951971)
(-0.785398163,0.284838733)
(-0.654498469,0.256274994)
(-0.523598776,0.222204711)
(-0.392699082,0.179721585)
(-0.261799388,0.130722728)
(-0.130899694,0.070556337)
(0.130899694,0.070192827)
(0.261799388,0.129774794)
(0.392699082,0.180098516)
(0.523598776,0.221475597)
(0.654498469,0.255983535)
(0.785398163,0.284407362)
(0.916297857,0.307341371)
(1.047197551,0.326624861)
(1.178097245,0.343031497)
(1.308996939,0.356206848)
(1.439896633,0.367351313)
(1.570796327,0.375236845)
(1.701696021,0.379251716)
(1.832595715,0.379447475)
(1.963495408,0.376162971)
(2.094395102,0.369707883)
(2.225294796,0.359305997)
(2.35619449,0.345017062)
(2.487094184,0.327219142)
(2.617993878,0.303781874)
(2.748893572,0.270787093)
(2.879793266,0.219648163)
(3.01069296,0.133376089)
};
    
\addplot[
    only marks,
    color=green,
    mark=*,
    mark size=1pt,
]
coordinates {
(-3.01069296,0.345964131)
(-2.879793266,0.324559375)
(-2.748893572,0.318852051)
(-2.617993878,0.322185251)
(-2.487094184,0.330354643)
(-2.35619449,0.341095381)
(-2.225294796,0.35429926)
(-2.094395102,0.36911532)
(-1.963495408,0.385625985)
(-1.832595715,0.404696862)
(-1.701696021,0.426443623)
(-1.570796327,0.450986585)
(-1.439896633,0.477853835)
(-1.308996939,0.508324459)
(-1.178097245,0.540680805)
(-1.047197551,0.574524144)
(-0.916297857,0.611619001)
(-0.785398163,0.652554371)
(-0.654498469,0.69639041)
(-0.523598776,0.74511708)
(-0.392699082,0.80084541)
(-0.261799388,0.861657508)
(-0.130899694,0.927377635)
(0.130899694,0.926855401)
(0.261799388,0.860799237)
(0.392699082,0.799062153)
(0.523598776,0.746327509)
(0.654498469,0.697607049)
(0.785398163,0.653145984)
(0.916297857,0.612244121)
(1.047197551,0.575084505)
(1.178097245,0.541054779)
(1.308996939,0.508143944)
(1.439896633,0.478697262)
(1.570796327,0.451697683)
(1.701696021,0.427019651)
(1.832595715,0.404933227)
(1.963495408,0.386056615)
(2.094395102,0.369185575)
(2.225294796,0.354408892)
(2.35619449,0.341608136)
(2.487094184,0.33065677)
(2.617993878,0.322796512)
(2.748893572,0.319222658)
(2.879793266,0.324677912)
(3.01069296,0.346839796)
};
    
\end{axis}

\end{tikzpicture}\caption{}\label{M_1O_s1s2_2s3}
     \end{subfigure} &
     \begin{subfigure}[t]{\linewidth}\centering
       \begin{tikzpicture}
\begin{axis}[
    height=5cm,
    width=5.5cm,
    thick,
    xlabel={\textbf{\textit{k}L}},
    ylabel={\textbf{Modulus}},
    xmin=-3.1416, xmax=3.1416,
    ymin=0, ymax=1,
    every x tick/.style={color=black, thick},
    xtick={-3.1416,-1.5708,0,1.5708,3.1416},
    xticklabels={-$\pi$,-$\frac{\pi}{2}$,0,$\frac{\pi}{2}$,$\pi$},
    every y tick/.style={color=black, thick},
    ytick={0,0.25,0.5,0.75,1},
    yticklabels={$0$,$\frac{1}{4}$,$\frac{1}{2}$,$\frac{3}{4}$,$1$},
    legend style={at={(1.05,1.25)}, anchor=east}, 
    legend cell align={left},
]

\addplot[
    only marks,
    color=red,
    mark=*,
    mark size=1pt,
]
coordinates {
(-3.01069296,0.518435312)
(-2.879793266,0.456240319)
(-2.748893572,0.409950395)
(-2.617993878,0.372800626)
(-2.487094184,0.341303099)
(-2.35619449,0.312764697)
(-2.225294796,0.286582761)
(-2.094395102,0.26100386)
(-1.963495408,0.237178381)
(-1.832595715,0.214954902)
(-1.701696021,0.194072236)
(-1.570796327,0.174016053)
(-1.439896633,0.155078118)
(-1.308996939,0.134621908)
(-1.178097245,0.115426561)
(-1.047197551,0.098387822)
(-0.916297857,0.080429028)
(-0.785398163,0.062606896)
(-0.654498469,0.047334596)
(-0.523598776,0.032678209)
(-0.392699082,0.019433005)
(-0.261799388,0.007619764)
(-0.130899694,0.002066028)
(0.130899694,0.002951772)
(0.261799388,0.00942597)
(0.392699082,0.020839331)
(0.523598776,0.032196894)
(0.654498469,0.046409416)
(0.785398163,0.062446654)
(0.916297857,0.080414508)
(1.047197551,0.098290635)
(1.178097245,0.115913724)
(1.308996939,0.135649208)
(1.439896633,0.153951425)
(1.570796327,0.173065472)
(1.701696021,0.193728633)
(1.832595715,0.215619299)
(1.963495408,0.237780414)
(2.094395102,0.261106542)
(2.225294796,0.286285111)
(2.35619449,0.313374802)
(2.487094184,0.342124088)
(2.617993878,0.373421614)
(2.748893572,0.409990249)
(2.879793266,0.455673925)
(3.01069296,0.519784115)
};

\addplot[
    only marks,
    color=blue,
    mark=*,
    mark size=1pt,
]
coordinates {
(-3.01069296,0.037219734)
(-2.879793266,0.071918933)
(-2.748893572,0.104705232)
(-2.617993878,0.135964415)
(-2.487094184,0.166148451)
(-2.35619449,0.1952545)
(-2.225294796,0.224197064)
(-2.094395102,0.252200615)
(-1.963495408,0.280286486)
(-1.832595715,0.308510964)
(-1.701696021,0.336825796)
(-1.570796327,0.365619127)
(-1.439896633,0.395334411)
(-1.308996939,0.426658304)
(-1.178097245,0.459063844)
(-1.047197551,0.494221917)
(-0.916297857,0.532587606)
(-0.785398163,0.574797299)
(-0.654498469,0.621141949)
(-0.523598776,0.674775166)
(-0.392699082,0.736264826)
(-0.261799388,0.809304249)
(-0.130899694,0.895239313)
(0.130899694,0.894797474)
(0.261799388,0.807350029)
(0.392699082,0.736076874)
(0.523598776,0.67511954)
(0.654498469,0.621998495)
(0.785398163,0.574553758)
(0.916297857,0.532751844)
(1.047197551,0.494628371)
(1.178097245,0.459225783)
(1.308996939,0.426525252)
(1.439896633,0.39569347)
(1.570796327,0.365896134)
(1.701696021,0.336693705)
(1.832595715,0.308376136)
(1.963495408,0.280271882)
(2.094395102,0.252198627)
(2.225294796,0.223812677)
(2.35619449,0.195334865)
(2.487094184,0.165969567)
(2.617993878,0.136061192)
(2.748893572,0.10469081)
(2.879793266,0.071680963)
(3.01069296,0.03712665)
};
    
\addplot[
    only marks,
    color=green,
    mark=*,
    mark size=1pt,
]
coordinates {
(-3.01069296,0.583093927)
(-2.879793266,0.56238943)
(-2.748893572,0.542823926)
(-2.617993878,0.524523391)
(-2.487094184,0.506888971)
(-2.35619449,0.489859433)
(-2.225294796,0.473160945)
(-2.094395102,0.457164981)
(-1.963495408,0.441016023)
(-1.832595715,0.424854247)
(-1.701696021,0.408737157)
(-1.570796327,0.39210599)
(-1.439896633,0.375066643)
(-1.308996939,0.357012573)
(-1.178097245,0.337578451)
(-1.047197551,0.316980975)
(-0.916297857,0.293893875)
(-0.785398163,0.268377633)
(-0.654498469,0.239372421)
(-0.523598776,0.20612877)
(-0.392699082,0.16766741)
(-0.261799388,0.12219328)
(-0.130899694,0.067610005)
(0.130899694,0.06774567)
(0.261799388,0.122616487)
(0.392699082,0.167973317)
(0.523598776,0.206586124)
(0.654498469,0.23988886)
(0.785398163,0.268755122)
(0.916297857,0.294354831)
(1.047197551,0.317182249)
(1.178097245,0.33806055)
(1.308996939,0.357288235)
(1.439896633,0.375175979)
(1.570796327,0.392111962)
(1.701696021,0.408881917)
(1.832595715,0.424964503)
(1.963495408,0.441047129)
(2.094395102,0.457171497)
(2.225294796,0.473455255)
(2.35619449,0.489901419)
(2.487094184,0.506937787)
(2.617993878,0.524577966)
(2.748893572,0.542944275)
(2.879793266,0.562533275)
(3.01069296,0.583204278)
};
    
\end{axis}

\end{tikzpicture}\caption{}\label{M_2O_s1s2_2s3}
     \end{subfigure} \\[1em]

   \rotatebox{90}{\makebox[0pt][l]{\hspace{2.7em}$\boldsymbol{\psi_1 = \psi_2,\ \psi_3 = 0.5\psi_1}$}} &
     \begin{subfigure}[t]{\linewidth}\centering
       \begin{tikzpicture}
\begin{axis}[
    height=5cm,
    width=5.5cm,
    thick,
    xlabel={\textbf{\textit{k}L}},
    ylabel={\textbf{Modulus}},
    xmin=-3.1416, xmax=3.1416,
    ymin=0, ymax=1,
    every x tick/.style={color=black, thick},
    xtick={-3.1416,-1.5708,0,1.5708,3.1416},
    xticklabels={-$\pi$,-$\frac{\pi}{2}$,0,$\frac{\pi}{2}$,$\pi$},
    every y tick/.style={color=black, thick},
    ytick={0,0.25,0.5,0.75,1},
    yticklabels={$0$,$\frac{1}{4}$,$\frac{1}{2}$,$\frac{3}{4}$,$1$},
    legend style={at={(2.5,0.3)}, anchor=east}, 
    legend cell align={left},
]

\addplot[
    only marks,
    color=red,
    mark=*,
    mark size=1pt,
]
coordinates
{
(-3.01069296,	0.781821875)
(-2.879793266,	0.731939668)
(-2.748893572,	0.701034759)
(-2.617993878,	0.683932959)
(-2.487094184,	0.677062544)
(-2.35619449,	0.676222434)
(-2.225294796,	0.680147679)
(-2.094395102,	0.687775403)
(-1.963495408,	0.69795665)
(-1.832595715,	0.710057901)
(-1.701696021,	0.723973437)
(-1.570796327,	0.739219898)
(-1.439896633,	0.756016447)
(-1.308996939,	0.773537195)
(-1.178097245,	0.79231792)
(-1.047197551,	0.81212759)
(-0.916297857,	0.83262579)
(-0.785398163,	0.854042997)
(-0.654498469,	0.876364946)
(-0.523598776,	0.899641117)
(-0.392699082,	0.92343175)
(-0.261799388,	0.948120952)
(-0.130899694,	0.973754874)
(0.130899694,	0.973675583)
(0.261799388,	0.94786713)
(0.392699082,	0.923300357)
(0.523598776,	0.899337256)
(0.654498469,	0.876210508)
(0.785398163,	0.853920867)
(0.916297857,	0.832368964)
(1.047197551,	0.812114396)
(1.178097245,	0.792257698)
(1.308996939,	0.77356387)
(1.439896633,	0.756026277)
(1.570796327,	0.739397848)
(1.701696021,	0.724145218)
(1.832595715,	0.710335571)
(1.963495408,	0.69800598)
(2.094395102,	0.688050846)
(2.225294796,	0.680578883)
(2.35619449,	0.676519288)
(2.487094184,	0.677170755)
(2.617993878,	0.684541999)
(2.748893572,	0.701690304)
(2.879793266,	0.731979019)
(3.01069296,	0.781720421)
};

\addplot[
    only marks,
    color=blue,
    mark=*,
    mark size=1pt,
]
coordinates {
(-3.01069296,	0.086063543)
(-2.879793266,	0.149553487)
(-2.748893572,	0.192778223)
(-2.617993878,	0.220423006)
(-2.487094184,	0.236904948)
(-2.35619449,	0.246407727)
(-2.225294796,	0.250462486)
(-2.094395102,	0.250203063)
(-1.963495408,	0.246819274)
(-1.832595715,	0.241263271)
(-1.701696021,	0.233379556)
(-1.570796327,	0.223866269)
(-1.439896633,	0.212373384)
(-1.308996939,	0.19988919)
(-1.178097245,	0.185748512)
(-1.047197551,	0.1701043)
(-0.916297857,	0.153559346)
(-0.785398163,	0.135440468)
(-0.654498469,	0.116180035)
(-0.523598776,	0.095606629)
(-0.392699082,	0.073762286)
(-0.261799388,	0.050598831)
(-0.130899694,	0.025937395)
(0.130899694,	0.026005101)
(0.261799388,	0.050851171)
(0.392699082,	0.073892825)
(0.523598776,	0.095796314)
(0.654498469,	0.116450242)
(0.785398163,	0.135747909)
(0.916297857,	0.153745926)
(1.047197551,	0.170199751)
(1.178097245,	0.18572967)
(1.308996939,	0.199791327)
(1.439896633,	0.21235411)
(1.570796327,	0.223654968)
(1.701696021,	0.233189939)
(1.832595715,	0.240970158)
(1.963495408,	0.246810032)
(2.094395102,	0.249851761)
(2.225294796,	0.249899939)
(2.35619449,	0.246026277)
(2.487094184,	0.236812291)
(2.617993878,	0.21972968)
(2.748893572,	0.191984328)
(2.879793266,	0.149508452)
(3.01069296,	0.086177103)
};

    
\addplot[
    only marks,
    color=green,
    mark=*,
    mark size=1pt,
]
coordinates {
(-3.01069296,	0.132114582)
(-2.879793266,	0.118506845)
(-2.748893572,	0.106187017)
(-2.617993878,	0.095644035)
(-2.487094184,	0.086032509)
(-2.35619449,	0.077369839)
(-2.225294796,	0.069389835)
(-2.094395102,	0.062021534)
(-1.963495408,	0.055224077)
(-1.832595715,	0.048678828)
(-1.701696021,	0.042647007)
(-1.570796327,	0.036913833)
(-1.439896633,	0.031610169)
(-1.308996939,	0.026573615)
(-1.178097245,	0.021933568)
(-1.047197551,	0.017768109)
(-0.916297857,	0.013814865)
(-0.785398163,	0.010516535)
(-0.654498469,	0.007455019)
(-0.523598776,	0.004752254)
(-0.392699082,	0.002805963)
(-0.261799388,	0.001280217)
(-0.130899694,	0.000307732)
(0.130899694,	0.000319317)
(0.261799388,	0.001281699)
(0.392699082,	0.002806818)
(0.523598776,	0.00486643)
(0.654498469,	0.00733925)
(0.785398163,	0.010331225)
(0.916297857,	0.013885109)
(1.047197551,	0.017685854)
(1.178097245,	0.022012632)
(1.308996939,	0.026644803)
(1.439896633,	0.031619613)
(1.570796327,	0.036947184)
(1.701696021,	0.042664843)
(1.832595715,	0.048694271)
(1.963495408,	0.055183988)
(2.094395102,	0.062097392)
(2.225294796,	0.069521178)
(2.35619449,	0.077454435)
(2.487094184,	0.086016954)
(2.617993878,	0.095728321)
(2.748893572,	0.106325368)
(2.879793266,	0.118512529)
(3.01069296,	0.132102477)
};
    
\end{axis}

\end{tikzpicture}\caption{}\label{M_AC_s1s2_05s3}
     \end{subfigure} &
     \begin{subfigure}[t]{\linewidth}\centering
       \begin{tikzpicture}
\begin{axis}[
    height=5cm,
    width=5.5cm,
    thick,
    xlabel={\textbf{\textit{k}L}},
    ylabel={\textbf{Modulus}},
    xmin=-3.1416, xmax=3.1416,
    ymin=0, ymax=1,
    every x tick/.style={color=black, thick},
    xtick={-3.1416,-1.5708,0,1.5708,3.1416},
    xticklabels={-$\pi$,-$\frac{\pi}{2}$,0,$\frac{\pi}{2}$,$\pi$},
    every y tick/.style={color=black, thick},
    ytick={0,0.25,0.5,0.75,1},
    yticklabels={$0$,$\frac{1}{4}$,$\frac{1}{2}$,$\frac{3}{4}$,$1$},
    legend style={at={(2.5,0.3)}, anchor=east}, 
    legend cell align={left},
]

\addplot[
    only marks,
    color=red,
    mark=*,
    mark size=1pt,
]
coordinates {
(-3.01069296,	0.091934492)
(-2.879793266,	0.154499866)
(-2.748893572,	0.192790003)
(-2.617993878,	0.215129421)
(-2.487094184,	0.226745409)
(-2.35619449,	0.23105435)
(-2.225294796,	0.229281916)
(-2.094395102,	0.225135083)
(-1.963495408,	0.218677707)
(-1.832595715,	0.210795674)
(-1.701696021,	0.200978601)
(-1.570796327,	0.189853359)
(-1.439896633,	0.179744424)
(-1.308996939,	0.167336112)
(-1.178097245,	0.156147155)
(-1.047197551,	0.144298145)
(-0.916297857,	0.1320514)
(-0.785398163,	0.116857994)
(-0.654498469,	0.101544252)
(-0.523598776,	0.085651618)
(-0.392699082,	0.068278534)
(-0.261799388,	0.046208707)
(-0.130899694,	0.024890272)
(0.130899694,	0.025365607)
(0.261799388,	0.046136406)
(0.392699082,	0.067828738)
(0.523598776,	0.086683688)
(0.654498469,	0.102251265)
(0.785398163,	0.117274373)
(0.916297857,	0.130491974)
(1.047197551,	0.143525654)
(1.178097245,	0.157243139)
(1.308996939,	0.169551646)
(1.439896633,	0.179953206)
(1.570796327,	0.19117493)
(1.701696021,	0.200554264)
(1.832595715,	0.209635979)
(1.963495408,	0.218574795)
(2.094395102,	0.225766507)
(2.225294796,	0.230288471)
(2.35619449,	0.230273632)
(2.487094184,	0.226497721)
(2.617993878,	0.215894368)
(2.748893572,	0.193378268)
(2.879793266,	0.154530621)
(3.01069296,	0.092102745)
};

\addplot[
    only marks,
    color=blue,
    mark=*,
    mark size=1pt,
]
coordinates {
(-3.01069296,	0.875198061)
(-2.879793266,	0.788100966)
(-2.748893572,	0.731998977)
(-2.617993878,	0.695587593)
(-2.487094184,	0.672433797)
(-2.35619449,	0.658526349)
(-2.225294796,	0.651983598)
(-2.094395102,	0.649056518)
(-1.963495408,	0.649583734)
(-1.832595715,	0.652138029)
(-1.701696021,	0.65779112)
(-1.570796327,	0.665030687)
(-1.439896633,	0.672693532)
(-1.308996939,	0.683742195)
(-1.178097245,	0.695379411)
(-1.047197551,	0.709271586)
(-0.916297857,	0.725834113)
(-0.785398163,	0.748281638)
(-0.654498469,	0.774357988)
(-0.523598776,	0.805192338)
(-0.392699082,	0.841711319)
(-0.261799388,	0.888332731)
(-0.130899694,	0.939999514)
(0.130899694,	0.939148968)
(0.261799388,	0.888422589)
(0.392699082,	0.842127461)
(0.523598776,	0.803537361)
(0.654498469,	0.772886357)
(0.785398163,	0.747197835)
(0.916297857,	0.727254123)
(1.047197551,	0.709515443)
(1.178097245,	0.694033961)
(1.308996939,	0.681511947)
(1.439896633,	0.67250796)
(1.570796327,	0.663603348)
(1.701696021,	0.657912358)
(1.832595715,	0.653407556)
(1.963495408,	0.649282712)
(2.094395102,	0.64843369)
(2.225294796,	0.650956411)
(2.35619449,	0.659020654)
(2.487094184,	0.672413834)
(2.617993878,	0.694345347)
(2.748893572,	0.731000147)
(2.879793266,	0.788297607)
(3.01069296,	0.8748531)
};
    
\addplot[
    only marks,
    color=green,
    mark=*,
    mark size=1pt,
]
coordinates {
(-3.01069296,	0.032867447)
(-2.879793266,	0.057399169)
(-2.748893572,	0.075211021)
(-2.617993878,	0.089282986)
(-2.487094184,	0.100820794)
(-2.35619449,	0.110419302)
(-2.225294796,	0.118734486)
(-2.094395102,	0.125808399)
(-1.963495408,	0.131738559)
(-1.832595715,	0.137066297)
(-1.701696021,	0.141230278)
(-1.570796327,	0.145115954)
(-1.439896633,	0.147562044)
(-1.308996939,	0.148921694)
(-1.178097245,	0.148473434)
(-1.047197551,	0.146430268)
(-0.916297857,	0.142114487)
(-0.785398163,	0.134860368)
(-0.654498469,	0.12409776)
(-0.523598776,	0.109156044)
(-0.392699082,	0.090010147)
(-0.261799388,	0.065458562)
(-0.130899694,	0.035110214)
(0.130899694,	0.035485425)
(0.261799388,	0.065441005)
(0.392699082,	0.090043801)
(0.523598776,	0.109778951)
(0.654498469,	0.124862378)
(0.785398163,	0.135527792)
(0.916297857,	0.142253903)
(1.047197551,	0.146958902)
(1.178097245,	0.1487229)
(1.308996939,	0.148936407)
(1.439896633,	0.147538833)
(1.570796327,	0.145221721)
(1.701696021,	0.141533378)
(1.832595715,	0.136956465)
(1.963495408,	0.132142493)
(2.094395102,	0.125799803)
(2.225294796,	0.118755118)
(2.35619449,	0.110705714)
(2.487094184,	0.101088445)
(2.617993878,	0.089760285)
(2.748893572,	0.075621584)
(2.879793266,	0.057171772)
(3.01069296,	0.033044155)
};
    
\end{axis}

\end{tikzpicture}\caption{}\label{M_1O_s1s2_05s3}
     \end{subfigure} &
     \begin{subfigure}[t]{\linewidth}\centering
       \begin{tikzpicture}
\begin{axis}[
    height=5cm,
    width=5.5cm,
    thick,
    xlabel={\textbf{\textit{k}L}},
    ylabel={\textbf{Modulus}},
    xmin=-3.1416, xmax=3.1416,
    ymin=0, ymax=1,
    every x tick/.style={color=black, thick},
    xtick={-3.1416,-1.5708,0,1.5708,3.1416},
    xticklabels={-$\pi$,-$\frac{\pi}{2}$,0,$\frac{\pi}{2}$,$\pi$},
    every y tick/.style={color=black, thick},
    ytick={0,0.25,0.5,0.75,1},
    yticklabels={$0$,$\frac{1}{4}$,$\frac{1}{2}$,$\frac{3}{4}$,$1$},
    legend style={at={(2.5,0.3)}, anchor=east}, 
    legend cell align={left},
]

\addplot[
    only marks,
    color=red,
    mark=*,
    mark size=1pt,
]
coordinates
{
(-3.01069296,	0.143718597)
(-2.879793266,	0.140604668)
(-2.748893572,	0.136622313)
(-2.617993878,	0.132608606)
(-2.487094184,	0.127998518)
(-2.35619449,	0.122570932)
(-2.225294796,	0.116728594)
(-2.094395102,	0.110990015)
(-1.963495408,	0.099041667)
(-1.832595715,	0.096905321)
(-1.701696021,	0.090359699)
(-1.570796327,	0.082851713)
(-1.439896633,	0.072236255)
(-1.308996939,	0.066296474)
(-1.178097245,	0.058129626)
(-1.047197551,	0.050224296)
(-0.916297857,	0.041434844)
(-0.785398163,	0.032856076)
(-0.654498469,	0.025006979)
(-0.523598776,	0.017390865)
(-0.392699082,	0.009944119)
(-0.261799388,	0.005635011)
(-0.130899694,	0.001806971)
(0.130899694,	0.000985028)
(0.261799388,	0.005192046)
(0.392699082,	0.010098251)
(0.523598776,	0.015242069)
(0.654498469,	0.025492395)
(0.785398163,	0.033563739)
(0.916297857,	0.039678871)
(1.047197551,	0.050953946)
(1.178097245,	0.057409973)
(1.308996939,	0.065751971)
(1.439896633,	0.072227024)
(1.570796327,	0.083340304)
(1.701696021,	0.09029229)
(1.832595715,	0.096962923)
(1.963495408,	0.099084498)
(2.094395102,	0.110962798)
(2.225294796,	0.116774598)
(2.35619449,	0.122368169)
(2.487094184,	0.127610818)
(2.617993878,	0.132603327)
(2.748893572,	0.136583098)
(2.879793266,	0.140720826)
(3.01069296,	0.143810637)
};

\addplot[
    only marks,
    color=blue,
    mark=*,
    mark size=1pt,
]
coordinates 
{
(-3.01069296,	0.01665557)
(-2.879793266,	0.032220485)
(-2.748893572,	0.047487622)
(-2.617993878,	0.061722903)
(-2.487094184,	0.075733116)
(-2.35619449,	0.088935026)
(-2.225294796,	0.100724345)
(-2.094395102,	0.11207241)
(-1.963495408,	0.127453599)
(-1.832595715,	0.132132866)
(-1.701696021,	0.140797069)
(-1.570796327,	0.147057189)
(-1.439896633,	0.154634135)
(-1.308996939,	0.156166547)
(-1.178097245,	0.156514518)
(-1.047197551,	0.155984565)
(-0.916297857,	0.15141909)
(-0.785398163,	0.144114009)
(-0.654498469,	0.132290902)
(-0.523598776,	0.11614973)
(-0.392699082,	0.094949854)
(-0.261799388,	0.056601943)
(-0.130899694,	0.030137873)
(0.130899694,	0.029478141)
(0.261799388,	0.05621379)
(0.392699082,	0.093982802)
(0.523598776,	0.11591631)
(0.654498469,	0.131388046)
(0.785398163,	0.143077018)
(0.916297857,	0.151394717)
(1.047197551,	0.155300333)
(1.178097245,	0.156929394)
(1.308996939,	0.155467706)
(1.439896633,	0.154589778)
(1.570796327,	0.147528648)
(1.701696021,	0.140116379)
(1.832595715,	0.131914873)
(1.963495408,	0.12737804)
(2.094395102,	0.111955754)
(2.225294796,	0.100575779)
(2.35619449,	0.088175391)
(2.487094184,	0.075074792)
(2.617993878,	0.061374706)
(2.748893572,	0.046875348)
(2.879793266,	0.03171072)
(3.01069296,	0.015979472)
};
    
\addplot[
    only marks,
    color=green,
    mark=*,
    mark size=1pt,
]
coordinates {
(-3.01069296,	0.839625832)
(-2.879793266,	0.827174846)
(-2.748893572,	0.815890065)
(-2.617993878,	0.805668491)
(-2.487094184,	0.796268367)
(-2.35619449,	0.788494042)
(-2.225294796,	0.782547061)
(-2.094395102,	0.776937574)
(-1.963495408,	0.773504734)
(-1.832595715,	0.770961812)
(-1.701696021,	0.770843232)
(-1.570796327,	0.770091099)
(-1.439896633,	0.77012961)
(-1.308996939,	0.777536979)
(-1.178097245,	0.784355856)
(-1.047197551,	0.793791138)
(-0.916297857,	0.807146065)
(-0.785398163,	0.823029915)
(-0.654498469,	0.842702119)
(-0.523598776,	0.866459405)
(-0.392699082,	0.895106027)
(-0.261799388,	0.937763045)
(-0.130899694,	0.968055156)
(0.130899694,	0.969536831)
(0.261799388,	0.938594164)
(0.392699082,	0.895918947)
(0.523598776,	0.868841621)
(0.654498469,	0.843119559)
(0.785398163,	0.823359243)
(0.916297857,	0.808926412)
(1.047197551,	0.793745721)
(1.178097245,	0.785660632)
(1.308996939,	0.778780324)
(1.439896633,	0.770183198)
(1.570796327,	0.769131048)
(1.701696021,	0.769591331)
(1.832595715,	0.771122204)
(1.963495408,	0.773537462)
(2.094395102,	0.777081447)
(2.225294796,	0.782649623)
(2.35619449,	0.789456441)
(2.487094184,	0.79731439)
(2.617993878,	0.806021966)
(2.748893572,	0.816541553)
(2.879793266,	0.827568455)
(3.01069296,	0.840209891)
};
    
\end{axis}

\end{tikzpicture}\caption{}\label{M_2O_s1s2_05s3}
     \end{subfigure} \\[1em]

   \rotatebox{90}{\makebox[0pt][l]{\hspace{3.2em}$\boldsymbol{\psi_2 = \psi_3,\ \psi_1 = 2\psi_2}$}} &
     \begin{subfigure}[t]{\linewidth}\centering
       \begin{tikzpicture}
\begin{axis}[
    height=5cm,
    width=5.5cm,
    thick,
    xlabel={\textbf{\textit{k}L}},
    ylabel={\textbf{Modulus}},
    xmin=-3.1416, xmax=3.1416,
    ymin=0, ymax=1,
    every x tick/.style={color=black, thick},
    xtick={-3.1416,-1.5708,0,1.5708,3.1416},
    xticklabels={-$\pi$,-$\frac{\pi}{2}$,0,$\frac{\pi}{2}$,$\pi$},
    every y tick/.style={color=black, thick},
    ytick={0,0.25,0.5,0.75,1},
    yticklabels={$0$,$\frac{1}{4}$,$\frac{1}{2}$,$\frac{3}{4}$,$1$},
    legend style={at={(2.5,0.3)}, anchor=east}, 
    legend cell align={left},
]

\addplot[
    only marks,
    color=red,
    mark=*,
    mark size=1pt,
]
coordinates {
(-3.01069296,	0.494682827)
(-2.879793266,	0.47477892)
(-2.748893572,	0.466463241)
(-2.617993878,	0.470626062)
(-2.487094184,	0.485488429)
(-2.35619449,	0.508847293)
(-2.225294796,	0.538158415)
(-2.094395102,	0.572034508)
(-1.963495408,	0.607858124)
(-1.832595715,	0.643667023)
(-1.701696021,	0.677877213)
(-1.570796327,	0.708691812)
(-1.439896633,	0.735040857)
(-1.308996939,	0.75692129)
(-1.178097245,	0.77405041)
(-1.047197551,	0.788928459)
(-0.916297857,	0.802966144)
(-0.785398163,	0.818244345)
(-0.654498469,	0.83776276)
(-0.523598776,	0.864399828)
(-0.392699082,	0.899961438)
(-0.261799388,	0.944406959)
(-0.130899694,	0.94960855)
(0.130899694,	0.949163787)
(0.261799388,	0.944593422)
(0.392699082,	0.899385382)
(0.523598776,	0.864083015)
(0.654498469,	0.837534748)
(0.785398163,	0.817950124)
(0.916297857,	0.802902859)
(1.047197551,	0.788878633)
(1.178097245,	0.774082077)
(1.308996939,	0.756963938)
(1.439896633,	0.735129337)
(1.570796327,	0.708954965)
(1.701696021,	0.678213179)
(1.832595715,	0.644027174)
(1.963495408,	0.608111938)
(2.094395102,	0.572152905)
(2.225294796,	0.538544594)
(2.35619449,	0.508941438)
(2.487094184,	0.485483947)
(2.617993878,	0.470748251)
(2.748893572,	0.466737071)
(2.879793266,	0.475017858)
(3.01069296,	0.494467971)
};

\addplot[
    only marks,
    color=blue,
    mark=*,
    mark size=1pt,
]
coordinates {
(-3.01069296,	0.332617213)
(-2.879793266,	0.364861072)
(-2.748893572,	0.383340982)
(-2.617993878,	0.387417325)
(-2.487094184,	0.380325339)
(-2.35619449,	0.364877624)
(-2.225294796,	0.344076527)
(-2.094395102,	0.319745235)
(-1.963495408,	0.293922063)
(-1.832595715,	0.268526742)
(-1.701696021,	0.244808308)
(-1.570796327,	0.224084134)
(-1.439896633,	0.207017484)
(-1.308996939,	0.193266327)
(-1.178097245,	0.18276494)
(-1.047197551,	0.173200953)
(-0.916297857,	0.163229829)
(-0.785398163,	0.15138304)
(-0.654498469,	0.135438257)
(-0.523598776,	0.11311556)
(-0.392699082,	0.083573221)
(-0.261799388,	0.046198116)
(-0.130899694,	0.042365443)
(0.130899694,	0.042529339)
(0.261799388,	0.046237399)
(0.392699082,	0.083885378)
(0.523598776,	0.113444092)
(0.654498469,	0.1356143)
(0.785398163,	0.151634951)
(0.916297857,	0.163290345)
(1.047197551,	0.173227292)
(1.178097245,	0.182753172)
(1.308996939,	0.193260813)
(1.439896633,	0.206974057)
(1.570796327,	0.223860555)
(1.701696021,	0.244491671)
(1.832595715,	0.268160102)
(1.963495408,	0.29370765)
(2.094395102,	0.31956166)
(2.225294796,	0.34380401)
(2.35619449,	0.364850758)
(2.487094184,	0.380247254)
(2.617993878,	0.387286118)
(2.748893572,	0.383074397)
(2.879793266,	0.364711444)
(3.01069296,	0.332863107)
};
    
\addplot[
    only marks,
    color=green,
    mark=*,
    mark size=1pt,
]
coordinates {
(-3.01069296,	0.17269996)
(-2.879793266,	0.160360008)
(-2.748893572,	0.150195776)
(-2.617993878,	0.141956612)
(-2.487094184,	0.134186232)
(-2.35619449,	0.126275083)
(-2.225294796,	0.117765058)
(-2.094395102,	0.108220257)
(-1.963495408,	0.098219813)
(-1.832595715,	0.087806234)
(-1.701696021,	0.077314479)
(-1.570796327,	0.067224054)
(-1.439896633,	0.05794166)
(-1.308996939,	0.049812383)
(-1.178097245,	0.04318465)
(-1.047197551,	0.037870588)
(-0.916297857,	0.033804027)
(-0.785398163,	0.030372615)
(-0.654498469,	0.026798983)
(-0.523598776,	0.022484612)
(-0.392699082,	0.016465341)
(-0.261799388,	0.009394925)
(-0.130899694,	0.008026007)
(0.130899694,	0.008306874)
(0.261799388,	0.009169179)
(0.392699082,	0.01672924)
(0.523598776,	0.022472893)
(0.654498469,	0.026850953)
(0.785398163,	0.030414925)
(0.916297857,	0.033806796)
(1.047197551,	0.037894075)
(1.178097245,	0.043164752)
(1.308996939,	0.049775249)
(1.439896633,	0.057896606)
(1.570796327,	0.06718448)
(1.701696021,	0.07729515)
(1.832595715,	0.087812724)
(1.963495408,	0.098180412)
(2.094395102,	0.108285435)
(2.225294796,	0.117651397)
(2.35619449,	0.126207804)
(2.487094184,	0.134268799)
(2.617993878,	0.141965631)
(2.748893572,	0.150188532)
(2.879793266,	0.160270698)
(3.01069296,	0.172668922)
};
    
\end{axis}

\end{tikzpicture}\caption{}\label{M_AC_2s1_s2s3}
     \end{subfigure} &
     \begin{subfigure}[t]{\linewidth}\centering
       \begin{tikzpicture}
\begin{axis}[
    height=5cm,
    width=5.5cm,
    thick,
    xlabel={\textbf{\textit{k}L}},
    ylabel={\textbf{Modulus}},
    xmin=-3.1416, xmax=3.1416,
    ymin=0, ymax=1,
    every x tick/.style={color=black, thick},
    xtick={-3.1416,-1.5708,0,1.5708,3.1416},
    xticklabels={-$\pi$,-$\frac{\pi}{2}$,0,$\frac{\pi}{2}$,$\pi$},
    every y tick/.style={color=black, thick},
    ytick={0,0.25,0.5,0.75,1},
    yticklabels={$0$,$\frac{1}{4}$,$\frac{1}{2}$,$\frac{3}{4}$,$1$},
    legend style={at={(2.5,0.3)}, anchor=east}, 
    legend cell align={left},
]

\addplot[
    only marks,
    color=red,
    mark=*,
    mark size=1pt,
]
coordinates {
(-3.01069296,	0.372655103)
(-2.879793266,	0.382564929)
(-2.748893572,	0.391677146)
(-2.617993878,	0.393842031)
(-2.487094184,	0.387511166)
(-2.35619449,	0.369750414)
(-2.225294796,	0.342721495)
(-2.094395102,	0.308344973)
(-1.963495408,	0.272020038)
(-1.832595715,	0.234278622)
(-1.701696021,	0.201367905)
(-1.570796327,	0.174310357)
(-1.439896633,	0.155521826)
(-1.308996939,	0.140616084)
(-1.178097245,	0.132074344)
(-1.047197551,	0.127818475)
(-0.916297857,	0.122959558)
(-0.785398163,	0.114006748)
(-0.654498469,	0.105795596)
(-0.523598776,	0.09718964)
(-0.392699082,	0.078014391)
(-0.261799388,	0.056803578)
(-0.130899694,	0.033933538)
(0.130899694,	0.034279863)
(0.261799388,	0.058086477)
(0.392699082,	0.077225352)
(0.523598776,	0.093979324)
(0.654498469,	0.1092187)
(0.785398163,	0.117298205)
(0.916297857,	0.123100723)
(1.047197551,	0.127751932)
(1.178097245,	0.13385671)
(1.308996939,	0.142541009)
(1.439896633,	0.155544604)
(1.570796327,	0.176939094)
(1.701696021,	0.203505671)
(1.832595715,	0.235956772)
(1.963495408,	0.270631451)
(2.094395102,	0.308888255)
(2.225294796,	0.344326112)
(2.35619449,	0.370602318)
(2.487094184,	0.387547676)
(2.617993878,	0.393592098)
(2.748893572,	0.39192059)
(2.879793266,	0.386964344)
(3.01069296,	0.381148297)
};

\addplot[
    only marks,
    color=blue,
    mark=*,
    mark size=1pt,
]
coordinates {
(-3.01069296,	0.575087623)
(-2.879793266,	0.543484073)
(-2.748893572,	0.509766337)
(-2.617993878,	0.480390752)
(-2.487094184,	0.457558338)
(-2.35619449,	0.44686353)
(-2.225294796,	0.447851389)
(-2.094395102,	0.461160471)
(-1.963495408,	0.482582921)
(-1.832595715,	0.509597688)
(-1.701696021,	0.535733835)
(-1.570796327,	0.558393583)
(-1.439896633,	0.576711798)
(-1.308996939,	0.591322252)
(-1.178097245,	0.600078284)
(-1.047197551,	0.603810777)
(-0.916297857,	0.606011755)
(-0.785398163,	0.606552175)
(-0.654498469,	0.604929592)
(-0.523598776,	0.600062917)
(-0.392699082,	0.60320699)
(-0.261799388,	0.606489184)
(-0.130899694,	0.614751716)
(0.130899694,	0.614813103)
(0.261799388,	0.605497584)
(0.392699082,	0.603894021)
(0.523598776,	0.602293489)
(0.654498469,	0.602446109)
(0.785398163,	0.604198001)
(0.916297857,	0.605812928)
(1.047197551,	0.603777973)
(1.178097245,	0.598757834)
(1.308996939,	0.5899296)
(1.439896633,	0.576664708)
(1.570796327,	0.556531329)
(1.701696021,	0.53418972)
(1.832595715,	0.508284616)
(1.963495408,	0.483913917)
(2.094395102,	0.460688363)
(2.225294796,	0.446315036)
(2.35619449,	0.446162346)
(2.487094184,	0.457540006)
(2.617993878,	0.480527096)
(2.748893572,	0.509602095)
(2.879793266,	0.540318062)
(3.01069296,	0.568950543)
};
    
\addplot[
    only marks,
    color=green,
    mark=*,
    mark size=1pt,
]
coordinates {
(-3.01069296,	0.052257274)
(-2.879793266,	0.073950998)
(-2.748893572,	0.098556517)
(-2.617993878,	0.125767216)
(-2.487094184,	0.154930496)
(-2.35619449,	0.183386056)
(-2.225294796,	0.209427116)
(-2.094395102,	0.230494556)
(-1.963495408,	0.245397041)
(-1.832595715,	0.25612369)
(-1.701696021,	0.262898259)
(-1.570796327,	0.26729606)
(-1.439896633,	0.267766377)
(-1.308996939,	0.268061664)
(-1.178097245,	0.267847373)
(-1.047197551,	0.268370748)
(-0.916297857,	0.271028687)
(-0.785398163,	0.279441077)
(-0.654498469,	0.289274812)
(-0.523598776,	0.302747443)
(-0.392699082,	0.318778619)
(-0.261799388,	0.336707238)
(-0.130899694,	0.351314747)
(0.130899694,	0.350907034)
(0.261799388,	0.336415939)
(0.392699082,	0.318880626)
(0.523598776,	0.303727187)
(0.654498469,	0.288335191)
(0.785398163,	0.278503794)
(0.916297857,	0.271086349)
(1.047197551,	0.268470095)
(1.178097245,	0.267385456)
(1.308996939,	0.267529391)
(1.439896633,	0.267790688)
(1.570796327,	0.266529577)
(1.701696021,	0.262304609)
(1.832595715,	0.255758611)
(1.963495408,	0.245454632)
(2.094395102,	0.230423383)
(2.225294796,	0.209358852)
(2.35619449,	0.183235336)
(2.487094184,	0.154912318)
(2.617993878,	0.125880806)
(2.748893572,	0.098477315)
(2.879793266,	0.072717594)
(3.01069296,	0.04990116)
};
    
\end{axis}

\end{tikzpicture}\caption{}\label{M_1O_2s1_s2s3}
     \end{subfigure} &
     \begin{subfigure}[t]{\linewidth}\centering
       \begin{tikzpicture}
\begin{axis}[
    height=5cm,
    width=5.5cm,
    thick,
    xlabel={\textbf{\textit{k}L}},
    ylabel={\textbf{Modulus}},
    xmin=-3.1416, xmax=3.1416,
    ymin=0, ymax=1,
    every x tick/.style={color=black, thick},
    xtick={-3.1416,-1.5708,0,1.5708,3.1416},
    xticklabels={-$\pi$,-$\frac{\pi}{2}$,0,$\frac{\pi}{2}$,$\pi$},
    every y tick/.style={color=black, thick},
    ytick={0,0.25,0.5,0.75,1},
    yticklabels={$0$,$\frac{1}{4}$,$\frac{1}{2}$,$\frac{3}{4}$,$1$},
    legend style={at={(2.5,0.3)}, anchor=east}, 
    legend cell align={left},
]

\addplot[
    only marks,
    color=red,
    mark=*,
    mark size=1pt,
]
coordinates {
(-3.01069296,	0.155346272)
(-2.879793266,	0.155661525)
(-2.748893572,	0.155801647)
(-2.617993878,	0.154786042)
(-2.487094184,	0.152555108)
(-2.35619449,	0.148325818)
(-2.225294796,	0.143385296)
(-2.094395102,	0.136055996)
(-1.963495408,	0.127427464)
(-1.832595715,	0.117852071)
(-1.701696021,	0.107507276)
(-1.570796327,	0.09578562)
(-1.439896633,	0.084866551)
(-1.308996939,	0.073355985)
(-1.178097245,	0.062185379)
(-1.047197551,	0.053206805)
(-0.916297857,	0.045472204)
(-0.785398163,	0.039723738)
(-0.654498469,	0.035131235)
(-0.523598776,	0.03220696)
(-0.392699082,	0.025369325)
(-0.261799388,	0.019893311)
(-0.130899694,	0.011647309)
(0.130899694,	0.012357727)
(0.261799388,	0.019922176)
(0.392699082,	0.026304566)
(0.523598776,	0.030494871)
(0.654498469,	0.035670341)
(0.785398163,	0.040234382)
(0.916297857,	0.045344746)
(1.047197551,	0.052660567)
(1.178097245,	0.062609029)
(1.308996939,	0.073689647)
(1.439896633,	0.084980795)
(1.570796327,	0.096185899)
(1.701696021,	0.107746553)
(1.832595715,	0.118061329)
(1.963495408,	0.127392067)
(2.094395102,	0.13610541)
(2.225294796,	0.143466967)
(2.35619449,	0.148486494)
(2.487094184,	0.152517832)
(2.617993878,	0.154645913)
(2.748893572,	0.155807638)
(2.879793266,	0.155654764)
(3.01069296,	0.155444068)
};

\addplot[
    only marks,
    color=blue,
    mark=*,
    mark size=1pt,
]
coordinates {
(-3.01069296,	0.148443227)
(-2.879793266,	0.148412962)
(-2.748893572,	0.14930932)
(-2.617993878,	0.151131944)
(-2.487094184,	0.15495325)
(-2.35619449,	0.160450258)
(-2.225294796,	0.168996365)
(-2.094395102,	0.18016911)
(-1.963495408,	0.193660537)
(-1.832595715,	0.210331492)
(-1.701696021,	0.228901485)
(-1.570796327,	0.24960829)
(-1.439896633,	0.270821896)
(-1.308996939,	0.292285708)
(-1.178097245,	0.313292281)
(-1.047197551,	0.332278564)
(-0.916297857,	0.348171358)
(-0.785398163,	0.36022882)
(-0.654498469,	0.367668413)
(-0.523598776,	0.370208372)
(-0.392699082,	0.369881407)
(-0.261799388,	0.366212837)
(-0.130899694,	0.364320525)
(0.130899694,	0.363653699)
(0.261799388,	0.366253911)
(0.392699082,	0.369373251)
(0.523598776,	0.37100083)
(0.654498469,	0.367370382)
(0.785398163,	0.359955158)
(0.916297857,	0.348201786)
(1.047197551,	0.332483311)
(1.178097245,	0.313053606)
(1.308996939,	0.292118129)
(1.439896633,	0.270707705)
(1.570796327,	0.249414629)
(1.701696021,	0.228744733)
(1.832595715,	0.210147372)
(1.963495408,	0.193844825)
(2.094395102,	0.179851242)
(2.225294796,	0.169058777)
(2.35619449,	0.160628553)
(2.487094184,	0.154666778)
(2.617993878,	0.151230331)
(2.748893572,	0.149296313)
(2.879793266,	0.148561878)
(3.01069296,	0.148337501)
};
    
\addplot[
    only marks,
    color=green,
    mark=*,
    mark size=1pt,
]
coordinates {
(-3.01069296,	0.696210501)
(-2.879793266,	0.695925513)
(-2.748893572,	0.694889033)
(-2.617993878,	0.694082014)
(-2.487094184,	0.692491642)
(-2.35619449,	0.691223924)
(-2.225294796,	0.687618339)
(-2.094395102,	0.683774894)
(-1.963495408,	0.678911998)
(-1.832595715,	0.671816437)
(-1.701696021,	0.663591239)
(-1.570796327,	0.65460609)
(-1.439896633,	0.644311553)
(-1.308996939,	0.634358307)
(-1.178097245,	0.62452234)
(-1.047197551,	0.61451463)
(-0.916297857,	0.606356438)
(-0.785398163,	0.600047442)
(-0.654498469,	0.597200352)
(-0.523598776,	0.597584668)
(-0.392699082,	0.604749267)
(-0.261799388,	0.613893852)
(-0.130899694,	0.624032166)
(0.130899694,	0.623988575)
(0.261799388,	0.613823914)
(0.392699082,	0.604322183)
(0.523598776,	0.598504299)
(0.654498469,	0.596959277)
(0.785398163,	0.599810459)
(0.916297857,	0.606453468)
(1.047197551,	0.614856122)
(1.178097245,	0.624337365)
(1.308996939,	0.634192224)
(1.439896633,	0.644311501)
(1.570796327,	0.654399473)
(1.701696021,	0.663508714)
(1.832595715,	0.671791299)
(1.963495408,	0.678763108)
(2.094395102,	0.684043348)
(2.225294796,	0.687474256)
(2.35619449,	0.690884953)
(2.487094184,	0.69281539)
(2.617993878,	0.694123755)
(2.748893572,	0.694896049)
(2.879793266,	0.695783358)
(3.01069296,	0.696218431)
};
    
\end{axis}

\end{tikzpicture}\caption{}\label{M_2O_2s1_s2s3}
     \end{subfigure} \\[1.2em]

   & \multicolumn{3}{c}{
     \begin{tikzpicture}
       \begin{axis}[
         hide axis, xmin=0,xmax=1,ymin=0,ymax=1,
         legend columns=3, legend cell align=left,
         legend style={
           /tikz/every even column/.append style={column sep=1.5em}
         }
       ]
         \addlegendimage{only marks, mark=*, mark size=2.5pt, red}
         \addlegendentry{$\lvert\hat\alpha\rvert$}
         \addlegendimage{only marks, mark=*, mark size=2.5pt, blue}
         \addlegendentry{$\lvert\hat\beta\rvert$}
         \addlegendimage{only marks, mark=*, mark size=2.5pt, green}
         \addlegendentry{$\lvert\hat\gamma\rvert$}
       \end{axis}
     \end{tikzpicture}
   }
\end{tabular}

\caption{Evolution of the three normalized modal amplitudes across the Brillouin zone for the three-mass lattice under different stiffness orderings. The inversion-symmetric cases (i) $\psi_1=\psi_2,\ \psi_3=2\psi_1$ and (ii) $\psi_1 = \psi_2,\ \psi_3 = 0.5\psi_1$  produce coefficients ($\hat\alpha,\hat\beta,\hat\gamma$) trends, with each band exhibiting a characteristic dominant component that cleanly distinguishes the acoustic and optical branches. When inversion symmetry is broken (iii) $\psi_2 = \psi_3,\ \psi_1 = 2\psi_2$, this ordering disappears and the bands display smoothly varying mixtures of $\hat\alpha$, $\hat\beta$, and $\hat\gamma$, reflecting the loss of symmetry-imposed structure.}

\label{Fig:Modulation_3Mass}
\end{figure}

From Fig.~\ref{Fig:Modulation_3Mass}, we see that the domination of the complex coefficients depends on the set of parameters. Across all three parameter sets, the acoustic branch is dominated by $\hat\alpha$. We can see an interchangeable effect in the 1st and 2nd optical branches in the case of the parameter set case (i) $\psi_1=\psi_2, \psi_3=2\psi_1$ and (ii) $\psi_1=\psi_2, \psi_3=0.5\psi_1$. These two conditions of stiffness show a shift in the domination of the complex coefficients $\lvert\hat\beta\rvert$ and $\lvert\hat\gamma\rvert$. This inversion in the 1st and 2nd optical shows a topological transition within the branches in the Hilbert space. And, across different parameters, we can see the clear separation of complex coefficients $\lvert\hat\alpha\rvert$, $\lvert\hat\beta\rvert$ and $\lvert\hat\gamma\rvert$. This clear separation---visible in Fig.~\ref{Fig:Modulation_3Mass} for cases (\ref{M_2O_s1s2_2s3})--(\ref{M_2O_2s1_s2s3})---provides an unambiguous, stiffness-agnostic labeling of the three branches that we will use to compare the phase evolution below. The continuity of these modulus trends through the Brillouin zone is also consistent with the smooth evolution of the Bloch eigenvectors away from high-symmetry points. We will focus on the acoustic branch for the set parameters.

\subsection{Spectral Analysis}
Normal modes describe synchronous oscillations in which all parts of the system move sinusoidally at a common frequency with fixed phase relations. An elastic system, consisting of masses coupled by harmonic springs, possesses characteristic resonance frequencies. When initial displacements and velocities are applied, the system first exhibits transient dynamics before settling into a steady state over time.  

We define the displacement from equilibrium as a time-dependent function across $N_c$ unit cells, each containing $N_m$ masses. For a given wave number $k$, the system can support multiple frequencies, denoted $\omega_j(k)$, where the index $j$ labels the bands in ascending order, with $j=1$ corresponding to the lowest frequency. The complex amplitude of oscillation of the $n$-th mass on the $j$-th band, $A_{n,j}(k)$, is obtained by projecting the displacement $u_{n,N_i}(t)$ onto that band.
\begin{equation}
    A_n(k,\omega_j(k))=\frac{1}{N_c}\sum\limits _{N_{i} =1}^{N_{c}}\frac{1}{\tau _{0}}\int _{0}^{\tau _{0}} u_{n,N_{i}}( t) e^{\mathrm{i}kLN_{i}} e^{-\mathrm{i}\omega _{j}( k) t} dt
\end{equation}
$L$ is the length of the sample and $\tau_0$ is the time over the sample of elastic waves. Also, the phase of the complex amplitude.
\begin{equation}
    \phi_{n,j}(k)=angle(A_{n,j}(k))
\end{equation}

The dispersion relation is computed by molecular dynamics (MD) simulation with initial conditions $u_{n,N_i}(0)=\cos(k N_i L)$ and $\dot{u}_{n,N_i}(0)=0$, together with periodic boundary conditions $e^{ik L N_c}=1$, which imply the discrete grid $kL = 2\pi m/N_c$ (spacing $\Delta(kL)=2\pi/N_c$). In the second Brillouin zone, we take $kL\in[-2\pi,\,2\pi]$. Taking into account all springs and integrating the equations of motion using a Runge–Kutta scheme, we obtain the trajectories $u_{n,N_i}(t)$. The frequency spectrum is then computed via the finite-time Fourier transform from which the peak locations produce the band frequencies $\omega_j(k)$.

\begin{equation}
    \ddot{u}_{n,N_i}(\omega)=\frac{1}{\tau_0}\int_0^{\tau_0}u_{n,N_i}(t)e^{-\mathrm{i}\omega t}dt
\end{equation}

Within the second Brillouin zone, the elastic band structure is generated for all values of the wave number $k$. Using the Fourier transform, the frequency of the band structure is calculated at the peak position at a given k, giving $\omega_j(k)$.

In the second set of MD simulations, the band structure is used to prescribe the initial conditions. We take $u_{n,N_i}(0)=\cos(k N_i L)$ and $\dot{u}_{n,N_i}(0)=-\,\omega_j(k)\,\sin(k N_i L)$. This sets the velocity consistent with the computed band structure. Each MD simulation yields the displacement $u_{n,N_i}(t)$, which is then used to compute the complex amplitude $A_{n,j}(k)$ of the mass $n-th$ and its phase $\phi_{n,j}(k)$.

\subsection{Berry Vector and Berry Phase}
Berry vector for a finite number of unit cells $N_c$, can be characterized by,

\begin{equation}
    B_V(k)=\sum_{n=1}^{N_m}\tilde{A}^*_{n,j}(k)\tilde{A}_{n,j}(k+\Delta k)\equiv\hat\alpha^*(k)\hat\alpha(k+\Delta k)+\hat\beta^*(k)\hat\beta(k+\Delta k)+...
\end{equation}
Where $\Delta kL=2\pi/N_c$, and $\tilde{A}_{n,j} (k)$ is the normalized complex amplitude of the unit cell mass $n-th$ in band $j$. This amplitude vector creates a variation of the Berry connection along some path in the complex space as $k$ varies. For possible $k$ values of some specific band in the closed path, summing the Berry connection defined by the second Brillouin zone gives the Berry or Zak phase. And each band Berry phases are defined by
\begin{equation}
   \chi=-Im\left[\ln\left(\prod_{i=1}^{N_c-1}B_v(k_i)\right)\right]mod(2\pi)
\end{equation}

The system is discretized at a finite length and $k$. In the case of an infinite length of continuous $k$, a differential form of the expression is used. The Berry phase is the sum of the unit vector amplitude over the entire manifold for a closed path. The evolution of the unit vector of amplitude in the $N_m$ dimensional space parametrized by the wave number $k$ generates a multiplicity.

The Berry phase is dependent on the existence of the closed path in the $k$-space along with the amplitude $\tilde{A}_{n,j}$ being periodic in reciprocal space, which can be affected by the ansatz required to define the system dynamics.

\subsection{Spectrum Analysis \& Molecular Dynamics Simulation}

For a numerical solution, the initial conditions are set as displacement $u_{n,N_i}(0)=\cos(kN_i L)$ and velocity $\ddot{u}_{n,N_i}=0$. for the case of masses. In the first molecular dynamics simulation using the prescribed initial conditions, the dispersion relation is calculated using the temporal integral of the displacement $u_{n,N_i}(t)$ and the band structure $\omega_j(k)$ is measured at the span of the wave number $k$. The wave number $k$ has a linear range in $[-\pi,\pi]$ with an increase of $2\pi/N_c$ in Brillouin zone 1.

\begin{figure}[H]
    \centering
    \begin{tikzpicture}
  \node (start) at (-1.5,0) [startstop] {Start};
  \node (in1)   at (3,0) [io, text width=2.2cm] {Insert $N_c$, $L$, $k$};
  \node (in2)   at (9.5,0) [io, text width=5.7cm] {Set the Initial Condition\\[1ex]$u_{n,N_i}(0)=\cos(kN_iL)$\\$\dot{u}_{n,N_i}(0)=0$};

  \node (pro1) at (11.3,-3)  [process, text width=2cm] {1st MD Simulation};
  \node (pro2) at (7,-3)  [process, text width=4cm] {Obtain Dispersion Relation \\[1ex] Find $\omega_j$ for each $k$};
  \node (in3)  at (0.8,-3)  [io, text width=5.2cm] {Set the Initial Conditions \\[1ex] $u_{n,N_i}(0)=cos(kN_iL)$\\$\dot{u}_{n,N_i}(0)=-\omega_j(k)sin(kN_iL)$};
  \node (pro3) at (0,-6)  [process, text width=3cm] {2nd MD Simulation};
  \node (pro4) at (7.8,-6)  [process, text width=9.5cm] {Projection of Displacement Wave \\[1ex]  $A_n(k,\omega_j(k))=\frac{1}{N_c}\sum\limits _{N_{i} =1}^{N_{c}}\frac{1}{\tau _{0}}\int _{0}^{\tau _{0}} u_{n,N_{i}}( t) e^{\mathrm{i}kLN_{i}} e^{-\mathrm{i}\omega _{j}( k) t} dt$};
  \node (pro5) at (9,-10)  [process, text width=6.5cm] {Compute Coefficient of Superposition of States\\[1ex]$\begin{array}{{>{\displaystyle}l}}
\begin{pmatrix}
u_{1,N_{i}}( k)\\
u_{2,N_{i}}( k)\\
\vdots 
\end{pmatrix} =\begin{pmatrix}
A_{1}( k,\omega _{j}( k))\\
A_{2}( k,\omega _{j}( k))\\
\vdots 
\end{pmatrix} e^{\mathrm{i}\omega t}\\
\qquad\equiv (\hat\alpha E_{1} +\hat\beta E_{2}+...) e^{\mathrm{i}\omega t}
\end{array}$};
  \node (pro6) at (0.8,-10)  [process, text width=6.5cm] {Berry Vector and Berry Phase Calculation \\[1ex] $B_v(k_i)= \hat\alpha^*(k)\hat\alpha(k+\Delta k)$\\$+ \hat\beta^*(k)\hat\beta(k+\Delta k)+ \dots$ \\[1ex] $\chi =-Im\left\{\ln\left[\prod\limits _{i=1}^{N_{c} -1}( B_{v}( k_{i}))\right]\right\}$};
  \node (stop) [startstop, below=of pro6] {Stop};
  
  \draw [arrow] (start) -- (in1);
  \draw [arrow] (in1) -- (in2);
  \draw [arrow] (in2) -| ([xshift=2cm, yshift=1cm]pro1.south) |- (pro1);
  \draw [arrow] (pro1) -- (pro2);
  \draw [arrow] (pro2) -- (in3);
  \draw [arrow] (in3.west) -| ([xshift=-2.9cm, yshift=1cm]pro3.south) |- (pro3);
  \draw [arrow] (pro3) -- (pro4);
  \draw [arrow] (pro4) -| ([xshift=5.5cm, yshift=1cm]pro4.south) |- (pro5);
  \draw [arrow] (pro5) -- (pro6);
  \draw [arrow] (pro6) -- (stop);
  
\end{tikzpicture}
    \caption{Flowchart for 2-Step Molecular Dynamics (MD) simulation and calculations of coefficients of the superposition of states.}
    \label{fig:FlowChart}
\end{figure}
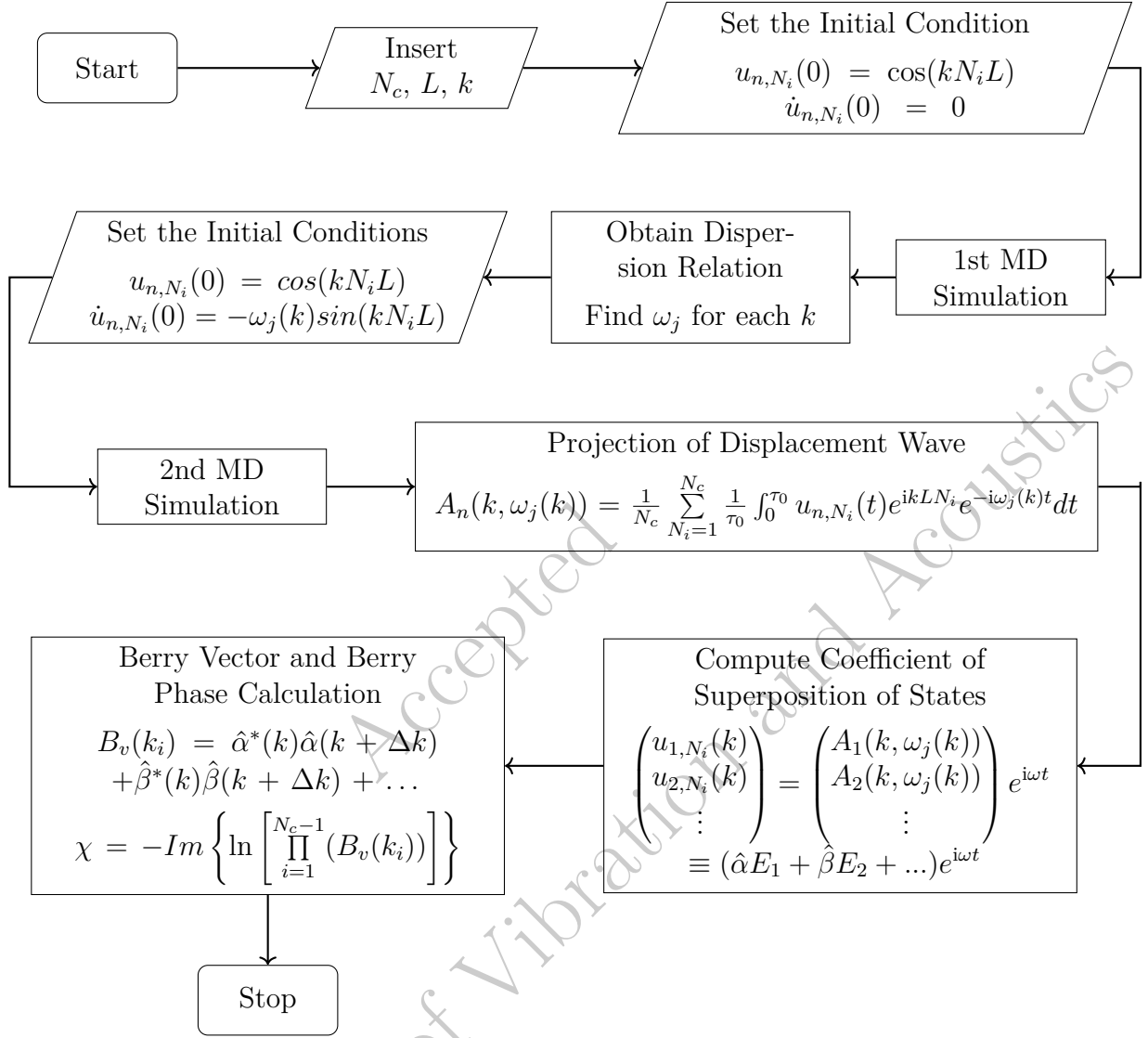

The second set of molecular dynamics simulations has a different initial condition as we have developed the band structure for the system. Using the initial condition as $u_{n,N_i}(0)=\cos(kN_i L)$ and $\dot{u}_{n,N_i}(0)=-\,\omega_j(k)\,\sin(k N_i L)$, the displacement of the traveling wave is obtained through the 2nd MD simulation necessary for the solution of the modulus. This initial condition sets the velocity value to the one prescribed by the band structure. One must use a specified traveling elastic wave initial condition to obtain the wave function band structure and the amplitude and phase of the masses. After the second set of MD simulations, we use the displacement $u_{n,N_i}(t)$ to calculate the complex amplitude, and the phase of the nth mass and the letter use those to calculate the modulus and the modulus's phase to calculate the system state. Geometrical representation by the Bloch sphere is demonstrated for the superposition of the state (Fig. \ref{fig:FlowChart}). 


\end{document}